\documentclass{article}

\usepackage{microtype}
\usepackage{graphicx}
\usepackage{subfigure}
\usepackage{booktabs} 
\usepackage{microtype}
\usepackage{graphicx}
\usepackage{booktabs} 
\usepackage[noend]{algorithmic}
\usepackage{mathrsfs}
\usepackage{amssymb}
\usepackage{amsmath}
\usepackage{wrapfig}
\usepackage{multirow}
\usepackage{hyperref}

\newtheorem{mydef}{Definition}
\newtheorem{theorem}{Theorem}
\newtheorem{lemma}{Lemma}

\newtheorem{proof}{Proof}

\newtheorem{corollary}{Corollary}

\usepackage{tikz}

\usepackage[accepted]{icml2020}
\usepackage{balance}

\icmltitlerunning{Stochastic Batch Mechanism: Scalable Differential Privacy with Certified Robustness in Adversarial Learning}

\begin{document}

\twocolumn[
\icmltitle{Scalable Differential Privacy with Certified Robustness in Adversarial Learning}




\begin{icmlauthorlist}
\icmlauthor{NhatHai Phan}{to}
\icmlauthor{My T. Thai}{goo}
\icmlauthor{Han Hu}{to}
\icmlauthor{Ruoming Jin}{ed}
\icmlauthor{Tong Sun}{adobe}
\icmlauthor{Dejing Dou}{uo,baidu}
\end{icmlauthorlist}

\icmlaffiliation{to}{Ying Wu College of Computing, New Jersey Institute of Technology, Newark, New Jersey, USA}
\icmlaffiliation{goo}{Department of Computer \& Information Sciences \& Engineering, University of Florida, Gainesville, Florida, USA}
\icmlaffiliation{ed}{Computer Science Department, Kent State University, Kent, Ohio, USA}
\icmlaffiliation{adobe}{Adobe Research, San Jose, California, USA}
\icmlaffiliation{uo}{Computer and Information Science Department, University of Oregon, Eugene, Oregon, USA}
\icmlaffiliation{baidu}{(Sabbatical leave from University of Oregon to) Baidu Research, Beijing, China}

\icmlcorrespondingauthor{NhatHai Phan}{phan@njit.edu}

\icmlkeywords{Machine Learning, ICML}

\vskip 0.3in
]



\printAffiliationsAndNotice{}  

\begin{abstract}
In this paper, we aim to develop a scalable algorithm to preserve differential privacy (\textbf{DP}) in adversarial learning for deep neural networks (\textbf{DNNs}), with certified robustness to adversarial examples. By leveraging the sequential composition theory in DP, we randomize both input and latent spaces to strengthen our certified robustness bounds. To address the trade-off among model utility, privacy loss, and robustness, we design an original adversarial objective function, based on the post-processing property in DP, to tighten the sensitivity of our model. A new \textit{stochastic batch training} is proposed to apply our mechanism on large DNNs and datasets, by bypassing the vanilla iterative batch-by-batch training in DP DNNs. An end-to-end theoretical analysis and evaluations show that our mechanism notably improves the robustness and scalability of DP DNNs.
\end{abstract} 

\section{Introduction}

The pervasiveness of machine learning exposes new vulnerabilities in software systems, in which deployed machine learning models can be used (a) to reveal sensitive information in private training data \citep{Fredrikson:2015:MIA}, and/or (b) to make the models misclassify, such as \textit{adversarial examples} \citep{7958570}. 
Efforts to prevent such attacks typically seek one of three solutions: \textbf{(1)} Models which preserve differential privacy (\textbf{DP}) \citep{dwork2006calibrating}, a rigorous formulation of privacy in probabilistic terms; \textbf{(2)} Adversarial training algorithms, which augment training data to consist of benign examples and adversarial examples crafted during the training process, thereby empirically increasing the classification accuracy given adversarial examples \citep{7965897,7965869}; and \textbf{(3)} Certified robustness, in which the model classification given adversarial examples is theoretically guaranteed to be consistent, i.e., a small perturbation in the input does not change the predicted label \citep{pmlr-v70-cisse17a,DBLP:journals/corr/abs-1711-00851,DBLP:journals/corr/abs-1906-04584}.

On the one hand, \textit{private models}, trained with existing privacy-preserving mechanisms \citep{Abadi,ShokriVitaly2015,Phan0WD16,PhanMLJ2017,NHPhanICDM17,Yu2019,Lee:2018:CDP:3219819.3220076}, are unshielded under adversarial examples. On the other hand, \textit{robust models}, trained with adversarial learning (with or without certified robustness to adversarial examples), do not offer privacy protections to the training data \citep{2019arXiv190510291S}. 
That one-sided approach poses serious risks to machine learning-based systems; since adversaries can attack a deployed model by using both privacy inference attacks and adversarial examples. To be safe, a model must be \textit{i) private to protect the training data}, {\bf and} \textit{ii) robust to adversarial examples}. 
Unfortunately, there still lacks of study on how to develop such a model, which thus remains a largely open challenge \citep{PhanIJCAI}.

Simply combining existing DP-preserving mechanisms and certified robustness conditions \citep{pmlr-v70-cisse17a,DBLP:journals/corr/abs-1711-00851,DBLP:journals/corr/abs-1801-09344} cannot solve the problem, for many reasons. \textbf{(a)} Existing sensitivity bounds \citep{Phan0WD16,PhanMLJ2017,NHPhanICDM17} and designs \citep{Yu2019,Lee:2018:CDP:3219819.3220076,PhanIJCAI,DBLP:journals/corr/abs-1905-12883,ADADP} have not been developed to protect the training data in adversarial training. It is obvious that using adversarial examples crafted from the private training data to train our models introduces a previously unknown privacy risk, disclosing the participation of the benign examples \citep{2019arXiv190510291S}. \textbf{(b)} There is an unrevealed interplay among DP preservation, adversarial learning, and robustness bounds. \textbf{(c)} Existing algorithms cannot be readily applied to address the trade-off among model utility, privacy loss, and robustness. \textbf{(d)} It is challenging in applying existing algorithms to train large DNNs given large data (i.e., scalability); since, they employ the vanilla \textit{iterative batch-by-batch training}, in which only a single batch of data instances can be used at each training step, such that the \textit{privacy loss} can be estimated \citep{Lee:2018:CDP:3219819.3220076,PhanIJCAI,Yu2019,DBLP:journals/corr/abs-1905-12883,ADADP}. That prevents us from applying scalable methods, e.g., \textit{distributed adversarial training} \citep{DBLP:journals/corr/GoyalDGNWKTJH17}, to achieve the same level of DP on large DNNs and datasets.
Therefore, bounding the robustness of a model (which both protects the privacy and is robust against adversarial examples) \textit{at scale} is nontrivial.

\textbf{Contributions.} Motivated by this open problem, we develop a novel \textit{stochastic batch (\textbf{StoBatch}) mechanism} to: \textbf{1)} preserve DP of the training data, \textbf{2)} be provably and practically robust to adversarial examples, \textbf{3)} retain high model utility, and \textbf{4)} be scalable to large DNNs and datasets.

$\bullet$ In StoBatch, privacy-preserving noise is injected into inputs and hidden layers to achieve DP in learning private model parameters (\textbf{Theorem \ref{lemma2}}). Then, we incorporate ensemble adversarial learning into our mechanism to improve the decision boundary under DP protections, by introducing a concept of \textit{DP adversarial examples} crafted using benign examples in the private training data (\textbf{Eq. \ref{I-DPAS}}). To address the trade-off between model utility and privacy loss, we propose a new DP adversarial objective function to tighten the model's global sensitivity (\textbf{Theorem \ref{BenignLoss}}); thus, we reduce the amount of noise injected into our function, compared with existing works \citep{Phan0WD16,PhanMLJ2017,NHPhanICDM17}. An end-to-end privacy analysis shows that, by slitting the private training data into \textit{disjoint} and \textit{fixed} batches across epochs, the privacy budget in our StoBatch is not accumulated across \textit{gradient descent}-based training steps (\textbf{Theorems \ref{BenignLoss}, \ref{OverallDP}}).

$\bullet$ After preserving DP in learning model parameters, we establish a new connection between DP preservation in adversarial learning and certified robustness.
Noise injected into different layers is considered as a sequence of randomizing mechanisms, providing different levels of robustness. By leveraging the \textit{sequential composition theory} in DP \citep{Dwork:2014:AFD:2693052.2693053}, we derive a generalized robustness bound, which is a composition of these levels of robustness in both input and latent spaces (\textbf{Theorem \ref{SRCC}} and \textbf{Corollary \ref{prop2}}), compared with only in the input space \citep{DBLP:journals/corr/abs-1906-04584} or only in the latent space \citep{Lecuyer2018}.

$\bullet$ To bypass the iterative batch-by-batch training, we develop a \textit{stochastic batch training}. In our algorithm, disjoint and fixed batches are distributed to local trainers, each of which learns DP parameters given its local data batches. A synchronous scheme can be leveraged to aggregate gradients observed from local trainers; thus enabling us to efficiently compute adversarial examples from multiple data batches at each iteration. This allows us to scale our mechanism to large DNNs and datasets, under the same DP guarantee. 
Rigorous experiments conducted on MNIST, CIFAR-10 \citep{Lecun726791,krizhevsky2009learning}, and \cite{Imagenet} datasets show that our mechanism notably enhances the robustness and scalability of DP DNNs. 


\section{Background}  

In this section, we revisit DP, adversarial learning, and certified robustness.
Let $D$ be a database that contains $N$ tuples, each of which contains data $x \in [-1, 1]^d$ and a \textit{ground-truth label} $y \in \mathbb{Z}_K$ (one-hot vector), with $K$ possible categorical outcomes $y = \{y_{1}, \ldots, y_{K}\}$. A single \textit{true class label} $y_x \in y$ given $x \in D$ is assigned to only one of the $K$ categories.
On input $x$ and parameters $\theta$, a model outputs class scores $f: \mathbb{R}^d \rightarrow \mathbb{R}^K$ that maps $x$ to a vector of scores $f(x) = \{f_1(x), \ldots, f_K(x)\}$ s.t. $\forall k \in [1, K]: f_k(x) \in [0, 1]$ and $\sum_{k = 1}^K f_k(x) = 1$. The class with the highest score value is selected as the \textit{predicted label} for $x$, denoted as $y(x) = \max_{k \in K} f_k(x)$. 
A loss function $L(f(x), y)$ presents the penalty for mismatching between the predicted values $f(x)$ and original values $y$. The notations and terminologies used in this paper are summarized in Table \ref{Notations} (\textbf{Appendix A}).
Let us briefly revisit DP DNNs, starting with the definition of DP. 
\begin{mydef}{$(\epsilon, \delta)$-DP \citep{dwork2006calibrating}.} A randomized algorithm $A$ fulfills $(\epsilon, \delta)$-DP, if for any two databases $D$ and $D'$ differing at most one tuple, and for all $O \subseteq Range(A)$, we have: 
\begin{equation} 
Pr[A(D) = O] \leq e^\epsilon Pr[A(D') = O] + \delta  
\end{equation}
$\epsilon$ controls the amount by which the distributions induced by $D$ and $D'$ may differ, $\delta$ is a broken probability. 
\label{Different Privacy}  
\end{mydef}

DP also applies to general metrics $\rho(D, D') \leq 1$, where $\rho$ can be $l_p$-norms \citep{Chatzikokolakis}. 
DP-preserving algorithms in DNNs can be categorized into three lines: 1) introducing noise into \textit{parameter gradients} \citep{Abadi,abadi2017protection,ShokriVitaly2015,Yu2019,Lee:2018:CDP:3219819.3220076,PhanIJCAI}, 2) injecting noise into objective functions \citep{Phan0WD16,PhanMLJ2017,NHPhanICDM17}, and 3) injecting noise into labels \citep{papernot2018scalable}. 

\textbf{Adversarial Learning.}
For some target model $f$ and inputs $(x, y_{x})$, the adversary's goal is to find an \textit{adversarial example} $x^{\text{adv}} = x + \alpha$, where $\alpha$ is the perturbation introduced by the attacker, such that: \textbf{(1)} $x^{\text{adv}}$ and $x$ are close, and \textbf{(2)} the model misclassifies $x^{\text{adv}}$, i.e., $y(x^{\text{adv}}) \neq y(x)$. In this paper, \textit{we consider well-known $l_{p \in \{1, 2, \infty\}}(\mu)$-norm bounded attacks \citep{DBLP:journals/corr/GoodfellowSS14}, where $\mu$ is the radius of the $p$-norm ball}.  
To improve the robustness of models, prior work focused on two directions: 1) Producing correct predictions on adversarial examples, while not compromising the accuracy on legitimate inputs \citep{7965897,7965869,DBLP:journals/corr/WangGZOXGL16,7546524,7467366,DBLP:journals/corr/GuR14,papernot2017extending,hosseini2017blocking}; and 2) Detecting adversarial examples \citep{metzen2017detecting,DBLP:journals/corr/GrosseMP0M17,DBLP:journals/corr/XuEQ17,DBLP:journals/corr/AbbasiG17,DBLP:journals/corr/GaoWQ17}.
Among existing solutions, adversarial training appears to hold the greatest promise for learning robust models \citep{tramer2017ensemble}. 
A well-known algorithm was proposed in \citep{DBLP:journals/corr/KurakinGB16a}. At every training step, new adversarial examples are generated and injected into batches containing both benign and adversarial examples (Alg. \ref{TAT}, \textbf{Appendix C}). 

\begin{figure*}[t]
\centering
$\begin{array}{c@{\hspace{0.4in}}c}
\includegraphics[width=0.35\textwidth]{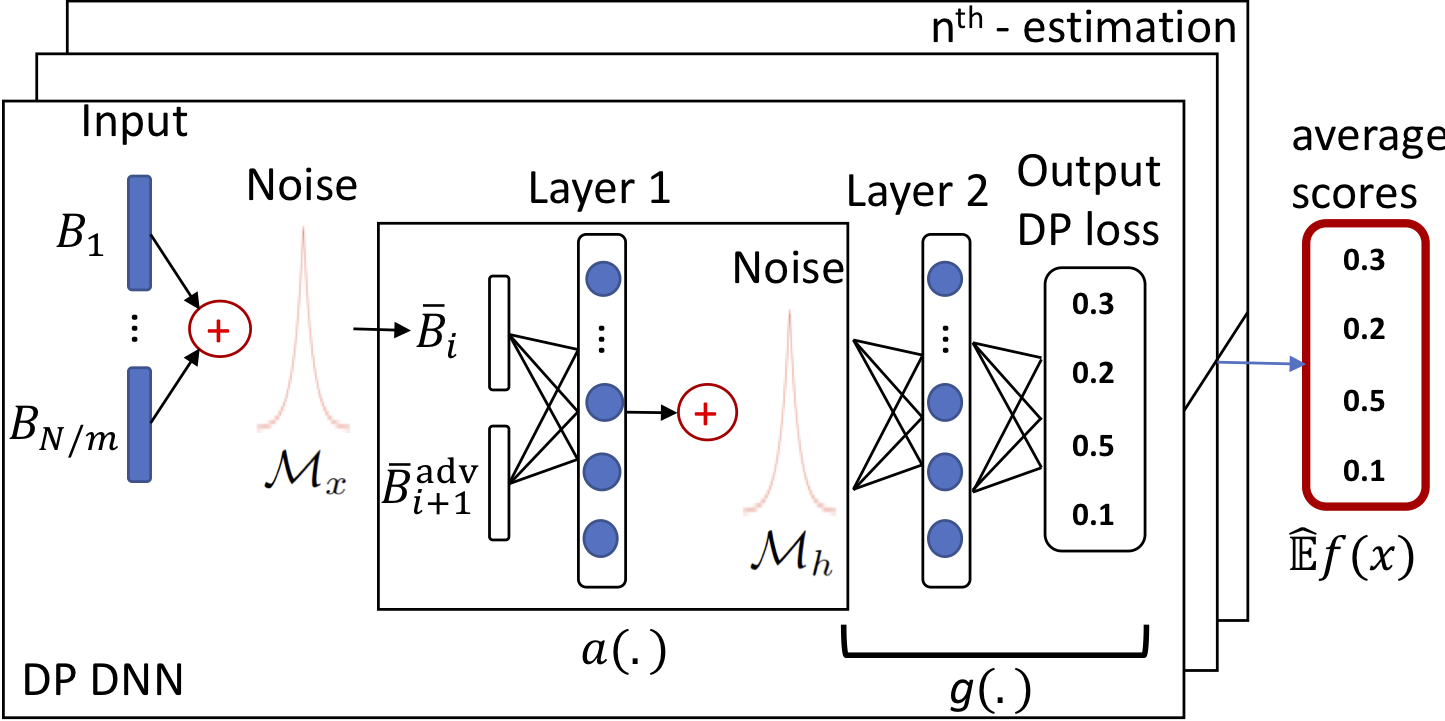} & \includegraphics[width=0.375\textwidth]{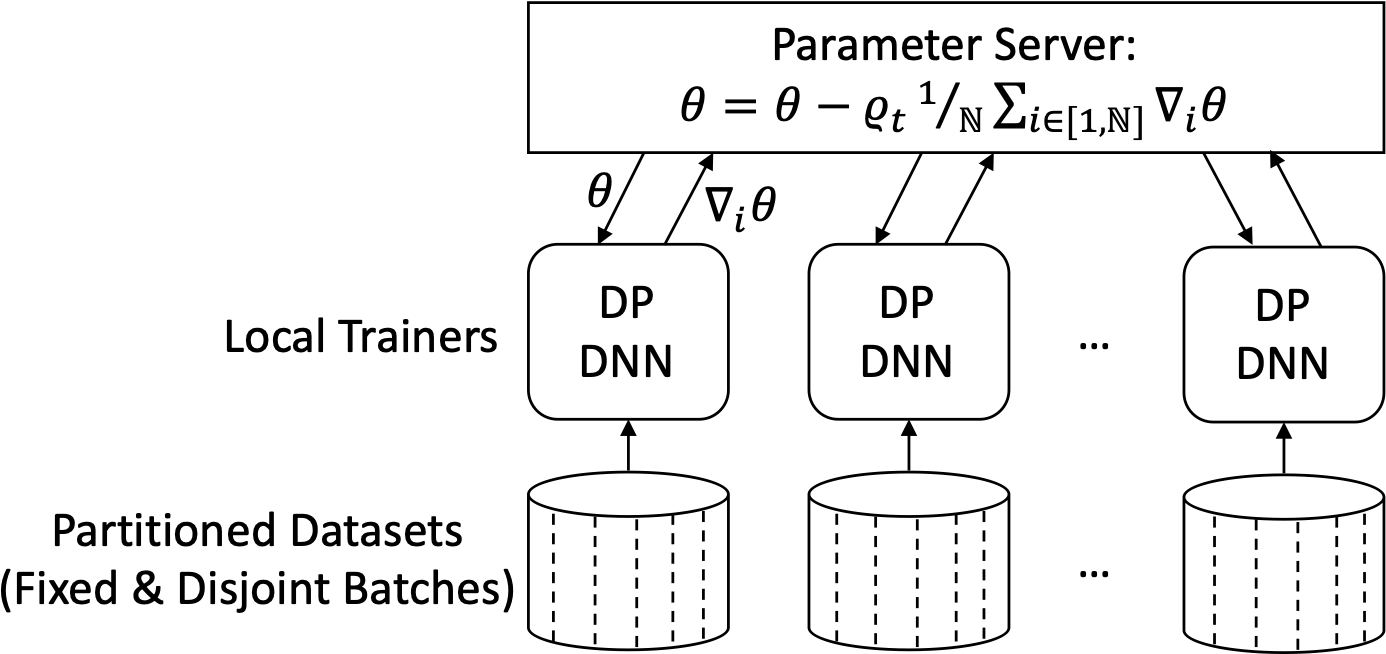} \\ [0.0cm] 
\mbox{(a) An instance of DP DNNs and verified inference} & \mbox{(b) An instance of stochastic batch training}
\end{array}$
\caption{Stochastic Batch mechanism.}
\label{DNN}
\end{figure*}

\textbf{Certified Robustness and DP.}
Recently, some algorithms \citep{pmlr-v70-cisse17a,DBLP:journals/corr/abs-1711-00851,DBLP:journals/corr/abs-1801-09344,pmlr-v97-cohen19c,DBLP:journals/corr/abs-1809-03113,DBLP:journals/corr/abs-1906-04584} have been proposed to derive certified robustness, in which each prediction is guaranteed to be consistent under the perturbation $\alpha$, if a robustness condition is held. Given a benign example $x$, we focus on achieving a robustness condition to $l_p(\mu)$-norm attacks, as follows: 
\begin{equation}
\forall \alpha \in l_p(\mu): f_k(x + \alpha) > \max_{i: i\neq k} f_{i}(x + \alpha) 
\label{RobustCond1}
\end{equation}
where $k$ = $y(x)$, indicating that a small perturbation $\alpha$ in the input does not change the predicted label $y(x)$.
To achieve the robustness condition in Eq. \ref{RobustCond1}, \cite{Lecuyer2018} introduce an algorithm, called \textbf{PixelDP}. By considering an input $x$ (e.g., images) as databases in DP parlance, and individual features (e.g., pixels) as tuples, PixelDP shows that randomizing the scoring function $f(x)$ to enforce DP on a small number of pixels in an image guarantees robustness of predictions. To randomize $f(x)$, \textit{random noise} $\sigma_r$ is injected into either input $x$ or an arbitrary hidden layer, resulting in the following $(\epsilon_r, \delta_r)$-PixelDP condition: 
\begin{lemma} $(\epsilon_r, \delta_r)$-PixelDP \citep{Lecuyer2018}. Given a randomized scoring function $f(x)$ satisfying $(\epsilon_r, \delta_r)$-PixelDP w.r.t. a $l_p$-norm metric, we have:
\begin{equation}
\forall k, \forall \alpha \in l_p(1): \mathbb{E} f_k(x) \leq e^{\epsilon_r} \mathbb{E} f_k(x + \alpha) + \delta_r
\label{PixelDPDef}
\end{equation}
where $\mathbb{E} f_k(x)$ is the expected value of $f_k(x)$, $\epsilon_r$ is a predefined budget, $\delta_r$ is a broken probability. 
\label{LemmaPixelDP} 
\end{lemma}

At the prediction time, a certified robustness check is implemented for each prediction, as follows: 
\begin{equation}
\hat{\mathbb{E}}_{lb} f_k(x) > e^{2\epsilon_r} \max_{i: i\neq k} \hat{\mathbb{E}}_{ub} f_{i}(x) + (1 + e^{\epsilon_r})\delta_r 
\label{RobustCon2} 
\end{equation}
where $\hat{\mathbb{E}}_{lb}$ and $\hat{\mathbb{E}}_{ub}$ are the lower and upper bounds of the expected value $\hat{\mathbb{E}} f(x) = \frac{1}{n} \sum_n f(x)_n$, derived from the Monte Carlo estimation with an $\eta$-confidence, given $n$ is the number of invocations of $f(x)$ with independent draws in the noise $\sigma_r$.
Passing the check for a given input guarantees that no perturbation up to $l_p(1)$-norm can change the model's prediction. 
PixelDP \textit{does not} preserve DP in learning private parameters $\theta$ to protect the training data. 


\section{Stochastic Batch (StoBatch) Mechanism}

StoBatch is presented in Alg. \ref{StoBatch} (\textbf{Appendix D}). Our DNN (Fig. \ref{DNN}a) is presented as: $f(x) = g(a(x, \theta_1), \theta_2)$, where $a(x, \theta_1)$ is a feature representation learning model with $x$ as an input, and $g$ will take the output of $a(x, \theta_1)$ and return the class scores $f(x)$.
At a high level, there are four key components: \textbf{(1)} DP $a(x, \theta_1)$, which is to preserve DP in learning the feature representation model $a(x, \theta_1)$; \textbf{(2)} DP Adversarial Learning, which focuses on preserving DP in adversarial learning, given DP $a(x, \theta_1)$; \textbf{(3)} Certified Robustness and Verified Inferring, which are to compute robustness bounds given an input at the inference time; and \textbf{(4)} Stochastic batch training (Fig. \ref{DNN}b).
To establish theoretical results in DP preservation and in deriving robustness bounds, let us first present our mechanism in the vanilla iterative batch-by-batch training (\textbf{Alg. \ref{DPAT}}). The network $f$ (Lines 2-3, Alg. \ref{DPAT}) is trained over $T$ training steps. In each step, a disjoint and fixed batch of $m$ perturbed training examples and a disjoint and fixed batch of $m$ DP adversarial examples, derived from $D$, are used to train our network (Lines 4-12, Alg. \ref{DPAT}).

\subsection{DP Feature Representation Learning}

Our idea is to use auto-encoder to simultaneously learn DP parameters $\theta_1$ and ensure that the output of $a(x, \theta_1)$ is DP, since: (1) It is easier to train, given its small size; and (2) It can be reused for different predictive models. 
A typical data reconstruction function (cross-entropy), given a batch $B_t$ at the training step $t$ of the input $x_i$, is as follows: $
\mathcal{R}_{B_t}(\theta_1) = \sum_{x_i \in B_t} \sum_{j = 1}^{d} \big[x_{ij}\log (1+e^{-\theta_{1j} h_i}) + (1-x_{ij}) \log (1+e^{\theta_{1j} h_i}) \big]
$,
where $h_i = \theta_1^T x_i$, the hidden layer $\mathbf{h}_1$ of $a(x, \theta_1)$ given the batch $B_t$ is denoted as $\mathbf{h}_{1B_t} = \{\theta_1^T x_i\}_{x_i \in B_t}$, and $\widetilde{x}_i = \theta_1 h_i$ is the reconstruction of $x_i$.

To preserve $\epsilon_1$-DP in learning $\theta_1$ where $\epsilon_1$ is a privacy budget, we first derive the 1st-order polynomial approximation of $\mathcal{R}_{B_t}(\theta_1)$ by applying Taylor Expansion \citep{tagkey1985}, denoted as $\widetilde{\mathcal{R}}_{B_t}(\theta_1)$. Then, \textit{Functional Mechanism} \citep{zhang2012functional} \textbf{(revisited in Appendix B)} is adapted to inject noise into coefficients of the approximated function $\widetilde{\mathcal{R}}_{B_t}(\theta_1) = \sum_{x_i \in B_t} \sum_{j = 1}^d \sum_{l=1}^{2} \sum_{r = 0}^{1} \frac{\mathbf{F}^{(r)}_{lj}(0)}{r!}\big(\theta_{1j}h_i\big)^r$,
where $\mathbf{F}_{1j}(z) = x_{ij}\log(1 + e^{-z})$, $\mathbf{F}_{2j}(z) = (1-x_{ij})\log (1 + e^z)$, we have that: $\widetilde{\mathcal{R}}_{B_t}(\theta_1) = \sum_{x_i \in B_t} \sum_{j = 1}^d \Big[ \log 2  +  \theta_{1j}\big(\frac{1}{2} - x_{ij}\big)h_i \Big]$.
In $\widetilde{\mathcal{R}}_{B_t}(\theta_1)$, parameters $\theta_{1j}$ derived from the function optimization need to be $\epsilon_1$-DP. To achieve that, Laplace noise $\frac{1}{m}Lap(\frac{\Delta_{\mathcal{R}}}{\epsilon_1})$ is injected into coefficients $\big(\frac{1}{2} - x_{ij}\big)h_i$, where $\Delta_{\mathcal{R}}$ is the sensitivity of $\widetilde{\mathcal{R}}_{B_t}(\theta_1)$, as follows: 
\begin{align}
\widetilde{\mathcal{R}}_{B_t}(\theta_1) & = \sum_{x_i \in B_t} \sum_{j = 1}^d \Big[ \theta_{1j}\Big(\big(\frac{1}{2} - x_{ij}\big)h_i + \frac{1}{m} Lap(\frac{\Delta_{\mathcal{R}}}{\epsilon_1})\Big) \Big] \nonumber \\
& = \sum_{x_i \in B_t} \Big[ \sum_{j = 1}^d (\frac{1}{2}\theta_{1j}\overline{h}_i) - x_{i} \widetilde{x}_{i} \Big] 
\label{poly1}
\end{align}
To ensure that the computation of $\widetilde{x}_{i}$ does not access the original data, we further inject Laplace noise $\frac{1}{m}Lap(\frac{\Delta_{\mathcal{R}}}{\epsilon_1})$ into $x_i$. This can be done as a preprocessing step for all the benign examples in $D$ to construct a set of \textit{disjoint} batches $\overline{\mathbf{B}}$ of perturbed benign examples (Lines 2 and 5, Alg. \ref{DPAT}). The perturbed function now becomes:
\begin{equation}
\overline{\mathcal{R}}_{\overline{B}_t}(\theta_1) = \sum_{\overline{x}_i \in \overline{B}_t} \Big[\sum_{j = 1}^d (\frac{1}{2}\theta_{1j}\overline{h}_i) - \overline{x}_{i} \widetilde{x}_{i} \Big]
\label{PerturbAutoencoder} 
\end{equation}
where $\overline{x}_{i} = x_{i} + \frac{1}{m}Lap(\frac{\Delta_{\mathcal{R}}}{\epsilon_1}), h_i = \theta_1^T \overline{x}_i, \overline{h}_i = h_i + \frac{2}{m}Lap(\frac{\Delta_{\mathcal{R}}}{\epsilon_1})$, and $\widetilde{x}_{i} = \theta_{1}\overline{h}_i$.
Let us denote $\beta$ as the number of neurons in $\mathbf{h}_{1}$, and $h_i$ is bounded in $[-1, 1]$, the global sensitivity $\Delta_{\mathcal{R}}$ is as follows:
\begin{lemma} The global sensitivity of $\widetilde{\mathcal{R}}$ over any two neighboring batches, $B_t$ and $B'_t$, is: $\Delta_{\mathcal{R}} \leq d(\beta + 2)$.
\label{lemma3}
\end{lemma}
All the proofs are in \textbf{Appendix}. By setting $\Delta_{\mathcal{R}} = d(\beta + 2)$, we show that the output of $a(\cdot)$, which is the perturbed affine transformation $\overline{\mathbf{h}}_{1\overline{B}_t} = \{\theta_1^T \overline{x}_i + \frac{2}{m}Lap(\frac{\Delta_{\mathcal{R}}}{\epsilon_1}) \}_{\overline{x}_i \in \overline{B}_t}$, is $(\epsilon_1/\gamma)$-DP, given $\gamma = \frac{2\Delta_{\mathcal{R}}}{m \lVert \overline{\theta}_1 \rVert_{1,1}}$ and $\lVert \overline{\theta}_1 \rVert_{1,1}$ is the maximum 1-norm of $\theta_1$'s columns \citep{WikipediaOperatornorm}. This is important to tighten the privacy budget consumption in computing the remaining hidden layers $g(a(x, \theta_1), \theta_2)$. In fact, without using additional information from the original data, the computation of $g(a(x, \theta_1), \theta_2)$ is also $(\epsilon_1/\gamma)$-DP. 

\begin{algorithm}[t]
\textbf{Input:} Database $D$, loss function $L$, parameters $\theta$, batch size $m$, learning rate $\varrho_t$, privacy budgets: $\epsilon_1$ and $\epsilon_2$, robustness parameters: $\epsilon_r$, $\Delta^x_r$, and $\Delta^h_r$, adversarial attack size $\mu_a$, the number of invocations $n$, ensemble attacks $A$, parameters $\psi$ and $\xi$, and the size $|\mathbf{h}_\pi|$ of $\mathbf{h}_\pi$
\begin{algorithmic}[1]
\STATE \textbf{Draw Noise} $\chi_1 \leftarrow [Lap(\frac{\Delta_{\mathcal{R}}}{\epsilon_1})]^{d}$, $\chi_2 \leftarrow [Lap(\frac{\Delta_{\mathcal{R}}}{\epsilon_1})]^{\beta}$, $\chi_3 \leftarrow [Lap(\frac{\Delta_{\mathcal{L}2}}{\epsilon_2})]^{|\mathbf{h}_\pi|}$
\STATE \textbf{Randomly Initialize} $\theta = \{\theta_1, \theta_2\}$, $\mathbf{B} = \{B_1, \ldots, B_{N/m}\}$ s.t. $\forall B \in \mathbf{B}: B$ is a batch with the size $m$, $B_1 \cap \ldots \cap B_{N/m} = \emptyset$, and $B_1 \cup \ldots \cup B_{N/m} = D$, $\overline{\mathbf{B}} = \{\overline{B}_1, \ldots, \overline{B}_{N/m}\}$ where $\forall i \in [1, N/m]: \overline{B}_i = \{\overline{x} \leftarrow x + \frac{\chi_1}{m}\}_{x \in B_i}$
\STATE \textbf{Construct} a deep network $f$ with \textbf{hidden layers} $\{\mathbf{h}_1 + \frac{2\chi_2}{m},\dots, \mathbf{h}_\pi\}$, where $\mathbf{h}_\pi$ is the last hidden layer
\FOR{$t \in [T]$}
	\STATE \textbf{Take} \text{a batch $\overline{B}_i \in \overline{\mathbf{B}}$ where $i = t\%(N/m)$}, $\overline{B}_t \leftarrow \overline{B}_i$ \\
	\STATE \textbf{Ensemble DP Adversarial Examples:}\\
	\STATE \textbf{Draw Random Perturbation Value} $\mu_t \in (0, 1]$ \\
	\STATE \textbf{Take} \text{a batch $\overline{B}_{i+1} \in \overline{\mathbf{B}}$}, \textbf{Assign} $\overline{B}^{\text{adv}}_t \leftarrow \emptyset$ \\
	\FOR{$l \in A$}
		\STATE \textbf{Take} \text{the next batch $\overline{B}_a \subset \overline{B}_{i+1}$ with the size $m/|A|$} \\
		\STATE $\forall \overline{x}_j \in \overline{B}_a$: \textbf{Craft } $\overline{x}^{\text{adv}}_j$ by using attack algorithm $A[l]$  with $l_\infty(\mu_t)$, $\overline{B}^{\text{adv}}_t \leftarrow \overline{B}^{\text{adv}}_t \cup \overline{x}^{\text{adv}}_j$\\
	\ENDFOR
	\STATE \textbf{Descent: } $\theta_1 \leftarrow \theta_1 - \varrho_t \nabla_{\theta_1}\overline{\mathcal{R}}_{\overline{B}_t \cup \overline{B}^{\text{adv}}_t}(\theta_1)$; $\theta_2 \leftarrow \theta_2 - \varrho_t \nabla_{\theta_2} \overline{L}_{\overline{B}_t \cup \overline{B}^{\text{adv}}_t}(\theta_2)$ with the noise $\frac{\chi_3}{m}$
\ENDFOR
\textbf{Output:} $\epsilon = (\epsilon_1 + \epsilon_1/\gamma_\mathbf{x} + \epsilon_1/\gamma+\epsilon_2)$-DP parameters $\theta = \{\theta_1, \theta_2\}$, robust model with an $\epsilon_r$ budget
\end{algorithmic}
\caption{\textbf{Adversarial Learning with DP}}
\label{DPAT} 
\end{algorithm}

Similarly, the perturbation of each benign example $x$ turns $\overline{B}_t = \{\overline{x}_i \leftarrow x_{i} + \frac{1}{m}Lap(\frac{\Delta_{\mathcal{R}}}{\epsilon_1})\}_{x_i \in B_t}$ into a $(\epsilon_1/\gamma_\mathbf{x})$-DP batch, with $\gamma_\mathbf{x}=\Delta_\mathcal{R} / m$. We do not use the post-processing property of DP to estimate the DP guarantee of $\overline{\mathbf{h}}_{1\overline{B}_t}$ based upon the DP guarantee of $\overline{B}_t$, since $\epsilon_1/\gamma < \epsilon_1/\gamma_\mathbf{x}$ in practice. So, the $(\epsilon_1/\gamma)$-DP $\overline{\mathbf{h}}_{1\overline{B}_t}$ provides a more rigorous DP protection to the computation of $g(\cdot)$ and to the output layer.
\begin{lemma} The computation of the batch $\overline{B}_t$ as the input layer is $(\epsilon_1/\gamma_\mathbf{x})$-DP, and the computation of the affine transformation $\overline{\mathbf{h}}_{1\overline{B}_t}$ is $(\epsilon_1/\gamma)$-DP.
\label{Operatornorm}
\end{lemma}
Departing from the vanilla Functional Mechanism, in which only \textit{grid search}-based approaches can be applied to find DP-preserving $\theta_1$ with a low loss $\overline{\mathcal{R}}_{\overline{B}_t}(\theta_1)$, our following Theorem \ref{lemma2} shows that \textit{gradient descent}-based optimizing $\overline{\mathcal{R}}_{\overline{B}_t}(\theta_1)$ is $(\epsilon_1/\gamma_\mathbf{x} + \epsilon_1)$-DP in learning $\theta_1$ given an $(\epsilon_1/\gamma_\mathbf{x})$-DP $\overline{B}_t$ batch. In fact, in addition to $h_i$, $\overline{h}_i$, $\widetilde{x}_i$, based on Lemma \ref{Operatornorm}, we further show that the computation of gradients, i.e., $\forall j \in [1, d]: \frac{\delta \overline{\mathcal{R}}_{\overline{B}_t}(\theta_1)}{\delta \theta_{1j}} = \sum_{i = 1}^{m}\overline{h}_i(\frac{1}{2} - \overline{x}_{ij})$, and descent operations given the $(\epsilon_1/\gamma_\mathbf{x})$-DP $\overline{B}_t$ batch are $(\epsilon_1/\gamma_\mathbf{x})$-DP, without incurring any additional information from the original data. As a result, gradient descent-based approaches can be applied to optimize $\overline{\mathcal{R}}_{\overline{B}_t}(\theta_1)$ in Alg. \ref{DPAT}, since all the computations on top of $\overline{B}_t$ are DP, without using any additional information from the original data. 
\begin{theorem} The gradient descent-based optimization of $\overline{\mathcal{R}}_{\overline{B}_t}(\theta_1)$ preserves $(\epsilon_1/\gamma_\mathbf{x} + \epsilon_1)$-DP in learning $\theta_1$. 
\label{lemma2}
\end{theorem}

\subsection{Adversarial Learning with Differential Privacy} \vspace{-2.5pt}

To integrate adversarial learning, we first draft DP adversarial examples $\overline{x}^{\text{adv}}_j$ using perturbed benign examples $\overline{x}_j$, with an ensemble of attack algorithms $A$ and a random perturbation budget $\mu_t \in (0, 1]$, at each step $t$ (Lines 6-11, Alg. \ref{DPAT}). This will significantly enhances the robustness of our models under different types of adversarial examples with an unknown adversarial attack size $\mu$.
\begin{equation}
\overline{x}^{\text{adv}}_j = \overline{x}_j + \mu \cdot \text{sign}\Big(\nabla_{\overline{x}_j} \mathcal{L}\big(f(\overline{x}_j, \theta), y (\overline{x}_j)\big)\Big)
\label{DPAS}
\end{equation}
with $y(\overline{x}_j)$ is the class prediction result of $f(\overline{x}_j)$ to avoid label leaking of $x_j$ during the adversarial example crafting. Given a set of DP adversarial examples $\overline{B}^{\text{adv}}_t$, training the auto-encoder with $\overline{B}^{\text{adv}}_t$ preserves $(\epsilon_1/\gamma_{\mathbf{x}} + \epsilon_1)$-DP.
\begin{theorem} The gradient descent-based optimization of $\overline{\mathcal{R}}_{\overline{B}^{\text{adv}}_t}(\theta_1)$ preserves $(\epsilon_1/\gamma_{\mathbf{x}} + \epsilon_1)$-DP in learning $\theta_1$.
\label{LemmaDPAE}
\end{theorem}
The proof of Theorem \ref{LemmaDPAE} is in \textbf{Appendix J, Result 4}.
It can be extended to iterative attacks as: $\overline{x}^{\text{adv}}_{j,0} = \overline{x}_j$,
\begin{equation}
\small
\overline{x}^{\text{adv}}_{j, \mathsf{t}+1} = \overline{x}^{\text{adv}}_{j, \mathsf{t}} + \frac{\mu}{T_{\mu}} \cdot \text{sign}\Big(\nabla_{\overline{x}^{\text{adv}}_{j, \mathsf{t}}} \mathcal{L}\big(f(\overline{x}^{\text{adv}}_{j, \mathsf{t}}, \theta), y(\overline{x}^{\text{adv}}_{j, \mathsf{t}})\big)\Big) 
\label{I-DPAS}
\end{equation}
where $y(\overline{x}^{\text{adv}}_{j, \mathsf{t}})$ is the prediction of $f(\overline{x}^{\text{adv}}_{j, \mathsf{t}}, \theta)$, $\mathsf{t} \in [0, T_\mu - 1]$.

Second, we propose a novel DP adversarial objective function $L_{B_t}(\theta_2)$, in which the loss function $\mathcal{L}$ for benign examples is combined with an additional loss function $\Upsilon$ for DP adversarial examples, to optimize the parameters $\theta_2$. 
The objective function $L_{B_t}(\theta_2)$ is defined as follows:
\begin{multline}
L_{\overline{B}_t \cup \overline{B}^{\text{adv}}_t}(\theta_2) = \frac{1}{m (1 + \xi)} \Big(\sum_{\overline{x}_i \in \overline{B}_t} \mathcal{L}\big(f(\overline{x}_i, \theta_2), y_i\big) \\ + \xi \sum_{\overline{x}^{\text{adv}}_j \in \overline{B}^{\text{adv}}_t} \Upsilon \big(f(\overline{x}^{\text{adv}}_j, \theta_2), y_j\big)\Big)
\label{DPLossAT}
\end{multline} 
where $\xi$ is a hyper-parameter. 
For the sake of clarity, in Eq. \ref{DPLossAT}, we denote $y_i$ and $y_j$ as the true class labels $y_{x_i}$ and $y_{x_j}$ of examples $x_i$ and $x_j$. $\overline{x}^{\text{adv}}_j$ and $x_j$ share the same label $y_{x_j}$. 

Now we are ready to preserve DP in objective functions $\mathcal{L}\big(f(\overline{x}_i, \theta_2), y_i\big)$ and $\Upsilon\big(f(\overline{x}^{\text{adv}}_j, \theta_2), y_j\big)$ in order to achieve DP in learning $\theta_2$. Since the objective functions use the true class labels $y_i$ and $y_j$, we need to protect the labels at the output layer. Let us first present our approach to preserve DP in the objective function $\mathcal{L}$ for benign examples. 
Given $\mathbf{h}_{\pi i}$ computed from the $\overline{x}_i$ through the network with $W_{\pi}$ is the parameter at the last hidden layer $\mathbf{h}_{\pi}$. Cross-entropy function is approximated as:
$
\mathcal{L}_{\overline{B}_t}\big(\theta_2\big) \approxeq \sum_{k = 1}^K \sum_{\overline{x}_i} \big[\mathbf{h}_{\pi i} W_{\pi k} - (\mathbf{h}_{\pi i} W_{\pi k}) y_{ik} - \frac{1}{2}|\mathbf{h}_{\pi i} W_{\pi k}|
+ \frac{1}{8} (\mathbf{h}_{\pi i} W_{\pi k})^2 \big] \approxeq \mathcal{L}_{1\overline{B}_t}\big(\theta_2\big) - \mathcal{L}_{2\overline{B}_t}\big(\theta_2\big)$,
where $\mathcal{L}_{1\overline{B}_t}\big(\theta_2\big) = \sum_{k = 1}^K \sum_{\overline{x}_i} \big[ \mathbf{h}_{\pi i} W_{\pi k} - \frac{1}{2}|\mathbf{h}_{\pi i} W_{\pi k}| + \frac{1}{8} (\mathbf{h}_{\pi i} W_{\pi k})^2 \big]$, and $\mathcal{L}_{2\overline{B}_t}\big(\theta_2\big) = \sum_{k = 1}^K \sum_{\overline{x}_i} (\mathbf{h}_{\pi i} y_{ik}) W_{\pi k}$.

Based on the \textit{post-processing property of DP} \citep{Dwork:2014:AFD:2693052.2693053}, $\mathbf{h}_{\pi \overline{B}_t} = \{\mathbf{h}_{\pi i}\}_{\overline{x}_i \in \overline{B}_t}$ is $(\epsilon_1/\gamma)$-DP, since the computation of $\overline{\mathbf{h}}_{1 \overline{B}_t}$ is $(\epsilon_1/\gamma)$-DP (Lemma \ref{Operatornorm}). Hence, the optimization of $\mathcal{L}_{1\overline{B}_t}\big(\theta_2\big)$ does not disclose any information from the training data, and $\frac{Pr(\mathcal{L}_{1\overline{B}_t}(\theta_2))}{Pr(\mathcal{L}_{1\overline{B}'_t}(\theta_2))} = \frac{Pr(\mathbf{h}_{\pi \overline{B}_t})}{Pr(\mathbf{h}_{\pi \overline{B}'_t})} \leq e^{\epsilon_1/\gamma}$, given neighboring batches $\overline{B}_t$ and $\overline{B}'_t$. Thus, we only need to preserve $\epsilon_2$-DP in the function $\mathcal{L}_{2\overline{B}_t}(\theta_2)$, which accesses the ground-truth label $y_{ik}$. Given coefficients $\mathbf{h}_{\pi i} y_{ik}$, the sensitivity $\Delta_{\mathcal{L}2}$ of $\mathcal{L}_{2\overline{B}_t}(\theta_2)$ is computed as:
\begin{lemma} Let $\overline{B}_t$ and $\overline{B}'_t$ be neighboring batches of benign examples, we have the following inequality: $\Delta_{\mathcal{L}2} \leq 2 |\mathbf{h}_\pi |$, where $|\mathbf{h}_\pi |$ is the number of hidden neurons in $\mathbf{h}_\pi$.
\label{NovelSensitivity}
\end{lemma} 

The sensitivity of our objective function is notably smaller than the state-of-the-art bound \citep{NHPhanICDM17}, which is crucial to improve our model utility. 
The perturbed functions become: $\overline{\mathcal{L}}_{\overline{B}_t}\big(\theta_2\big) = \mathcal{L}_{1\overline{B}_t}(\theta_2) - \overline{\mathcal{L}}_{2\overline{B}_t}(\theta_2),$
where $\overline{\mathcal{L}}_{2\overline{B}_t}(\theta_2) = \sum_{k = 1}^K \sum_{\overline{x}_i} \big(\mathbf{h}_{\pi i} y_{ik} + \frac{1}{m}Lap(\frac{\Delta_{\mathcal{L}2}}{\epsilon_2})\big) W_{\pi k}$.
\begin{theorem} Algorithm \ref{DPAT} preserves $(\epsilon_1/\gamma + \epsilon_2)$-DP in the gradient descent-based optimization of $\overline{\mathcal{L}}_{\overline{B}}\big(\theta_2\big)$.
\label{BenignLoss} 
\end{theorem}

We apply the same technique to preserve $(\epsilon_1/\gamma + \epsilon_2)$-DP in the optimization of the function $\Upsilon\big(f(\overline{x}^{\text{adv}}_j, \theta_2), y_j\big)$ over the DP adversarial examples $\overline{x}^{\text{adv}}_j \in \overline{B}^{\text{adv}}_t$. 
As the perturbed functions $\overline{\mathcal{L}}$ and $\overline{\Upsilon}$ are always optimized given two disjoint batches $\overline{B}_t$ and $\overline{B}^{\text{adv}}_t$, the privacy budget used to preserve DP in the adversarial objective function $L_{B_t}(\theta_2)$ is $(\epsilon_1/\gamma + \epsilon_2)$, following the \textit{parallel composition} property \citep{Dwork:2014:AFD:2693052.2693053}. 
The total budget to learn private parameters $\overline{\theta} = \{\overline{\theta}_1, \overline{\theta}_2\} = \arg \min_{\{\theta_1, \theta_2\}} (\overline{\mathcal{R}}_{\overline{B}_t \cup \overline{B}^{\text{adv}}_t}(\theta_1) + \overline{L}_{\overline{B}_t \cup \overline{B}^{\text{adv}}_t}(\theta_2))$ is $\epsilon = (\epsilon_1 + \epsilon_1/\gamma_{\mathbf{x}} + \epsilon_1/\gamma+\epsilon_2)$ (Line 12, Alg. \ref{DPAT}).

\paragraph{DP at the Dataset Level.}
Our mechanism achieves DP at the batch level $\overline{B}_t \cup \overline{B}^{\text{adv}}_t$ given a specific training step $t$. By constructing \textit{disjoint} and \textit{fixed} batches from $D$, we leverage both parallel composition and post-processing properties of DP to extend the result to $\epsilon$-DP in learning $\{\overline{\theta}_1, \overline{\theta}_2\}$ on $D$ across $T$ training steps. There are three key properties in our model: \textbf{(1)} It only reads perturbed inputs $\overline{B}_t$ and perturbed coefficients $\overline{\mathbf{h}}_1$, which are DP across $T$ training steps with \textit{a single draw of Laplace noise} (i.e., no further privacy leakage); \textbf{(2)} Given $N/m$ disjoint batches in each epoch, $\forall \overline{x}$, $\overline{x}$ is included in \textit{one and only one} batch, denoted $B_x \in \overline{\mathbf{B}}$. As a result, the DP guarantee to $\overline{x}$ in $D$ is equivalent to the DP guarantee to $\overline{x}$ in $B_x$; since the optimization using any other batches does not affect the DP guarantee of $\overline{x}$, even the objective function given $B_x$ can be slightly different from the objective function given any other batches in $\overline{\mathbf{B}}$; and \textbf{(3)} All the batches are fixed across $T$ training steps to prevent additional privacy leakage, caused by generating new and overlapping batches (which are considered overlapping datasets in the parlance of DP) in the typical training.

\begin{theorem} Algorithm \ref{DPAT} achieves $(\epsilon_1 + \epsilon_1/\gamma_{\mathbf{x}} + \epsilon_1/\gamma+\epsilon_2)$-DP parameters $\overline{\theta} = \{\overline{\theta}_1, \overline{\theta}_2\}$ on the private training data $D$ across $T$ gradient descent-based training steps.
\label{OverallDP}
\end{theorem}

\subsection{Certified Robustness}

Now, we establish the correlation between our mechanism and certified robustness. In the \textit{inference time}, to derive the certified robustness condition against adversarial examples $x + \alpha$, i.e., $\forall \alpha \in l_p(1)$, PixelDP randomizes the function $f(x)$ by injecting \textit{robustness noise} $\sigma_r$ into either input $x$ or a hidden layer, i.e., $x' = x + Lap(\frac{\Delta^x_r}{\epsilon_r})$ or $h' = h + Lap(\frac{\Delta^h_r}{\epsilon_r})$, where $\Delta^x_r$ and $\Delta^h_r$ are the sensitivities of $x$ and $h$, measuring how much $x$ and $h$ can be changed given the perturbation $\alpha \in l_p(1)$ in the input $x$. Monte Carlo estimation of the expected values $\hat{\mathbb{E}} f(x)$, $\hat{\mathbb{E}}_{lb} f_k(x)$, and $\hat{\mathbb{E}}_{ub} f_k(x)$ are used to derive the robustness condition in Eq. \ref{RobustCon2}. 

On the other hand, in our mechanism, the privacy noise $\sigma_p$ includes Laplace noise injected into both input $x$, i.e., $\frac{1}{m}Lap(\frac{\Delta_{\mathcal{R}}}{\epsilon_1})$, and its affine transformation $h$, i.e., $\frac{2}{m}Lap(\frac{\Delta_{\mathcal{R}}}{\epsilon_1})$. Note that the perturbation of $\overline{\mathcal{L}}_{2\overline{B}_t}\big(\theta_2\big)$ is equivalent to $\overline{\mathcal{L}}_{2\overline{B}_t}(\theta_2) = \sum_{k = 1}^K \sum_{\overline{x}_i} (\mathbf{h}_{\pi i} y_{ik}W_{\pi k} + \frac{1}{m}Lap(\frac{\Delta_{\mathcal{L}2}}{\epsilon_2}) W_{\pi k})$. This helps us to avoid injecting the noise directly into the coefficients $\mathbf{h}_{\pi i} y_{ik}$.
The correlation between our DP preservation and certified robustness lies in the correlation between the privacy noise $\sigma_p$  and the robustness noise $\sigma_r$. 

\textit{We can derive a robustness bound by projecting the privacy noise $\sigma_p$ on the scale of the robustness noise $\sigma_r$}. 
Given the input $x$, let $\kappa = \frac{\Delta_{\mathcal{R}}}{m \epsilon_1} / \frac{\Delta^x_r}{\epsilon_r}$, in our mechanism we have that: $\overline{x} = x + Lap(\kappa \Delta^x_r / \epsilon_r)$.
By applying a group privacy size $\kappa$ \citep{Dwork:2014:AFD:2693052.2693053,Lecuyer2018}, the scoring function $f(x)$ satisfies $\epsilon_r$-PixelDP given $\alpha \in l_p(\kappa)$, or equivalently is $\epsilon_r/\kappa$-PixelDP given $\alpha \in l_p(1)$, $\delta_r = 0$. By applying Lemma \ref{LemmaPixelDP}, we have
\begin{align}
& \forall k, \forall \alpha \in l_p(\kappa): \mathbb{E} f_k(x) \leq e^{\epsilon_r} \mathbb{E} f_k(x + \alpha), \nonumber \\
\textit{or \ \ } & \forall k, \forall \alpha \in l_p(1): \mathbb{E} f_k(x) \leq e^{\frac{\epsilon_r}{\kappa}} \mathbb{E} f_k(x + \alpha) \nonumber
\end{align}
With that, we can achieve a robustness condition against $l_p(\kappa)$-norm attacks, as follows:
\begin{equation}
\hat{\mathbb{E}}_{lb} f_k(x) > e^{2\epsilon_r} \max_{i: i\neq k} \hat{\mathbb{E}}_{ub} f_{i}(x)
\label{Robust_x} 
\end{equation} 
with the probability $\geq \eta_x$-confidence, derived from the Monte Carlo estimation of $\hat{\mathbb{E}}f(x)$. 
Our mechanism also perturbs $h$ (Eq. \ref{PerturbAutoencoder}). Given $\varphi = \frac{2\Delta_\mathcal{R}}{m \epsilon_1}/\frac{\Delta^h_r}{\epsilon_r}$, we further have $\overline{h} = h + Lap(\frac{\varphi \Delta^h_r}{\epsilon_r})$.
Therefore, the scoring function $f(x)$ also satisfies $\epsilon_r$-PixelDP given the perturbation $\alpha \in l_p(\varphi)$. In addition to the robustness to the $l_p(\kappa)$-norm attacks, we achieve an additional robustness bound in Eq. \ref{Robust_x} against $l_p(\varphi)$-norm attacks. Similar to PixelDP, these robustness conditions can be achieved as randomization processes in the inference time. They can be considered as \textit{two independent and certified defensive mechanisms} applied against two $l_p$-norm attacks, i.e., $l_p(\kappa)$ and $l_p(\varphi)$.

One challenging question here is: \textit{``What is the general robustness bound, given $\kappa$ and $\varphi$?"} Intuitively, our model is robust to attacks with $\alpha \in l_p(\frac{\kappa \varphi}{\kappa + \varphi})$. We leverage the theory of \textit{sequential composition} in DP \citep{Dwork:2014:AFD:2693052.2693053} to theoretically answer this question. Given $S$ independent mechanisms $\mathcal{M}_1, \ldots, \mathcal{M}_S$, whose privacy guarantees are $\epsilon_1, \ldots, \epsilon_S$-DP with $\alpha \in l_p(1)$. Each mechanism $\mathcal{M}_s$, which takes the input $x$ and outputs the value of $f(x)$ with the Laplace noise only injected to randomize the layer $s$ (i.e., no randomization at any other layers), denoted as $f^s(x)$, is defined as: $\forall s \in [1, S], \mathcal{M}_s f(x): \mathbb{R}^d \rightarrow f^s(x) \in \mathbb{R}^K$. 
We aim to derive a generalized robustness of any composition scoring function $f(\mathcal{M}_1, \ldots, \mathcal{M}_s| x): \prod_{s = 1}^S \mathcal{M}_s f(x)$ bounded in $[0, 1]$, defined as follows:
\begin{equation}
f(\mathcal{M}_1, \ldots, \mathcal{M}_S| x): \mathbb{R}^d \rightarrow \prod_{s \in [1,S]} f^s(x) \in \mathbb{R}^K
\end{equation} 
Our setting follows the sequential composition in DP \citep{Dwork:2014:AFD:2693052.2693053}. Thus, we can prove that the expected value $\mathbb{E} f(\mathcal{M}_1, \ldots, \mathcal{M}_S| x)$ is insensitive to small perturbations $\alpha \in l_p(1)$ in Lemma \ref{OutputStability}, and we derive our composition of robustness in Theorem \ref{SRCC}, as follows:
\begin{lemma} 
Given $S$ independent mechanisms $\mathcal{M}_1, \ldots,$ $\mathcal{M}_S$, which are $\epsilon_1, \ldots, \epsilon_S$-DP w.r.t a $l_p$-norm metric, then the expected output value of any sequential function $f$ of them, i.e., $f(\mathcal{M}_1, \ldots, \mathcal{M}_S | x) \in [0, 1]$, satisfies:
\begin{multline}
\forall \alpha \in l_p(1): \mathbb{E} f(\mathcal{M}_1, \ldots, \mathcal{M}_S | x) \leq \\ 
e^{(\sum_{s = 1}^S \epsilon_s)} \mathbb{E} f(\mathcal{M}_1, \ldots, \mathcal{M}_S | x + \alpha) \nonumber 
\end{multline} 
\label{OutputStability}
\end{lemma}

\begin{theorem} (Composition of Robustness) 
Given $S$ independent mechanisms $\mathcal{M}_1, \ldots, \mathcal{M}_S$. Given any sequential function $f(\mathcal{M}_1, \ldots, \mathcal{M}_S | x)$, and let $\hat{\mathbb{E}}_{lb}$ and $\hat{\mathbb{E}}_{ub}$ are lower and upper bounds with an $\eta$-confidence, for the Monte Carlo estimation of $\hat{\mathbb{E}}f(\mathcal{M}_1, \ldots, \mathcal{M}_S| x) = \frac{1}{n}\sum_n f(\mathcal{M}_1, \ldots, \mathcal{M}_S| x)_n = \frac{1}{n}\sum_n (\prod_{s = 1}^S f^s(x)_n)$. 
\begin{multline}
\forall x, \textit {if } \exists k \in K: \hat{\mathbb{E}}_{lb} f_k(\mathcal{M}_1, \ldots, \mathcal{M}_S | x) > \\
 e^{2(\sum_{s=1}^S \epsilon_s)} \max_{i: i\neq k} \hat{\mathbb{E}}_{ub} f_{i}(\mathcal{M}_1, \ldots, \mathcal{M}_S | x), 
\label{SCR}
\end{multline}
then the predicted label $k = \arg \max_k \hat{\mathbb{E}} f_k(\mathcal{M}_1,$ $\ldots, \mathcal{M}_S | x)$, is robust to adversarial examples $x + \alpha$, $\forall \alpha \in l_p(1)$, with probability $\geq \eta$, by satisfying: $\hat{\mathbb{E}} f_k(\mathcal{M}_1, \ldots, \mathcal{M}_S | x + \alpha)
> \max_{i: i\neq k} \hat{\mathbb{E}} f_{i}(\mathcal{M}_1, \ldots, \mathcal{M}_S | x + \alpha)$,
which is the targeted robustness condition in Eq. \ref{RobustCond1}. 
\label{SRCC}
\end{theorem}

There is no $\eta_s$-confidence for each mechanism $s$, since we do not estimate the expected value $\hat{\mathbb{E}} f^s(x)$ independently. To apply the composition of robustness in our mechanism, the noise injections into the input $x$ and its affine transformation $h$ can be considered as two mechanisms $\mathcal{M}_x$ and $\mathcal{M}_h$, sequentially applied as $(\mathcal{M}_h(x), \mathcal{M}_x(x))$. When $\mathcal{M}_h(x)$ is applied by invoking $f(x)$ with independent draws in the noise $\chi_2$, the noise $\chi_1$ injected into $x$ is fixed; and vice-versa. By applying group privacy \citep{Dwork:2014:AFD:2693052.2693053} with sizes $\kappa$ and $\varphi$, the scoring functions $f^x(x)$ and $f^h(x)$, given $\mathcal{M}_x$ and $\mathcal{M}_h$, are $\epsilon_r / \kappa$-DP and $\epsilon_r / \varphi$-DP with $\alpha \in l_p(1)$. With Theorem \ref{SRCC}, we have a generalized bound as follows: 
\begin{corollary} (StoBatch Robustness). $\forall x$, if $\exists k \in K:
\hat{\mathbb{E}}_{lb} f_k(\mathcal{M}_h, \mathcal{M}_x | x) > e^{2 \epsilon_r} \max_{i: i\neq k} \hat{\mathbb{E}}_{ub} f_{i}(\mathcal{M}_h, \mathcal{M}_x | x)$ (i.e., Eq. \ref{SCR}),
then the predicted label $k$ of our function $f(\mathcal{M}_h, \mathcal{M}_x | x)$ is robust to perturbations $\alpha \in l_p(\frac{\kappa \varphi}{\kappa + \varphi})$ with the probability $\geq \eta$, by satisfying 
\begin{multline}
\forall \alpha \in l_p(\frac{\kappa \varphi}{\kappa + \varphi}): \hat{\mathbb{E}} f_k(\mathcal{M}_h, \mathcal{M}_x | x + \alpha) >  \\
\max_{i: i \neq k} \hat{\mathbb{E}} f_i(\mathcal{M}_h, \mathcal{M}_x | x + \alpha) \nonumber
\end{multline} \par \vspace{-2.5pt}
\label{prop2} 
\end{corollary}
Compared with state-of-the-art robustness analysis \citep{DBLP:journals/corr/abs-1906-04584,Lecuyer2018}, in which either the input space or the latent space are randomized, the advantage of our robustness bound is the composition of different levels of robustness in both input and latent spaces. 

\subsection{Verified Inference} 

At the inference time, we implement a \textit{verified inference} (Alg. \ref{Verified Inferring}, \textbf{Appendix D}) to return a \textit{robustness size guarantee} for each example $x$, i.e., the maximal value of $\frac{\kappa \varphi}{\kappa + \varphi}$, for which the robustness condition in Corollary \ref{prop2} holds.
Maximizing $\frac{\kappa \varphi}{\kappa + \varphi}$ is equivalent to maximizing the robustness epsilon $\epsilon_r$, which is the only parameter controlling the size of $\frac{\kappa \varphi}{\kappa + \varphi}$; since, all the other hyper-parameters, i.e., $\Delta_{\mathcal{R}}$, $m$, $\epsilon_1$, $\epsilon_2$, $\theta_1$, $\theta_2$, $\Delta^x_r$, and $\Delta^h_r$ are fixed given a well-trained model $f(x)$: 
\begin{align}
& (\frac{\kappa \varphi}{\kappa + \varphi})_{max} = \max_{\epsilon_r} 
\frac{\Delta_\mathcal{R} \epsilon_r}{m \epsilon_1 (\Delta^x_r + \Delta^h_r / 2)} \nonumber \\
& \text{s.t. } \hat{\mathbb{E}}_{lb} f_k(x) > e^{2 \epsilon_r} \max_{i: i\neq k} \hat{\mathbb{E}}_{ub} f_{i}(x) \text{ (i.e., Eq. \ref{SCR})}
\label{RobustnessSize} 
\end{align}
The prediction on an example $x$ is robust to attacks up to $(\frac{\kappa  \varphi}{\kappa + \varphi})_{max}$. The failure probability $1$-$\eta$ can be made arbitrarily small by increasing the number of invocations of $f(x)$, with independent draws in the noise. Similar to \citep{Lecuyer2018}, Hoeffding's inequality is applied to \textit{bound} the approximation error in $\hat{\mathbb{E}}f_k(x)$ and to \textit{search} for the robustness bound $(\frac{\kappa \varphi}{\kappa + \varphi})_{max}$. We use the following sensitivity bounds $\Delta^h_r = \beta \lVert \theta_1 \rVert_\infty$ where $\lVert \theta_1 \rVert_\infty$ is the maximum 1-norm of $\theta_1$'s rows, and $\Delta^x_r = \mu d$ for $l_\infty$ attacks.   
In the Monte Carlo Estimation of $\hat{\mathbb{E}}f(x)$, we also propose a new method to draw independent noise to control the \textit{distribution shifts} between training and inferring, in order to improve the verified inference effectiveness, without affecting the DP protection and the robustness bounds (\textbf{Appendix N}). 

\subsection{Distributed Training}

In the vanilla iterative batch-by-batch training for DP DNNs, at each step, only one batch of examples can be used to train our model, so that the privacy loss can be computed \citep{Lee:2018:CDP:3219819.3220076,Yu2019,DBLP:journals/corr/abs-1905-12883,ADADP}. Parameters $\theta_1$ and $\theta_2$ are independently updated (Lines 4-12, Alg. \ref{DPAT}). This prevents us from applying \textit{practical adversarial training} \citep{Xie_2019_CVPR,DBLP:journals/corr/GoyalDGNWKTJH17}, in which \textit{distributed training using synchronized SGD} on many GPUs (e.g., 128 GPUs) is used to scale adversarial training to large DNNs. Each GPU processes a mini-batch of 32 images (i.e., the total batch size is $128 \times 32 = 4,096$).

To overcome this, a well-applied technique \citep{Yu2019} is to fine-tune a limited number of layers, such as a fully connected layer and the output layer, under DP of a pre-trained model, i.e., VGG16, trained over a public and large dataset, e.g., ImageNet, in order to handle simpler tasks on smaller private datasets, e.g., CIFAR-10. Although this approach works well, there are several utility and security concerns: \textbf{(1)} Suitable public data may not always be available, especially for highly sensitive data; \textbf{(2)} Trojans can be implanted in the pre-trained model for backdoor attacks \citep{Trojannn}; and \textbf{(3)} Public data can be poisoned \citep{NIPS2018_7849}. Fine-tuning a limited number of layers may not be secure; while fine-tuning an entire of a large pre-trained model iteratively batch-by-batch is still inefficient.

To address this bottleneck, we leverage the training recipe of \citep{Xie_2019_CVPR,DBLP:journals/corr/GoyalDGNWKTJH17} to propose a distributed training algorithm, called \textbf{StoBatch} (Fig. \ref{DNN}b), in order to efficiently train large DP DNNs in adversarial learning, without affecting the DP protection (Alg. \ref{StoBatch}, \textbf{Appendix D}). In StoBatch, fixed and disjoint batches $\overline{\mathbf{B}}$ are distributed to $N/(2m)$ local trainers, each of which have two batches $\{\overline{B}_{i1}, \overline{B}_{i2}\}$ randomly picked from $\overline{\mathbf{B}}$ with $i \in [1, N/(2m)]$ (Line 4, Alg. \ref{StoBatch}). At each training step $t$, we randomly pick $\mathbb{N}$ local trainers, each of which gets the latest global parameters $\theta$ from the parameter server. A local trainer $i$ will compute the gradients $\nabla_i \theta_1$ and $\nabla_i \theta_2$ to optimize the DP objective functions $\overline{\mathcal{R}}$ and $\overline{L}$ using its local batch $\overline{B}_{i1}$ and ensemble DP adversarial examples crafted from $\overline{B}_{i2}$ (Lines 5-14, Alg. \ref{StoBatch}). The gradients will be sent back to the parameter server for a synchronized SGD (Lines 15-16, Alg. \ref{StoBatch}), as follows: $\theta_1 \leftarrow \theta_1 - \frac{\varrho_t}{\mathbb{N}} \sum_{i \in [1, \mathbb{N}]} \nabla_i \theta_1, \text{\ \ \ }
\theta_2 \leftarrow \theta_2 - \frac{\varrho_t}{\mathbb{N}} \sum_{i \in [1, \mathbb{N}]} \nabla_i \theta_2$. 
This enables us to train large DNNs with our DP adversarial learning, by training from multiple batches simultaneously with more adversarial examples, without affecting the DP guarantee in Theorem \ref{OverallDP}; since the optimization of one batch does not affect the DP protection at any other batch and at the dataset level $D$ across $T$ training steps (\textbf{Theorem \ref{OverallDP}}). 

In addition, the \textit{average errors of our approximation functions} are always \textit{bounded}, and are \textit{independent} of the number of data instances $N$ in $D$ (\textbf{Appendix O}). This further ensures that our functions can be applied in large datasets.

Our approach can be extended into two different complementary scenarios: \textbf{(1)} Distributed training for each local trainer $i$, in which the batches $\{\overline{B}_{i1}, \overline{B}_{i2}\}$ can be located across $\mathbb{M}$ GPUs to efficiently compute the gradients $\nabla_i \theta_1 = \frac{1}{\mathbb{M}}\sum_{j \in [1, \mathbb{M}]} \nabla_{i,j} \theta_1$ and $\nabla_i \theta_2 = \frac{1}{\mathbb{M}}\sum_{j \in [1, \mathbb{M}]} \nabla_{i,j} \theta_2$; and \textbf{(2)} Federated training, given each local trainer can be considered as an independent party. In this setting, an independent party can further have different sizes of batches. As long as the global sensitivities $\Delta_{\mathcal{R}}$ and $\Delta_{\mathcal{L}2}$ are the same for all the parties, the DP guarantee in Theorem \ref{OverallDP} does hold given $D$ be the union of all local datasets from all the parties. This can be achieved by nomalizing all the inputs $x$ to be in $[-1, 1]^d$. This is a step forward compared with the classical federated learning \cite{DBLP:journals/corr/McMahanMRA16}. We focus on the distributed training setting in this paper, and reserve the federated learning scenarios for future exploration.


\section{Experimental Results}

We have conducted an extensive experiment on the MNIST, CIFAR-10, and Tiny ImageNet datasets. Our validation focuses on shedding light into the interplay among model utility, privacy loss, and robustness bounds, by learning \textbf{1)} the impact of the privacy budget $\epsilon = (\epsilon_1 + \epsilon_1/\gamma_\mathbf{x} + \epsilon_1/\gamma+\epsilon_2)$, \textbf{2)} the impact of attack sizes $\mu_a$, and \textbf{3)} the scalability of our mechanism. We consider the class of $l_\infty$-bounded adversaries. \textit{All statistical tests are 2-tail t-tests.}
Please refer to the \textbf{Appendix Q} for a complete analysis of our experimental results, including \textbf{Figures \ref{MNIST02Full} - \ref{StoBatchImageNet}}. The implementation of our mechanism is available in TensorFlow\footnote{\url{https://github.com/haiphanNJIT/StoBatch}}.


\textbf{Baseline Approaches.} Our \textbf{StoBatch} mechanism is evaluated in comparison with state-of-the-art mechanisms in: (1) DP-preserving algorithms in deep learning, i.e., \textbf{DP-SGD} \citep{Abadi}, \textbf{AdLM} \citep{NHPhanICDM17}; in (2) Certified robustness, i.e., \textbf{PixelDP} \citep{Lecuyer2018}; and in (3) DP-preserving algorithms with certified robustness, i.e., \textbf{SecureSGD} given heterogeneous noise \citep{PhanIJCAI}, and \textbf{SecureSGD-AGM} \citep{PhanIJCAI} given the Analytic Gaussian Mechanism (AGM) \citep{pmlr-v80-balle18a}.
To preserve DP, DP-SGD injects random noise into gradients of parameters, while AdLM is a Functional Mechanism-based approach. PixelDP is one of the state-of-the-art mechanisms providing certified robustness using DP bounds. SecureSGD is a combination of PixelDP and DP-SGD with an advanced heterogeneous noise distribution; i.e., ``more noise'' is injected into ``more vulnerable'' latent features, to improve the robustness. The baseline models share the same design in our experiment.
Four white-box attacks were used, including \textbf{FGSM}, \textbf{I-FGSM}, Momentum Iterative Method (\textbf{MIM}) \citep{DBLP:journals/corr/abs-1710-06081}, and \textbf{MadryEtAl} \citep{madry2018towards}. 
Pure robust training and analysis can incur privacy leakage \citep{2019arXiv190510291S}; thus, in this study, similar algorithms to \citep{DBLP:journals/corr/abs-1906-04584} do not fit as comparable baselines, since they may not be directly applicable to DP DNNs.


\textbf{Model Configuration (Appendix P).} It is important to note that $x \in [-1, 1]^d$ in our setting, which is different from a common setting, $x \in [0, 1]^d$. Thus, a given attack size $\mu_a = 0.3$ in the setting of $x \in [0, 1]^d$ is equivalent to an attack size $2\mu_a = 0.6$ in our setting. The reason for using $x \in [-1, 1]^d$ is to achieve better model utility, while retaining the same global sensitivities to preserve DP, compared with $x \in [0, 1]^d$.
As in \citep{Lecuyer2018}, we apply two accuracy metrics: 
\begin{align}
&\textit{conventional acc} = \sum_{i = 1}^{|test|} \frac{isCorrect(x_i)}{|test|} \nonumber \\
&\textit{certified acc} = \sum_{i = 1}^{|test|} \frac{isCorrect(x_i) \textit{ \& } isRobust(x_i)}{|test|} \nonumber
\end{align}
where $|test|$ is the number of test cases, $isCorrect(\cdot)$ returns $1$ if the model makes a correct prediction (else, returns 0), and $isRobust(\cdot)$ returns $1$ if the robustness size is larger than a given attack size $\mu_a$ (else, returns 0).

\textbf{Results on the MNIST Dataset.} Figure \ref{MNIST02Full} illustrates the conventional accuracy of each model as a function of the privacy budget $\epsilon$ on the MNIST dataset under $l_\infty(\mu_a)$-norm attacks, with $\mu_a = 0.2$. Our StoBatch outperforms AdLM, DP-SGD, SecureSGD, and SecureSGD-AGM, in all cases, with $p < 1.32e-4$.
When the privacy budget $\epsilon = 0.2$ (a tight DP protection), there are significant drops, in terms of conventional accuracy, given the baseline approaches. By contrast, our StoBatch only shows a small degradation in the conventional accuracy. At $\epsilon = 0.2$, our StoBatch achieves 82.7\%, compared with 11.2\% and 41.64\% correspondingly for SecureSGD-AGM and SecureSGD. This shows the ability to offer tight DP protections under adversarial example attacks in our model, compared with existing algorithms.

$\bullet$ Figure \ref{MNISTAttack02Full} presents the conventional accuracy as a function of $\mu_a$, under a strong DP guarantee, $\epsilon = 0.2$. It is clear that our StoBatch mechanism outperforms the baseline approaches in all cases. On average, our StoBatch model improves 44.91\% over SecureSGD, a 61.13\% over SecureSGD-AGM, a 52.21\% over AdLM, and a 62.20\% over DP-SGD.
More importantly, thanks to the composition robustness bounds in both input and latent spaces, and the random perturbation size $\mu_t \in (0, 1]$, our StoBatch model is resistant to different attack algorithms with different attack sizes $\mu_a$, compared with baseline approaches. 

$\bullet$ Figure \ref{MNIST2Full} demonstrates the certified accuracy as a function of $\mu_a$. The privacy budget is set to $1.0$, offering a reasonable privacy protection. In PixelDP, the construction attack bound $\epsilon_r$ is set to $0.1$, which is a pretty reasonable defense. With (small perturbation) $\mu_a \leq 0.2$, PixelDP achieves better certified accuracies under all attacks; since PixelDP does not preserve DP to protect the training data, compared with other models. 
Meanwhile, our StoBatch model outperforms all the other models when $\mu_a \geq 0.3$, indicating a stronger defense to more aggressive attacks. 

\textbf{Results on the CIFAR-10 Dataset} further strengthen our observations. In Figure \ref{CIFAR02Full}, our StoBatch outperforms baseline models in all cases ($p < 6.17e-9$), especially with small privacy budget ($\epsilon < 4$), yielding strong DP protections. On average conventional accuracy, our StoBatch mechanism has an improvement of 10.42\% over SecureSGD, 14.08\% over SecureSGD-AGM, 29.22\% over AdLM, and 14.62\% over DP-SGD.
Furthermore, the accuracy of our model is consistent given different attacks with different adversarial perturbations $\mu_a$ under a rigorous DP protection ($\epsilon_t = 2.0$), compared with baseline approaches (Figure \ref{CIFARAttack02Full}). In fact, when the attack size $\mu_a$ increases from 0.05 to 0.5, the conventional accuracies of the baseline approaches are remarkably reduced, i.e., a drop of 25.26\% on average given the most effective baseline approach, SecureSGD. Meanwhile, there is a much smaller degradation (4.79\% on average) in terms of the conventional accuracy observed in our StoBatch model.
Figure \ref{CIFAR2Full} further shows that our StoBatch model is more accurate than baseline approaches (i.e., $\epsilon_r$ is set to 0.1 in PixelDP) in terms of certified accuracy in all cases, with a tight privacy budget set to 2.0 ($p < 2.04e-18$).

\textbf{Scalability and Strong Iterative Attacks.} First, we scale our model in terms of \textit{adversarial training} in the CIFAR-10 data, i.e., the number of iterative attack steps is increased to \textit{$T_\mu$=200 in training}, and to \textit{$T_a$=2,000 in testing}. The iterative batch-by-batch DP adversarial training (Alg. \ref{DPAT}) is infeasible in this setting, taking over 30 days for one training with 600 epochs. Thanks to the distributed training, our StoBatch takes $\approxeq$ 3 days to finish the training ($\mathbb{N}=1$, $\mathbb{M}=4$). More importantly, our StoBatch achieves consistent accuracies under strong iterative attacks with $T_a$=$\{1,000; 2,000\}$, compared with the best baseline, i.e., SecureSGD (Figure \ref{StoBatchCIFAR}). On average, across attack sizes $\mu_a \in \{0.05, 0.1, 0.2, 0.3, 0.4, 0.5\}$ and steps $T_a \in \{100, 500, 1000, 2000\}$, our StoBatch achieves 45.25$\pm$1.6\% and 42.59$\pm$1.58\% in conventional and certified accuracies, compared with 29.08$\pm$11.95\% and 19.58$\pm$5.0\% of SecureSGD ($p < 2.75e-20$). 

$\bullet$ We achieve a similar improvement over the \textbf{Tiny ImageNet} with a ResNet18 model, i.e., \textit{a larger dataset on a larger network}, ($\mathbb{N}=1$, $\mathbb{M}=20$) (Figure \ref{StoBatchImageNet}). On average, across attack sizes $\mu_a \in \{0.05, 0.1, 0.2, 0.3, 0.4, 0.5\}$ and steps $T_a \in \{100, 500, 1000, 2000\}$, our StoBatch achieves 29.78$\pm$4.8\% and 28.31$\pm$1.58\% in conventional and certified accuracies, compared with 8.99$\pm$5.95\% and 8.72$\pm$5.5\% of SecureSGD ($p < 1.55e-42$). 


\textbf{Key observations:} \textbf{(1)} Incorporating ensemble adversarial learning into DP preservation, tightened sensitivity bounds, a random perturbation size $\mu_t$ at each training step, and composition robustness bounds in both input and latent spaces does enhance the consistency, robustness, and accuracy of DP model against different attacks with different levels of perturbations. These are key advantages of our mechanism; \textbf{(2)} As a result, our StoBatch model outperforms baseline algorithms, 
in terms of conventional and certified accuracies in most of the cases. It is clear that existing DP-preserving approaches have not been designed to withstand against adversarial examples; and
\textbf{(3)} Our StoBatch training can help us to scale our mechanism to larger DP DNNs and datasets with distributed adversarial learning, without affecting the model accuracies and DP protections. 

\section{Conclusion} 

In this paper, we established a connection among DP preservation to protect the training data, adversarial learning, and certified robustness. A sequential composition robustness was introduced to generalize robustness given any sequential and bounded function of independent defensive mechanisms in both input and latent spaces. We addressed the trade-off among model utility, privacy loss, and robustness by tightening the global sensitivity bounds. We further developed a stochastic batch training mechanism to bypass the vanilla iterative batch-by-batch training in DP DNNs. The average errors of our approximation functions are always bounded by constant values. Last but not least, a new Monte Carlo Estimation was proposed to stabilize the estimation of the robustness bounds. 
Rigorous experiments conducted on benchmark datasets shown that our mechanism significantly enhances the robustness and scalability of DP DNNs.
In future work, we will test our algorithms and models in the Baidu Fedcube platform \citep{FedCubeBaidu}. In addition, we will evaluate our robustness bounds against synergistic attacks, in which adversarial examples can be combined with other attacks, such as Trojans \citep{DBLP:journals/corr/abs-1708-06733,Trojannn}, to create more lethal and stealthier threats \citep{pang2020tale}. 


\section*{Acknowledgements}

The authors gratefully acknowledge the support from the National Science Foundation (NSF) grants CNS-1850094, CNS-1747798, CNS-1935928 / 1935923, and Adobe Unrestricted Research Gift.



\balance
\nocite{langley00}



\newpage

\onecolumn

\appendix

\section{Notations and Terminologies}

\begin{table*}[h]
\centering 
\caption{Notations and Terminologies.}
\begin{tabular}{|c|l|}  \hline
$D$ and $x$ & Training data with benign examples $x \in [-1, 1]^d$ \\ \hline
$y = \{y_1, \ldots, y_K\}$ & One-hot label vector of $K$ categories \\ \hline
\multirow{2}{*}{$f:\mathbb{R}^d \rightarrow \mathbb{R}^K$} & Function/model $f$ that maps inputs $x$ \\
 & to a vector of scores $f(x) = \{f_1(x), \ldots, f_K(x)\}$ \\ \hline
$y_x \in y $ & A single true class label of example $x$ \\ \hline
$y(x) = \max_{k \in K} f_k(x)$ & Predicted label for the example $x$ given the function $f$ \\ \hline
$x^{\text{adv}} = x + \alpha$ & Adversarial example where $\alpha$ is the perturbation \\ \hline
$l_p (\mu) = \{\alpha \in \mathbb{R}^d : \lVert \alpha \rVert_p \leq \mu \}$ & The $l_p$-norm ball of attack radius $\mu$ \\ \hline
$(\epsilon_r, \delta_r)$ & Robustness budget $\epsilon_r$ and broken probability $\delta_r$ \\ \hline
$\mathbb{E} f_k(x)$ & The expected value of $f_k(x)$ \\ \hline
\multirow{2}{*}{$\hat{\mathbb{E}}_{lb}$ and $\hat{\mathbb{E}}_{ub}$} & Lower and upper bounds of \\
& the expected value $\hat{\mathbb{E}} f(x) = \frac{1}{n} \sum_n f(x)_n$ \\ \hline
$a(x, \theta_1)$ & Feature representation learning model with $x$ and parameters $\theta_1$ \\ \hline
$B_t$ & A batch of benign examples $x_i$ \\ \hline
$\mathcal{R}_{B_t}(\theta_1)$ & Data reconstruction function given $B_t$ in $a(x, \theta_1)$ \\ \hline
\multirow{2}{*}{$\mathbf{h}_{1B_t} = \{\theta_1^T x_i\}_{x_i \in B_t}$} & The values of all hidden neurons in the hidden layer $\mathbf{h}_1$ \\
 & of $a(x, \theta_1)$ given the batch $B_t$ \\ \hline
$\widetilde{\mathcal{R}}_{B_t}(\theta_1)$ and $\overline{\mathcal{R}}_{\overline{B}_t}(\theta_1)$ & Approximated and perturbed functions of $\mathcal{R}_{B_t}(\theta_1)$ \\ \hline
$\overline{x}_{i}$ and $\widetilde{x}_{i}$ & Perturbed and reconstructed inputs $x_i$ \\ \hline
$\Delta_{\mathcal{R}} = d(\beta + 2)$ & Sensitivity of the approximated function $\widetilde{\mathcal{R}}_{B_t}(\theta_1)$ \\ \hline
$\overline{\mathbf{h}}_{1B_t}$ & Perturbed affine transformation $\mathbf{h}_{1B_t}$ \\ \hline
$\overline{x}^{\text{adv}}_j = x^{\text{adv}}_j  + \frac{1}{m}Lap(\frac{\Delta_{\mathcal{R}}}{\epsilon_{1}})$ & DP adversarial examples crafting from benign example $x_j$ \\ \hline
$\overline{B}_t$ and $\overline{B}^{\text{adv}}_t$ & Sets of perturbed inputs $\overline{x}_i$ and DP adversarial examples $\overline{x}^{\text{adv}}_j$ \\ \hline
$\mathcal{L}_{\overline{B}_t}\big(\theta_2\big)$ & Loss function of perturbed benign examples in $\overline{B}_t$, given $\theta_2$ \\ \hline
$\Upsilon\big(f(\overline{x}^{\text{adv}}_j, \theta_2), y_j\big)$ & Loss function of DP adversarial examples $\overline{x}^{\text{adv}}_j$, given $\theta_2$ \\ \hline
$\overline{\mathcal{L}}_{\overline{B}_t}\big(\theta_2\big)$ & DP loss function for perturbed benign examples $\overline{B}_t$ \\ \hline
$\mathcal{L}_{2\overline{B}_t}(\theta_2)$ & A part of the loss function $\mathcal{L}_{\overline{B}_t}\big(\theta_2\big)$ that needs to be DP \\ \hline
\multirow{2}{*}{$f(\mathcal{M}_1, \ldots, \mathcal{M}_s| x)$} & Composition scoring function given \\ 
 & independent randomizing mechanisms $\mathcal{M}_1, \ldots, \mathcal{M}_s$ \\ \hline
$\Delta^x_r$ and $\Delta^h_r$ & Sensitivities of $x$ and $h$, given the perturbation $\alpha \in l_p(1)$ \\ \hline
$(\epsilon_1 + \epsilon_1/\gamma_\mathbf{x} + \epsilon_1/\gamma + \epsilon_2)$ & Privacy budget to protect the training data $D$ \\ \hline
$(\frac{\kappa \varphi}{\kappa + \varphi})_{max}$ & Robustness size guarantee given an input $x$ at the inference time \\ \hline
\end{tabular}
\label{Notations} 
\end{table*}

\section{Functional Mechanism \citep{zhang2012functional}}

Functional mechanism \cite{zhang2012functional} achieves $\epsilon$-DP by perturbing
the objective function $L_D(\theta)$ and then releasing the model parameter $\overline{\theta}$ minimizing the perturbed objective function $\overline{L}_D(\theta)$ instead of the original $\theta$, given a private training dataset $D$.
The mechanism exploits the polynomial representation of $L_D(\theta)$. The model parameter $\theta$ is a vector that contains $\textbf{d}$ values $\theta_1, \ldots, \theta_\mathbf{d}$. Let $\phi(\theta)$ denote a product of $\theta_1, \ldots, \theta_\mathbf{d}$, namely, $\phi(\theta) = \theta^{c_1}_1 \cdot \theta^{c_2}_2 \cdot \cdot \cdot \theta^{c_\mathbf{d}}_\mathbf{d}$ for some $c_1, \ldots, c_\mathbf{d} \in \mathbb{N}$. Let $\Phi_j (j \in \mathbb{N})$ denote the set of all products of $\theta_1, \ldots, \theta_\mathbf{d}$ with degree $j$, i.e., $\Phi_j = \big\{\theta^{c_1}_1 \cdot \theta^{c_2}_2 \cdot \cdot \cdot \theta^{c_\mathbf{d}}_\mathbf{d} \Big\vert \sum_{a = 1}^\mathbf{d} c_a = j \big\}$. 
By the Stone-Weierstrass Theorem \cite{WalterRudin}, any continuous and differentiable $L(x_i, \theta)$ can always be written as a polynomial of $\theta_1, \ldots, \theta_\mathbf{d}$, for some $J \in [0, \infty]$, i.e., $L(x_i, \theta) = \sum_{j = 0}^J\sum_{\phi \in \Phi_j}\lambda_{\phi x_i}\phi(\theta)$ where $\lambda_{\phi x_i} \in \mathbb{R}$ denotes the coefficient of $\phi(\theta)$ in the polynomial.

For instance, the polynomial expression of the loss function in the linear regression is as follows: $L(x_i, \theta) = (y_i - x_i^\top \theta)^2 =  y_i^2 - \sum_{j = 1}^{d}(2y_ix_{ij})\theta_j + \sum_{1 \leq j, a \leq d}(x_{ij}x_{ia})\theta_j \theta_a$, where $d$ $(= \mathbf{d})$ is the number of features in $x_i$. 
In fact, $L(x_i, \theta)$ only involves monomials in $\Phi_0 = \{1\}, \Phi_1 = \{\theta_1, \ldots, \theta_d\}$, and $\Phi_2 = \{\theta_i\theta_a \big\vert i, a \in [1, d]\}$. Each $\phi(\theta)$ has its own coefficient, e.g., for $\theta_j$, its polynomial coefficient $\lambda_{\phi_{x_i}} = -2y_ix_{ij}$. Similarly, $L_D(\theta)$ can be expressed as a polynomial of $\theta_1, \ldots, \theta_d$, as
\begin{equation}
L_D(\theta) = \sum_{x_i \in D} L(x_i, \theta) = \sum_{j = 0}^J\sum_{\phi \in \Phi_j}\sum_{x_i \in D}\lambda_{\phi x_i}\phi(\theta)  
\end{equation} 

To achieve $\epsilon$-DP, $L_D(\theta)$ is perturbed by injecting Laplace noise $Lap(\frac{\Delta}{\epsilon})$ into its polynomial coefficients $\lambda_{\phi}$, and then the model parameter $\overline{\theta}$ is derived to minimize the perturbed function $\overline{L}_D(\theta)$, where the global sensitivity $\Delta = 2 \max_x \sum_{j = 1}^J \sum_{\phi \in \Phi_j} \lVert \lambda_{\phi x}\rVert_1$ is derived given any two neighboring datasets. To guarantee that the optimization of $\overline{\theta} = \arg\min_\theta \overline{L}_D(\theta)$ achieves $\epsilon$-DP without accessing the original data, i.e., that may potentially incur additional privacy leakage, grid search-based approaches are applied to learn the $\epsilon$-DP parameters $\overline{\theta}$ with low loss $\overline{L}_D(\theta)$. Although this approach works well in simple tasks, i.e., logistic regression, it may not be optimal in large models, such as DNNs.

\section{Pseudo-code of Adversarial Training \citep{DBLP:journals/corr/KurakinGB16a}}

Let $l_p (\mu) = \{\alpha \in \mathbb{R}^d : \lVert \alpha \rVert_p \leq \mu \}$ be the $l_p$-norm ball of radius $\mu$. One of the goals in adversarial learning is to minimize the risk over adversarial examples: $\theta^* = \arg \min_{\theta} \mathbb{E}_{(x, y_{\text{true}}) \sim \mathcal{D}} \big[\max_{\lVert \alpha \rVert_p \leq \mu} L\big(f(x + \alpha, \theta), y_{x}\big) \big]$, where an attack is used to approximate solutions to the inner maximization problem, and the outer minimization problem corresponds to training the model $f$ with parameters $\theta$ over these adversarial examples $x^{\text{adv}}= x + \alpha$.
There are two basic adversarial example attacks. The first one is a \textit{single-step} algorithm, e.g., \textbf{FGSM} algorithm \citep{DBLP:journals/corr/GoodfellowSS14}, in which only a single gradient computation is required to find adversarial examples by solving the inner maximization $\max_{\lVert \alpha \rVert_p \leq \mu} L\big(f(x + \alpha, \theta), y_{x}\big)$. 
The second one is an \textit{iterative} algorithm, e.g., \textbf{Iterative-FGSM} algorithm \citep{DBLP:journals/corr/KurakinGB16}, in which multiple gradients are computed and updated in $T_{\mu}$ small steps, each of which has a size of $\mu / T_{\mu}$.

Given a loss function:
\begin{equation}
L(\theta) = \frac{1}{m_1 + \xi m_2} \Big(\sum_{{x}_i \in {B}_t} \mathcal{L}\big(f({x}_i, \theta), y_i\big) 
+ \xi \sum_{{x}^{\text{adv}}_j \in {B}^{\text{adv}}_t} \Upsilon \big(f({x}^{\text{adv}}_j, \theta), y_j\big)\Big)
\end{equation}
where $m_1$ and $m_2$ correspondingly are the numbers of examples in $B_t$ and ${B}^{\text{adv}}_t$ at each training step. Algorithm \ref{TAT} presents the vanilla adversarial training. 

\begin{algorithm}[h]
\textbf{Input:} Database $D$, loss function $L$, parameters $\theta$, batch sizes $m_1$ and $m_2$, learning rate $\varrho_t$, parameter $\xi$
\begin{algorithmic}[1]
\STATE \textbf{Initialize} $\theta$ randomly
\FOR{$t \in [T]$}
	\STATE \textbf{Take} a random batch $B_t$ with the size $m_1$, and a random batch $B_a$ with the size $m_2$ \\
	\STATE Craft adversarial examples ${B}^{\text{adv}}_t = \{{x}^{\text{adv}}_j\}_{j \in [1, m_2]}$ from corresponding benign examples $x_j \in B_a$ \\
	\STATE \textbf{Descent: } $\theta \leftarrow \theta - \varrho_t \nabla_{\theta} L(\theta)$
\ENDFOR
\end{algorithmic}
\caption{\textbf{Adversarial Training \citep{DBLP:journals/corr/KurakinGB16a}}}
\label{TAT} 
\end{algorithm}

\section{Pseudo-code of Verified Inferring and StoBatch Training}

\begin{algorithm}[!t!h]
\textbf{Input:} (an input $x$, attack size $\mu_a$)
\begin{algorithmic}[1]
\STATE \textbf{Compute} robustness size $(\frac{\kappa \varphi}{\kappa + \varphi})_{max}$ in Eq. \ref{RobustnessSize} of $x$ 
\IF{$(\frac{\kappa \varphi}{\kappa + \varphi})_{max} \geq \mu_a$}
	\STATE \textbf{Return} $isRobust(x) = True$, label $k$, $(\frac{\kappa \varphi}{\kappa + \varphi})_{max}$
\ELSE
	\STATE \textbf{Return} $isRobust(x) = False$, label $k$, $(\frac{\kappa \varphi}{\kappa + \varphi})_{max}$
\ENDIF
\end{algorithmic}
\caption{\textbf{Verified Inferring}}
\label{Verified Inferring} 
\end{algorithm}

\begin{algorithm}[!t!h]
\textbf{Input:} Database $D$, loss function $L$, parameters $\theta$, batch size $m$, learning rate $\varrho_t$, privacy budgets: $\epsilon_1$ and $\epsilon_2$, robustness parameters: $\epsilon_r$, $\Delta^x_r$, and $\Delta^h_r$, adversarial attack size $\mu_a$, the number of invocations $n$, ensemble attacks $A$, parameters $\psi$ and $\xi$, the size $|\mathbf{h}_\pi|$ of $\mathbf{h}_\pi$, a number of $\mathbb{N}$ random local trainers ($\mathbb{N} \leq N/(2m)$)
\begin{algorithmic}[1]
\STATE \textbf{Draw Noise} $\chi_1 \leftarrow [Lap(\frac{\Delta_{\mathcal{R}}}{\epsilon_1})]^{d}$, $\chi_2 \leftarrow [Lap(\frac{\Delta_{\mathcal{R}}}{\epsilon_1})]^{\beta}$, $\chi_3 \leftarrow [Lap(\frac{\Delta_{\mathcal{L}2}}{\epsilon_2})]^{|\mathbf{h}_\pi|}$
\STATE \textbf{Randomly Initialize} $\theta = \{\theta_1, \theta_2\}$, $\mathbf{B} = \{B_1, \ldots, B_{N/m}\}$ s.t. $\forall B \in \mathbf{B}: B$ is a batch with the size $m$, $B_1 \cap \ldots \cap B_{N/m} = \emptyset$, and $B_1 \cup \ldots \cup B_{N/m} = D$, $\overline{\mathbf{B}} = \{\overline{B}_1, \ldots, \overline{B}_{N/m}\}$ where $\forall i \in [1, N/m]: \overline{B}_i = \{\overline{x} \leftarrow x + \frac{\chi_1}{m}\}_{x \in B_i}$ \\ 
\STATE \textbf{Construct} a deep network $f$ with \textbf{hidden layers} $\{\mathbf{h}_1 + \frac{2\chi_2}{m},\dots, \mathbf{h}_\pi\}$, where $\mathbf{h}_\pi$ is the last hidden layer \\
\STATE \textbf{Distribute} fixed and disjoint batches $\overline{\mathbf{B}}$ to $N/(2m)$ local trainers, each of which have two batches $\{\overline{B}_{i1}, \overline{B}_{i2}\}$ randomly picked from $\overline{\mathbf{B}}$ with $i \in [1, N/(2m)]$ \\ 
\FOR{$t \in [T]$}
	\STATE \textbf{Randomly Pick} $\mathbb{N}$ local trainers, each of which \textbf{Gets} the latest global parameters $\theta$ from the parameter server \\
	\FOR{$i \in [1, \mathbb{N}]$}
		\STATE \textbf{Assign} $\overline{B}_{t,i} \leftarrow \overline{B}_{i1}$ \\
		\STATE \textbf{Ensemble DP Adversarial Examples:}\\
		\STATE \textbf{Draw Random Perturbation Value} $\mu_t \in (0, 1]$, \textbf{Assign} $\overline{B}^{\text{adv}}_{t,i} \leftarrow \emptyset$ \\
		\FOR{$l \in A$}
			\STATE \textbf{Take} \text{the next batch $\overline{B}_a \subset \overline{B}_{i2}$ with the size $m/|A|$} \\
			\STATE $\forall \overline{x}_j \in \overline{B}_a$: \textbf{Craft } $\overline{x}^{\text{adv}}_j$ by using attack algorithm $A[l]$  with $l_\infty(\mu_t)$, $\overline{B}^{\text{adv}}_{t,i} \leftarrow \overline{B}^{\text{adv}}_{t,i} \cup \overline{x}^{\text{adv}}_j$\\
		\ENDFOR
		\STATE \textbf{Compute} $\nabla_i \theta_1 \leftarrow \nabla_{\theta_1}\overline{\mathcal{R}}_{\overline{B}_{t,i} \cup \overline{B}^{\text{adv}}_{t,i}}(\theta_1)$, $\nabla_i \theta_2 \leftarrow \nabla_{\theta_2} \overline{L}_{\overline{B}_{t,i} \cup \overline{B}^{\text{adv}}_{t,i}}(\theta_2)$ with the noise $\frac{\chi_3}{m}$ \\
		\STATE \textbf{Send} $\nabla_i \theta_1$ and $\nabla_i \theta_2$ to the parameter server
	\ENDFOR
	\STATE \textbf{Descent: } $\theta_1 \leftarrow \theta_1 - \varrho_t \frac{1}{\mathbb{N}} \sum_{i \in [1, \mathbb{N}]} \nabla_i \theta_1$; $\theta_2 \leftarrow \theta_2 - \varrho_t \frac{1}{\mathbb{N}} \sum_{i \in [1, \mathbb{N}]} \nabla_i \theta_2$, on the parameter server
\ENDFOR
\textbf{Output:} $\epsilon = (\epsilon_1 + \epsilon_1/\gamma_\mathbf{x} + \epsilon_1/\gamma+\epsilon_2)$-DP parameters $\theta = \{\theta_1, \theta_2\}$, robust model with an $\epsilon_r$ budget
\end{algorithmic}
\caption{\textbf{StoBatch Training}}
\label{StoBatch} 
\end{algorithm}

\newpage

\section{Proof of Lemma \ref{lemma3}}

\begin{proof}
Assume that $B_t$ and $B'_t$ differ in the last tuple, $x_m$ ($x'_m$). Then, 
\begin{align}
\Delta_{\mathcal{R}} & = \sum_{j = 1}^d \Big[\big \lVert \sum_{x_i \in B_t} \frac{1}{2}h_i - \sum_{x'_i \in B'_t} \frac{1}{2}h'_i \big\rVert_1 + \big \lVert \sum_{x_i \in B_t} x_{ij} - \sum_{x'_i \in B'_t} x'_{ij} \big\rVert_1 \Big] \nonumber \\
& \leq 2\max_{x_i} \sum_{j = 1}^d (\lVert \frac{1}{2}h_i \rVert_1 + \lVert x_{ij}\rVert_1) \leq d(\beta + 2) \nonumber
\end{align}
\label{PLemma3}
\end{proof}

\section{Proof of Lemma \ref{Operatornorm}}

\begin{proof}
Regarding the computation of $\mathbf{h}_{1\overline{B}_t} = \{\overline{\theta}_1^T \overline{x}_i\}_{\overline{x}_i \in \overline{B}_t}$, we can see that $h_i = \overline{\theta}_1^T \overline{x}_i$ is a linear function of $x$. The sensitivity of a function $h$ is defined as the maximum change in output, that can be generated by a change in the input \citep{Lecuyer2018}. Therefore, the global sensitivity of $\mathbf{h}_{1}$ can be computed as follows: 
\begin{equation}
\Delta_{\mathbf{h}_{1}} = \frac{\lVert \sum_{\overline{x}_i \in \overline{B}_t} \overline{\theta}^T_1 \overline{x}_i - \sum_{\overline{x}'_i \in \overline{B}'_t} \overline{\theta}^T_1 \overline{x}'_i \rVert_1}{\lVert \sum_{\overline{x}_i \in \overline{B}_t} \overline{x}_i - \sum_{\overline{x}'_i \in \overline{B}'_t} \overline{x}'_i \rVert_1}  \leq \max_{x_i \in B_t} \frac{\lVert \overline{\theta}^T_1 \overline{x}_i \rVert_1}{\lVert \overline{x}_i \rVert_1} \leq \lVert \overline{\theta}^T_1 \rVert_{1, 1} \nonumber 
\end{equation}
following matrix norms \citep{WikipediaOperatornorm}: $\lVert \overline{\theta}^T_1 \rVert_{1, 1}$ is the maximum 1-norm of $\theta_1$'s columns. By injecting Laplace noise $Lap(\frac{\Delta_{\mathbf{h}_{1}}}{\epsilon_1})$ into $\mathbf{h}_{1B_t}$, i.e., $\overline{\mathbf{h}}_{1\overline{B}_t} = \{\overline{\theta}_1^T \overline{x}_i + Lap(\frac{\Delta_{\mathbf{h}_{1}}}{\epsilon_1})\}_{\overline{x}_i \in \overline{B}_t}$, we can preserve $\epsilon_1$-DP in the computation of $\overline{\mathbf{h}}_{1\overline{B}_t}$. Let us set $\Delta_{\mathbf{h}_{1}} = \lVert \overline{\theta}^T_1 \rVert_{1,1}$, $\gamma = \frac{2\Delta_{\mathcal{R}}}{m \Delta_{\mathbf{h}_{1}}}$, and $\chi_2$ drawn as a Laplace noise $[Lap(\frac{\Delta_{\mathcal{R}}}{\epsilon_1})]^\beta$, in our mechanism, the perturbed affine transformation $\overline{\mathbf{h}}_{1\overline{B}_t}$ is presented as: 
\begin{align}
\overline{\mathbf{h}}_{1\overline{B}_t} & = \{\overline{\theta}_1^T \overline{x}_i + \frac{2\chi_2}{m} \}_{\overline{x}_i \in \overline{B}_t} = \{\overline{\theta}_1^T \overline{x}_i + \frac{2}{m} [Lap(\frac{\Delta_{\mathcal{R}}}{\epsilon_1})]^\beta \}_{\overline{x}_i \in \overline{B}_t} \nonumber \\
& = \{\overline{\theta}_1^T \overline{x}_i + [Lap(\frac{\gamma \Delta_{\mathbf{h}_{1}}}{\epsilon_1})]^\beta \}_{\overline{x}_i \in \overline{B}_t}  = \{\overline{\theta}_1^T \overline{x}_i + [Lap(\frac{\Delta_{\mathbf{h}_{1}}}{\epsilon_1/\gamma})]^\beta \}_{\overline{x}_i \in \overline{B}_t} \nonumber
\end{align}
This results in an $(\epsilon_1/\gamma)$-DP affine transformation $\overline{\mathbf{h}}_{1B_t} = \{\overline{\theta}_1^T \overline{x}_i + [Lap(\frac{\Delta_{\mathbf{h}_{1}}}{\epsilon_1/\gamma})]^\beta \}_{\overline{x}_i \in \overline{B}_t}$.

Similarly, the perturbed inputs $\overline{B}_t = \{\overline{x}_i\}_{\overline{x}_i \in \overline{B}_t} = \{x_i + \frac{\chi_1}{m}\}_{x_i \in B_t} = \{x_i + [Lap(\frac{\Delta_\mathbf{x}}{\epsilon_1/\gamma_\mathbf{x}})]^d\}_{x_i \in B_t}$, where $\Delta_\mathbf{x}$ is the sensitivity measuring the maximum change in the input layer that can be generated by a change in the batch $B_t$ and $\gamma_\mathbf{x} = \frac{\Delta_\mathcal{R}}{m\Delta_\mathbf{x}}$. Following \citep{Lecuyer2018}, $\Delta_\mathbf{x}$ can be computed as follows: $\Delta_\mathbf{x} = \frac{\lVert \sum_{x_i \in B_t}  \overline{x}_i - \sum_{x'_i \in B'_t}  \overline{x}'_i \rVert_1}{\lVert \sum_{x_i \in B_t} \overline{x}_i - \sum_{x'_i \in B'_t} \overline{x}'_i \rVert_1} = 1$. As a result, the computation of $\overline{B}_t$ is $(\epsilon_1/\gamma_\mathbf{x})$-DP.
Consequently, Lemma \ref{Operatornorm} does hold.
\end{proof}

\section{Proof of Theorem \ref{lemma2}}

\begin{proof}
Given $\chi_1$ drawn as a Laplace noise $[Lap(\frac{\Delta_{\mathcal{R}}}{\epsilon_1})]^d$ and $\chi_2$ drawn as a Laplace noise $[Lap(\frac{\Delta_{\mathcal{R}}}{\epsilon_1})]^\beta$, the perturbation of the coefficient $\phi \in \Phi = \{\frac{1}{2}h_i, x_{i}\}$, denoted as $\overline{\phi}$, can be rewritten as follows:
\begin{align}
\textit{for } \phi \in \{x_i\}: \overline{\phi} &= \sum_{x_i \in B} (\phi_{x_i} + \frac{\chi_1}{m}) = \sum_{x_i \in B} \phi_{x_i} + \chi_1 = \sum_{x_i \in B} \phi_{x_i} + [Lap(\frac{\Delta_{\mathcal{R}}}{\epsilon_1})]^d \label{coeff1} \\
\textit{for } \phi \in \{\frac{1}{2}h_i\}: \overline{\phi} &= \sum_{x_i \in B} \frac{1}{2}(h_i + \frac{2\chi_2}{m}) = \sum_{x_i \in B} (\phi_{x_i} + \frac{\chi_2}{m})  = \sum_{x_i \in B} \phi_{x_i} + \chi_2 = \sum_{x_i \in B} \phi_{x_i} + [Lap(\frac{\Delta_{\mathcal{R}}}{\epsilon_1})]^\beta \label{coeff2}
\end{align}
we have
\begin{equation}
Pr\big(\overline{\mathcal{R}}_{\overline{B}_t}(\theta_1)\big) = \prod_{j =1}^{d} \prod_{\phi \in \Phi} \exp\big(- \frac{\epsilon_1 \lVert \sum_{x_i \in B_t} \phi_{x_{i}} -  \overline{\phi} \rVert_1}{\Delta_{\mathcal{R}}}\big) \nonumber 
\end{equation}
$\Delta_{\mathcal{R}}$ is set to $d(\beta + 2)$, we have that:
\begin{align}
\frac{Pr\big(\overline{\mathcal{R}}_{\overline{B}_t}(\theta_1)\big)}{Pr\big(\overline{\mathcal{R}}_{\overline{B}'_t}(\theta_1)\big)} & = \frac{\prod_{j =1}^{d} \prod_{\phi \in \Phi} \exp\big(-\frac{\epsilon_1 \lVert \sum_{x_i \in B_t}  \phi_{x_{i}} -  \overline{\phi} \rVert_1}{\Delta_{\mathcal{R}}}\big)}
{\prod_{j =1}^{d} \prod_{\phi \in \Phi} \exp\big(-\frac{\epsilon_1 \lVert \sum_{x'_i \in B'_t}  \phi_{x'_{i}} -  \overline{\phi} \rVert_1}{\Delta_{\mathcal{R}}}\big)} \nonumber 
\\
& \leq \prod_{j =1}^{d} \prod_{\phi \in \Phi} \exp(\frac{\epsilon_1}{\Delta_{\mathcal{R}}} \Big\lVert \sum_{x_i \in B_t} \phi_{x_{i}} -  \sum_{x'_i \in B'_t} \phi_{{x}'_{i}} \Big\rVert_1) \nonumber 
\\
& \leq \prod_{j =1}^{d} \prod_{\phi \in \Phi} \exp(\frac{\epsilon_1}{\Delta_{\mathcal{R}}} 2\max_{x_i \in B_t} \big\lVert \phi_{x_{i}} \big\rVert _1)  \leq \exp(\frac{\epsilon_1 d(\beta + 2)}{\Delta_{\mathcal{R}}}) = \exp(\epsilon_1) 
\label{RL3}
\end{align}
Consequently, the computation of $\overline{\mathcal{R}}_{\overline{B}_t}(\theta_1)$ preserves $\epsilon_1$-DP in Alg. \ref{DPAT} (\textbf{Result 1}). 
To show that gradient descent-based optimizers can be used to optimize the objective function $\overline{\mathcal{R}}_{\overline{B}_t}(\theta_1)$ in learning private parameters $\theta_1$, we prove that all the computations on top of the perturbed data $\overline{B}_t$, including $h_i$, $\overline{h}_i$, $\widetilde{x}_i$, gradients and descent, are DP without 
incurring any additional information from the original data, as follows.

First, by following the post-processing property in DP \citep{Dwork:2014:AFD:2693052.2693053}, it is clear that the computations of $\mathbf{h}_{1\overline{B}_t} = \{h_i\}_{\overline{x}_i \in \overline{B}_t} = \theta_1^T \{\overline{x}_i\}_{\overline{x}_i \in \overline{B}_t}$ is $(\epsilon_1/\gamma_\mathbf{x})$-DP. As in Lemma \ref{Operatornorm}, we also have that $\overline{\mathbf{h}}_{1\overline{B}_t} = \{h_i + \frac{2\chi_2}{m}\}_{\overline{x}_i \in \overline{B}_t}$ is $(\epsilon_1/\gamma)$-DP. Given this, it is obvious that $\widetilde{\mathbf{x}}_{i} = \{\widetilde{x}_{i}\}_{\overline{x}_i \in \overline{B}_t} = \theta_{1}\{\overline{h}_i\}_{\overline{x}_i \in \overline{B}_t}$ is $(\epsilon_1/\gamma)$-DP, i.e., the post-processing property in DP. In addition, the computations of $\mathbf{h}_{1\overline{B}_t}$, $\overline{\mathbf{h}}_{1\overline{B}_t}$, and $\widetilde{\mathbf{x}}_{i}$ do not access the original data $B_t$. Therefore, they do not incur any additional information from the private data, except the privacy loss measured by $(\epsilon_1/\gamma_\mathbf{x})$-DP, since the computations of $\overline{\mathbf{h}}_{1\overline{B}_t}$ and $\widetilde{\mathbf{x}}_{i}$ are based on the $(\epsilon_1/\gamma_\mathbf{x})$-DP $\mathbf{h}_{1\overline{B}_t}$. (\textbf{Result 2})

Second, the gradient of a particular parameter $\theta_{1j}$, with $\forall j \in [1, d]$, can be computed as follows:
\begin{align}
\forall j \in [1, d]: \nabla_{\theta_{1j}}\overline{\mathcal{R}}_{\overline{B}_t}(\theta_{1}) & = \frac{\delta \overline{\mathcal{R}}_{\overline{B}_t}(\theta_1)}{\delta \theta_{1j}} = \sum_{i = 1}^{m}\overline{h}_i(\frac{1}{2} - \overline{x}_{ij}) \label{GradSub} \\
& = \sum_{i = 1}^{m}(h_i + \frac{2\chi_2}{m}) (\frac{1}{2} - \overline{x}_{ij}) \\
& = \big[\sum_{i = 1}^{m} h_i(\frac{1}{2} - \overline{x}_{ij})\big] + \chi_2 - \big[\frac{2\chi_2}{m} \sum_{i = 1}^{m} \overline{x}_{ij} \big] 
\label{Grad1}
\end{align}
In Eq. \ref{Grad1}, we have that $\sum_{i = 1}^{m} \overline{x}_{ij} = (\sum_{i = 1}^{m} x_{ij}) + Lap(\frac{\Delta_{\mathcal{R}}}{\epsilon_1})$ (Eq. \ref{coeff1}), which is $(\epsilon_1/\gamma_{\mathbf{x}})$-DP. Therefore, the term $\frac{2\chi_2}{m} \sum_{i = 1}^{m} \overline{x}_{ij}$ also is $(\epsilon_1/\gamma_{\mathbf{x}})$-DP (the post-processing property in DP). (\textbf{Result 3})



Regarding the term $\sum_{i = 1}^{m} h_i(\frac{1}{2} - \overline{x}_{ij})$ in Eq. \ref{Grad1}, its global sensitivity given two arbitrary neighboring batches, denoted as $\Delta_g$, can be bounded as follows: $\Delta_g \leq 2 \max_{\overline{x}_i} \lVert h_i (\frac{1}{2} - \overline{x}_{ij})\rVert_1 = 3\beta$. As a result, we have that: 
\begin{equation}
\big[\sum_{i = 1}^{m} h_i(\frac{1}{2} - \overline{x}_{ij})\big] + \chi_2 = \big[\sum_{i = 1}^{m} h_i(\frac{1}{2} - \overline{x}_{ij})\big] + [Lap(\frac{\Delta_g}{\epsilon_1/\frac{\Delta_{\mathcal{R}}}{\Delta_g}})]^\beta
\end{equation}
which is $(\epsilon_1/\frac{\Delta_{\mathcal{R}}}{\Delta_g}$)-DP. (\textbf{Result 4})

From Results 3 and 4, the computation of gradients $\nabla_{\theta_{1j}}\overline{\mathcal{R}}_{\overline{B}_t}(\theta_{1})$ is $(\epsilon_1/\frac{\Delta_{\mathcal{R}}}{\Delta_g} + \epsilon_1/\gamma_{\mathbf{x}}$)-DP, since: \textbf{(1)} The computations of the two terms in Eq. \ref{Grad1} can be treated as two independent DP-preserving mechanisms applied on the perturbed batch $\overline{B}_t$; and \textbf{(2)} This is true for every dimension $j \in [1, d]$, each of which $\nabla_{\theta_{1j}}$ is independently computed and bounded. It is important to note that this result is different from the traditional DPSGD \citep{Abadi}, in which the parameter gradients are jointly clipped by a $l_2$-norm constant bound, such that Gaussian noise can be injected to achieve DP. In addition, as in Eq. \ref{GradSub}, the computation of $\nabla_{\theta_{1j}}\overline{\mathcal{R}}_{\overline{B}_t}(\theta_{1})$ only uses $(\epsilon_1/\gamma_{\mathbf{x}})$-DP $\overline{B}_t = \{\overline{x}_i\}_{\overline{x}_i \in \overline{B}_t}$ and $(\epsilon_1/\gamma)$-DP $\overline{\mathbf{h}}_{1\overline{B}_t}$, without accessing the original data. Basically, $\overline{\mathbf{h}}_{1\overline{B}_t}$ is computed on top of $\overline{B}_t$, without touching any benign example. Therefore, it does not incur any additional information from the private data, except the privacy loss $(\epsilon_1/\frac{\Delta_{\mathcal{R}}}{\Delta_g} + \epsilon_1/\gamma_{\mathbf{x}})$-DP.
In practice, we observed that $\epsilon_1/\gamma_{\mathbf{x}} \gg \epsilon_1/\frac{\Delta_{\mathcal{R}}}{\Delta_g} \approxeq \epsilon_1 \times 1e-3$, which is tiny. We can simply consider that the computation of gradients $\nabla_{\theta_{1j}}\overline{\mathcal{R}}_{\overline{B}_t}(\theta_{1})$ is $(\epsilon_1/\gamma_{\mathbf{x}}$)-DP without affecting the general DP protection. In addition to the gradient computation, the descent operations are simply post-processing steps without consuming any further privacy budget. (\textbf{Result 5})

From Results 1, 2, and 5, we have shown that all the computations on top of $(\epsilon_1/\gamma_{\mathbf{x}})$-DP $\overline{B}_t$, including parameter gradients and gradient descents, clearly are DP without accessing the original data; therefore, they do not incur any additional information from the private data (the post-processing property in DP). As a result, gradient descent-based approaches can be applied to optimize $\overline{\mathcal{R}}_{\overline{B}_t}(\theta_1)$ in Alg. \ref{DPAT}. The total privacy budget to learn the perturbed optimal parameters $\overline{\theta}_1$ in Alg. \ref{DPAT} is $(\epsilon_1/\gamma_{\mathbf{x}} + \epsilon_1)$-DP, where the $\epsilon_1/\gamma_{\mathbf{x}}$ is counted for the perturbation on the batch of benign examples $B_t$.

Consequently, Theorem \ref{lemma2} does hold.

\label{PTheorem1}



\end{proof}

\section{Proof of Lemma \ref{NovelSensitivity}}

\begin{proof} Assume that $\overline{B}_t$ and $\overline{B}'_t$
 differ in the last tuple, and $\overline{x}_m$ $(\overline{x}'_m)$ be the last tuple in $\overline{B}_t$ $(\overline{B}'_t)$, we have that
\begin{equation} 
\Delta_{\mathcal{L}2} = \sum_{k = 1}^K \Big\lVert \sum_{\overline{x}_i \in \overline{B}_t} (\mathbf{h}_{\pi i} y_{ik}) - \sum_{\overline{x}'_i \in \overline{B}'_t} (\mathbf{h}'_{\pi i} y'_{ik}) \Big\rVert_1  = \sum_{k = 1}^K \big\lVert \mathbf{h}_{\pi m} y_{mk} - \mathbf{h}'_{\pi m} y'_{mk} \big\rVert_1 \nonumber
\end{equation}
Since $y_{mk}$ and $y'_{mk}$ are one-hot encoding, we have that $\Delta_{\mathcal{L}2} \leq 2 \max_{\overline{x}_i}\lVert \mathbf{h}_{\pi i} \rVert_1$. Given $\mathbf{h}_{\pi i} \in [-1, 1]$, we have 
\begin{equation}
\Delta_{\mathcal{L}2} \leq 2|\mathbf{h}_{\pi}|
\end{equation} 
Lemma \ref{NovelSensitivity} does hold.
\end{proof}

\section{Proof of Theorem \ref{BenignLoss}}

\begin{proof}
Let $\overline{B}_t$ and $\overline{B}'_t$ be neighboring batches of benign examples, and $\chi_3$ drawn as Laplace noise $[Lap(\frac{\Delta_{\mathcal{L}2}}{\epsilon_2})]^{|\mathbf{h}_{\pi}|}$, the perturbations of the coefficients $\mathbf{h}_{\pi i} y_{ik}$ can be rewritten as:
\begin{equation}
\overline{\mathbf{h}}_{\pi i} \overline{y}_{ik} = \sum_{\overline{x}_i} (\mathbf{h}_{\pi i} y_{ik} + \frac{\chi_3}{m}) = \sum_{\overline{x}_i} (\mathbf{h}_{\pi i} y_{ik}) + [Lap(\frac{\Delta_{\mathcal{L}2}}{\epsilon_2})]^{|\mathbf{h}_{\pi}|} \nonumber
\end{equation}
Since all the coefficients are perturbed, and given $\Delta_{\mathcal{L}2} = 2|\mathbf{h}_{\pi}|$, we have that
\begin{align}
& \frac{Pr(\mathcal{L}_{\overline{B}_t}(\theta_2))}{Pr(\mathcal{L}_{\overline{B}'_t}(\theta_2))} = \frac{Pr(\mathcal{L}_{1\overline{B}_t}(\theta_2))}{Pr(\mathcal{L}_{1\overline{B}'_t}(\theta_2))} \times \frac{Pr(\overline{\mathcal{L}}_{2\overline{B}_t}(\theta_2))}{Pr(\overline{\mathcal{L}}_{2\overline{B}'_t}(\theta_2))} \nonumber \\
& \leq e^{\epsilon_1/\gamma} \sum_{k = 1}^K \frac{\exp(-\frac{\epsilon_2 \lVert \sum_{\overline{x}_i} \mathbf{h}_{\pi i} y_{ik} - \overline{\mathbf{h}}_{\pi i} \overline{y}_{ik} \rVert_1}{\Delta_{\mathcal{L}2}})}{\exp(-\frac{\epsilon_2 \lVert \sum_{\overline{x}'_i} \mathbf{h}_{\pi i} y_{ik} - \overline{\mathbf{h}}_{\pi i} \overline{y}_{ik} \rVert_1}{\Delta_{\mathcal{L}2}})} \nonumber \\
& \leq e^{\epsilon_1/\gamma} \sum_{k = 1}^K \exp(\frac{\epsilon_2}{\Delta_{\mathcal{L}2}} \big\lVert \sum_{\overline{x}_i} \mathbf{h}_{\pi i} y_{ik} - \sum_{\overline{x}'_i} \mathbf{h}_{\pi i} y_{ik}\big\rVert_1) \nonumber \\
& \leq e^{\epsilon_1/\gamma} \exp(\frac{\epsilon_2}{\Delta_{\mathcal{L}2}} 2 \max_{\overline{x}_i}\lVert \mathbf{h}_{\pi i} \rVert_1) = e^{\epsilon_1/\gamma + \epsilon_2} \nonumber
\end{align}
The computation of $\overline{\mathcal{L}}_{2\overline{B}_t}\big(\theta_2\big)$ preserves $(\epsilon_1/\gamma + \epsilon_2)$-differential privacy. Similar to Theorem \ref{lemma2}, the gradient descent-based optimization of $\overline{\mathcal{L}}_{2\overline{B}_t}\big(\theta_2\big)$ does not access additional information from the original input $x_i \in B_t$. It only reads the $(\epsilon_1/\gamma)$-DP $\overline{\mathbf{h}}_{1\overline{B}_t} = \{h_i + \frac{2\chi_2}{m}\}_{\overline{x}_i \in \overline{B}_t}$. Consequently, the optimal perturbed parameters $\overline{\theta}_2$ derived from $\overline{\mathcal{L}}_{2\overline{B}_t}\big(\theta_2\big)$ are $(\epsilon_1/\gamma + \epsilon_2)$-DP.
\end{proof}

\section{Proofs of Theorem \ref{LemmaDPAE}  and Theorem \ref{OverallDP}}

\begin{proof}
First, we optimize for a single draw of noise during training (Line 3, Alg. \ref{DPAT}) and all the batches of perturbed benign examples are disjoint and fixed across epochs. As a result, the computation of $\overline{x}_i$ is equivalent to a data preprocessing step with DP, which does not incur any additional privacy budget consumption over $T$ training steps (the post-processing property of DP) (\textbf{Result 1}). That is different from repeatedly applying a DP mechanism on either the same or overlapping datasets causing the accumulation of the privacy budget. 

Now, we show that our algorithm achieves DP at the dataset level $D$. Let us consider the computation of the first hidden layer, given any two neighboring datasets $D$ and $D'$ differing at most one tuple $\overline{x}_e \in D$ and $\overline{x}'_e \in D'$. For any $O = \prod_{i = 1}^{N/m}o_i \in \prod_{i = 1}^{N/m} \overline{\mathbf{h}}_{1\overline{B}_i} (\in \mathbb{R}^{\beta \times m})$, we have that
\begin{align}
\frac{P\big(\overline{\mathbf{h}}_{1D} = O \big)}{P\big(\overline{\mathbf{h}}_{1D'} = O \big)} = \frac{P(\overline{\mathbf{h}}_{1\overline{B}_1} = o_1) \ldots P(\overline{\mathbf{h}}_{1\overline{B}_{N/m}} = o_{N/m})}{P(\overline{\mathbf{h}}_{1\overline{B}'_1} = o_1) \ldots P(\overline{\mathbf{h}}_{1\overline{B}'_{N/m}} = o_{N/m})}
\label{Cond1}
\end{align}
By having disjoint and fixed batches, we have that: 
\begin{equation}
\exists! \tilde{B} \in \overline{\mathbf{B}} \textit{ s.t. } x_e \in \tilde{B} \textit{ and } \exists! \tilde{B}' \in \overline{\mathbf{B}}' \textit{ s.t. } x'_e \in \tilde{B}'
\label{Cond2}
\end{equation}

From Eqs. \ref{Cond1}, \ref{Cond2}, and Lemma \ref{Operatornorm}, we have that 
\begin{align}
& \forall \overline{B} \in \mathbf{B}, \overline{B} \neq \tilde{B}: \overline{B} = \overline{B}' \Rightarrow \frac{P\big(\overline{\mathbf{h}}_{1\overline{B}} = o \big)}{P\big(\overline{\mathbf{h}}_{1\overline{B}'} = o \big)} = 1 
\label{Cond2-1}
\\
& \textit{Eqs. \ref{Cond2} and \ref{Cond2-1}} \Rightarrow \frac{P\big(\overline{\mathbf{h}}_{1D} = O \big)}{P\big(\overline{\mathbf{h}}_{1D'} = O \big)} = \frac{P\big(\overline{\mathbf{h}}_{1\tilde{B}} = \tilde{o} \big)}{P\big(\overline{\mathbf{h}}_{1\tilde{B}'} = \tilde{o} \big)} \leq e^{\epsilon_1/\gamma}
\label{Cond3}
\end{align}
As a result, the computation of $\overline{\mathbf{h}}_{1D}$ is $(\epsilon_1/\gamma)$-DP given the data $D$, since the Eq. \ref{Cond3} does hold for any tuple $x_e \in D$. That is consistent with the parallel composition property of DP, in which batches can be considered disjoint datasets given $\overline{\mathbf{h}}_{1\overline{B}}$ as a DP mechanism \citep{Dwork:2014:AFD:2693052.2693053}.

This does hold across epochs, since batches $\overline{\mathbf{B}}$ are disjoint and fixed among epochs. At each training step $t \in [1, T]$, the computation of $\overline{\mathbf{h}}_{1\overline{B}_t}$ does not access the original data. It only reads the perturbed batch of inputs $\overline{B}_t$, which is $(\epsilon_1/\gamma_\mathbf{x})$-DP (Lemma \ref{Operatornorm}). Following the post-processing property in DP \citep{Dwork:2014:AFD:2693052.2693053}, the computation of $\overline{\mathbf{h}}_{1\overline{B}_t}$ does not incur any additional information from the original data across $T$ training steps. 
(\textbf{Result 2})

Similarly, we show that the optimization of the function $\overline{\mathcal{R}}_{\overline{B}_t}(\theta_1)$ is $(\epsilon_1/\gamma_{\mathbf{x}} + \epsilon_1)$-DP across $T$ training steps. As in Theorem \ref{lemma2} and Proof \ref{PTheorem1}, we have that $Pr\big(\overline{\mathcal{R}}_{\overline{B}}(\theta_1)\big) = \prod_{j =1}^{d} \prod_{\phi \in \Phi} \exp\big(- \frac{\epsilon_1 \lVert \sum_{x_i \in B} \phi_{x_{i}} -  \overline{\phi} \rVert_1}{\Delta_{\mathcal{R}}}\big)$, where $\overline{B} \in \overline{\mathbf{B}}$. Given any two perturbed neighboring datasets $\overline{D}$ and $\overline{D}'$ differing at most one tuple $\overline{x}_e \in \overline{D}$ and $\overline{x}'_e \in \overline{D}'$:
\begin{equation}
\frac{Pr\big(\overline{\mathcal{R}}_{\overline{D}}(\theta_1)\big)}{Pr\big(\overline{\mathcal{R}}_{\overline{D}'}(\theta_1)\big)} = \frac{
Pr\big(\overline{\mathcal{R}}_{\overline{B}_1}(\theta_1)\big) \ldots Pr\big(\overline{\mathcal{R}}_{\overline{B}_{N/m}}(\theta_1)\big)}{Pr\big(\overline{\mathcal{R}}_{\overline{B}'_1}(\theta_1)\big) \ldots Pr\big(\overline{\mathcal{R}}_{\overline{B}'_{N/m}}(\theta_1)\big)}
\label{Cond4}
\end{equation}
From Eqs. \ref{Cond2}, \ref{Cond4}, and Theorem \ref{lemma2}, we have that
\begin{align}
& \forall \overline{B} \in \mathbf{B}, \overline{B} \neq \tilde{B}:  \overline{B} = \overline{B}' \Rightarrow \frac{P\big(\overline{\mathcal{R}}_{\overline{B}}(\theta_1) \big)}{P\big(\overline{\mathcal{R}}_{\overline{B}'}(\theta_1) \big)} = 1 
\label{Cond4-1}
\\
& \textit{Eqs. \ref{Cond4} and \ref{Cond4-1}} \Rightarrow \frac{P\big(\overline{\mathcal{R}}_{\overline{D}}(\theta_1) \big)}{P\big(\overline{\mathcal{R}}_{\overline{D}'}(\theta_1) \big)} = \frac{P\big(\overline{\mathcal{R}}_{\tilde{B}}(\theta_1) \big)}{P\big(\overline{\mathcal{R}}_{\tilde{B}'}(\theta_1) \big)} \leq e^{\epsilon_1}
\label{Cond5}
\end{align}
As a result, the optimization of $\overline{\mathcal{R}}_{\overline{D}}(\theta_1)$ is $(\epsilon_1/\gamma_{\mathbf{x}} + \epsilon_1)$-DP given the data $\overline{D}$ (which is $\epsilon_1/\gamma_{\mathbf{x}}$-DP (Lemma \ref{Operatornorm})), since the Eq. \ref{Cond5} does hold for any tuple $\overline{x}_e \in \overline{D}$. This is consistent with the parallel composition property in DP \citep{Dwork:2014:AFD:2693052.2693053}, in which batches can be considered disjoint datasets and the optimization of the function on one batch does not affect the privacy guarantee in any other batch, even the objective function given one batch can be slightly different from the objective function given any other batch in $\overline{\mathbf{B}}$. In addition, $\forall t \in [1, T]$, the optimization of $\overline{\mathcal{R}}_{\overline{B}_t}(\theta_1)$ does not use any additional information from the original data $D$. Consequently, the privacy budget is $(\epsilon_1/\gamma_{\mathbf{x}} + \epsilon_1)$ across $T$ training steps, following the post-processing property in DP \citep{Dwork:2014:AFD:2693052.2693053} (\textbf{Result 3}). 

Similarly, we can also prove that  optimizing the data reconstruction function $\overline{\mathcal{R}}_{\overline{B}^{adv}_t}(\theta_1)$ given the DP adversarial examples crafted in Eqs. \ref{DPAS} and \ref{I-DPAS}, i.e., $\overline{x}^{\text{adv}}_j$, is also $(\epsilon_1/\gamma_{\mathbf{x}} + \epsilon_1)$-DP given $t \in [1, T]$ on the training data $D$. First, DP adversarial examples $\overline{x}^{\text{adv}}_j$ are crafted from perturbed benign examples $\overline{x}_j$. As a result, the computation of the batch $\overline{B}^{adv}_t$ of DP adversarial examples is 1) $(\epsilon_1/\gamma_\mathbf{x})$-DP (the post-processing property of DP \citep{Dwork:2014:AFD:2693052.2693053}), and 2) does not access the original data $\forall t \in [1, T]$. 
In addition, the computation of $\overline{\mathbf{h}}_{1\overline{B}^{adv}_t}$ and the optimization of $\overline{\mathcal{R}}_{\overline{B}^{adv}_t}(\theta_1)$ correspondingly are $\epsilon_1/\gamma$-DP and $\epsilon_1$-DP. In fact, the data reconstruction function $\overline{\mathcal{R}}_{\overline{B}^{adv}_t}$ is presented as follows:
\begin{align}
& \overline{\mathcal{R}}_{\overline{B}^{adv}_t}(\theta_1)  = \sum_{\overline{x}^{\text{adv}}_j \in \overline{B}^{adv}_t} \Big[\sum_{i = 1}^d (\frac{1}{2}\theta_{1i}\overline{h}^{\text{adv}}_j) - \overline{x}^{\text{adv}}_j \widetilde{x}^{\text{adv}}_{j} \Big] \nonumber \\
& = \sum_{\overline{x}^{\text{adv}}_j \in \overline{B}^{adv}_t} \Big[\sum_{i = 1}^d (\frac{1}{2}\theta_{1i}\overline{h}^{\text{adv}}_j) - \overline{x}_j\widetilde{x}^{\text{adv}}_{j} - \mu \cdot \text{sign}\Big(\nabla_{\overline{x}_j} \mathcal{L}\big(f(\overline{x}_j, \theta), y (\overline{x}_j)\big)\Big)\widetilde{x}^{\text{adv}}_{j}  \Big]  \nonumber \\
& = \sum_{\overline{x}^{\text{adv}}_j \in \overline{B}^{adv}_t} \Big[\sum_{i = 1}^d (\frac{1}{2}\theta_{1i}\overline{h}^{\text{adv}}_j) - \overline{x}_j\widetilde{x}^{\text{adv}}_{j} \Big] - \sum_{\overline{x}^{\text{adv}}_j \in \overline{B}^{adv}_t} \mu \cdot \text{sign}\Big(\nabla_{\overline{x}_j} \mathcal{L}\big(f(\overline{x}_j, \theta), y (\overline{x}_j)\big)\Big)\widetilde{x}^{\text{adv}}_{j} 
\label{PerturbDPAE} 
\end{align}
where $h^{\text{adv}}_j = \theta_1^T \overline{x}^{\text{adv}}_j, \overline{h}^{\text{adv}}_j = h^{\text{adv}}_j + \frac{2}{m}Lap(\frac{\Delta_{\mathcal{R}}}{\epsilon_1})$, and $\widetilde{x}^{\text{adv}}_{j} = \theta_{1}\overline{h}^{\text{adv}}_j$. The right summation component in Eq. \ref{PerturbDPAE} does not disclose any additional information, since the $sign(\cdot)$ function is computed from perturbed benign examples (the post-processing property in DP \citep{Dwork:2014:AFD:2693052.2693053}). Meanwhile, the left summation component has the same form with $\overline{\mathcal{R}}_{\overline{B}_t}(\theta_1)$ in Eq. \ref{PerturbAutoencoder}. Therefore, we can employ the Proof \ref{PTheorem1} in Theorem \ref{lemma2}, by replacing the coefficients $\Phi = \{\frac{1}{2}h_i, x_{i}\}$ with $\Phi = \{\frac{1}{2}h^{\text{adv}}_j, x_{j}\}$ to prove that the optimization of $\overline{\mathcal{R}}_{\overline{B}^{adv}_t}(\theta_1)$ is $(\epsilon_1/\gamma_{\mathbf{x}} + \epsilon_1)$-DP. As a result, Theorem \ref{LemmaDPAE} does hold. (\textbf{Result 4})

In addition to the Result 4, by applying the same analysis in Result 3, we can further show that the optimization of $\overline{\mathcal{R}}_{D^{\textit{adv}}}(\theta_1)$ is $(\epsilon_1/\gamma_{\mathbf{x}} + \epsilon_1)$-DP given the DP adversarial examples $D^{\text{adv}}$ crafted using the data $\overline{D}$ across $T$ training steps, since batches used to created DP adversarial examples are disjoint and fixed across epochs. It is also straightforward to conduct the same analysis in Result 2, in order to prove that the computation of the first affine transformation $\overline{\mathbf{h}}_{1\overline{B}^{adv}_t} = \{\overline{\theta}_1^T \overline{x}^{\text{adv}}_j + \frac{2}{m}Lap(\frac{\Delta_{\mathcal{R}}}{\epsilon_1}) \}_{\overline{x}^{\text{adv}}_j \in \overline{B}^{adv}_t}$ given the batch of DP adversarial examples $\overline{B}^{adv}_t$, is $(\epsilon_1/\gamma)$-DP with $t \in [1, T]$ training steps. This is also true given the data level $D^{\text{adv}}$. (\textbf{Result 5})

Regarding the output layer, the Algorithm \ref{DPAT} preserves $(\epsilon_1/\gamma + \epsilon_2)$-DP in optimizing the adversarial objective function $\overline{L}_{\overline{B}_t \cup \overline{B}^{\text{adv}}_t}(\theta_2)$ (Theorem \ref{BenignLoss}). We apply the same technique to preserve $(\epsilon_1/\gamma + \epsilon_2)$-DP across $T$ training steps given disjoint and fixed batches derived from the private training data $D$. In addition, as our objective functions $\overline{\mathcal{R}}$ and $\overline{L}$ are always optimized given two disjoint batches $\overline{B}_t$ and $\overline{B}^{\text{adv}}_t$, the  privacy budget used to preserve DP in these functions is $(\epsilon_1 + \epsilon_1/\gamma + \epsilon_2)$, following the \textit{parallel composition} property in DP \citep{Dwork:2014:AFD:2693052.2693053}. \textbf{(Result 6)}

With the \textbf{Results 1-6}, all the computations and optimizations in the Algorithm \ref{DPAT} are DP following the post-processing property in DP \citep{Dwork:2014:AFD:2693052.2693053}, by working on perturbed inputs and perturbed coefficients. The crafting and utilizing processes of DP adversarial examples based on the perturbed benign examples do not disclose any additional information. The optimization of our DP adversarial objective function at the output layer is DP to protect the ground-truth labels. More importantly, the DP guarantee in learning given the whole dataset level $\overline{D}$ is equivalent to the DP guarantee in learning on disjoint and fixed batches across epochs. Consequently, Algorithm \ref{DPAT} preserves $(\epsilon_1 + \epsilon_1/\gamma_{\mathbf{x}} + \epsilon_1/\gamma + \epsilon_2)$-DP in learning private parameters $\overline{\theta} = \{\overline{\theta}_1, \overline{\theta}_2\}$ given the training data $D$ across $T$ training steps. Note that the $\epsilon_1/\gamma_{\mathbf{x}}$ is counted for the perturbation on the benign examples. Theorem \ref{OverallDP} does hold.
\end{proof}

\section{Proof of Lemma \ref{OutputStability}}

\begin{proof}
Thanks to the sequential composition theory in DP \citep{Dwork:2014:AFD:2693052.2693053}, $f(\mathcal{M}_1, \ldots, \mathcal{M}_S | x)$ is $(\sum_s \epsilon_s)$-DP, since for any $O = \prod_{s = 1}^S o_s \in \prod_{s = 1}^S f^s(x) (\in \mathbb{R}^K)$, we have that
\begin{align}
\frac{P\big(f(\mathcal{M}_1, \ldots, \mathcal{M}_S | x)  = O \big)}{P\big(f(\mathcal{M}_1, \ldots, \mathcal{M}_S | x + \alpha) = O \big)} & = \frac{P(\mathcal{M}_1 f(x) = o_1) \ldots P(\mathcal{M}_S f(x) = o_S)}{P(\mathcal{M}_1 f(x + \alpha) = o_1) \ldots P(\mathcal{M}_S f(x + \alpha) = o_S)} \nonumber \\ 
& \leq \prod_{s = 1}^S \exp(\epsilon_s) = e^{(\sum_{s = 1}^S \epsilon_s)} \nonumber
\end{align}
As a result, we have 
\begin{equation}
P\big(f(\mathcal{M}_1, \ldots, \mathcal{M}_S | x) \big) \leq e^{(\sum_i \epsilon_i)} P\big( f(\mathcal{M}_1, \ldots, \mathcal{M}_S | x + \alpha)\big) \nonumber 
\end{equation}
The sequential composition of the expected output is as: 
\begin{align}
\mathbb{E} f(\mathcal{M}_1, \ldots, \mathcal{M}_S | x) & = \int_0^1 P\big( f(\mathcal{M}_1, \ldots, \mathcal{M}_S | x) > t\big) dt \nonumber \\
& \leq e^{(\sum_s \epsilon_s)} \int_0^1 P\big( f(\mathcal{M}_1, \ldots, \mathcal{M}_S | x + \alpha) > t\big) dt \nonumber \\
& = e^{(\sum_s \epsilon_s)} \mathbb{E} f(\mathcal{M}_1, \ldots, \mathcal{M}_S | x + \alpha) \nonumber
\end{align} 
Lemma \ref{OutputStability} does hold.
\end{proof}

\section{Proof of Theorem \ref{SRCC}}

\begin{proof}
$\forall \alpha \in l_p(1)$, from Lemma \ref{OutputStability}, with probability $\geq \eta$, we have that
\begin{equation}
\hat{\mathbb{E}} f_k(\mathcal{M}_1, \ldots, \mathcal{M}_S | x + \alpha) \geq \frac{\hat{\mathbb{E}} f_k(\mathcal{M}_1, \ldots, \mathcal{M}_S | x)}{e^{(\sum_{s = 1}^s\epsilon_s)}} \geq \frac{\hat{\mathbb{E}}_{lb} f_k(\mathcal{M}_1, \ldots, \mathcal{M}_S | x)}{e^{(\sum_{s = 1}^S\epsilon_s)}} \label{FirstInequality}
\end{equation}
In addition, we also have
\begin{equation}
\forall i \neq k: \hat{\mathbb{E}} f_{i: i \neq k}(\mathcal{M}_1, \ldots, \mathcal{M}_S | x + \alpha)
 \leq e^{{(\sum_{s = 1}^S\epsilon_s)}} \hat{\mathbb{E}} f_{i: i \neq k}(\mathcal{M}_1, \ldots, \mathcal{M}_S | x) \nonumber 
\end{equation}
\begin{equation}
\Rightarrow \forall i \neq k: \hat{\mathbb{E}} f_i(\mathcal{M}_1, \ldots, \mathcal{M}_S | x + \alpha)
\leq e^{{(\sum_{s = 1}^S\epsilon_s)}} \max_{i: i \neq k} \hat{\mathbb{E}}_{ub} f_i(\mathcal{M}_1, \ldots, \mathcal{M}_S | x) 
\label{thirdinequality}
\end{equation}

Using the hypothesis (Eq. \ref{SCR}) and the first inequality (Eq. \ref{FirstInequality}), we have that 
\begin{align}
\hat{\mathbb{E}} f_k(\mathcal{M}_1, \ldots, \mathcal{M}_S | x + \alpha) & > \frac{e^{2(\sum_{s=1}^S \epsilon_s)} \max_{i: i\neq k} \hat{\mathbb{E}}_{ub} f_{i}(\mathcal{M}_1, \ldots, \mathcal{M}_S | x)}{e^{(\sum_{s=1}^S \epsilon_s)}} \nonumber \\
& > e^{(\sum_{s=1}^S \epsilon_s)} \max_{i: i\neq k} \hat{\mathbb{E}}_{ub} f_{i}(\mathcal{M}_1, \ldots, \mathcal{M}_S | x) \nonumber
\end{align}
Now, we apply the third inequality (Eq. \ref{thirdinequality}), we have that
\begin{align}
& \forall i \neq k: \hat{\mathbb{E}} f_k(\mathcal{M}_1, \ldots, \mathcal{M}_S | x + \alpha) 
> \hat{\mathbb{E}} f_i(\mathcal{M}_1, \ldots, \mathcal{M}_S | x + \alpha) \nonumber \\
& \Leftrightarrow
\hat{\mathbb{E}} f_k(\mathcal{M}_1, \ldots, \mathcal{M}_S | x + \alpha) 
> \max_{i: i \neq k} \hat{\mathbb{E}} f_i(\mathcal{M}_1, \ldots, \mathcal{M}_S | x + \alpha) \nonumber
\end{align}
The Theorem \ref{SRCC} does hold.
\label{proof4}
\end{proof}

\section{Proof of Corollary \ref{prop2}}

\begin{proof} $\forall \alpha \in l_p(1)$, by applying Theorem \ref{SRCC}, we have 
\begin{align}
\hat{\mathbb{E}}_{lb} f_k(\mathcal{M}_h, \mathcal{M}_x | x) & > e^{2 (\frac{\epsilon_r}{\kappa} + \frac{\epsilon_r}{\varphi})} \max_{i: i\neq k} \hat{\mathbb{E}}_{ub} f_{i}(\mathcal{M}_h, \mathcal{M}_x | x) \nonumber \\
& > e^{2 (\frac{\kappa  + \varphi}{\kappa  \varphi}) \epsilon_r} \max_{i: i\neq k} \hat{\mathbb{E}}_{ub} f_{i}(\mathcal{M}_h, \mathcal{M}_x | x) = e^{2 (\epsilon_r / \frac{\kappa  \varphi}{\kappa  + \varphi}) } \max_{i: i\neq k} \hat{\mathbb{E}}_{ub} f_{i}(\mathcal{M}_h, \mathcal{M}_x | x) \nonumber
\end{align}
Furthermore, by applying group privacy, we have that
\begin{equation}
\forall \alpha \in l_p(\frac{\kappa  \varphi}{\kappa + \varphi}): \hat{\mathbb{E}}_{lb} f_k(\mathcal{M}_h, \mathcal{M}_x | x)
> e^{2\epsilon_r} \max_{i: i\neq k} \hat{\mathbb{E}}_{ub} f_{i}(\mathcal{M}_h, \mathcal{M}_x | x) 
\label{UpdatedCorollary}
\end{equation}
By applying Proof \ref{proof4}, it is straight to have 
\begin{equation}
\forall \alpha \in l_p(\frac{\kappa \varphi}{\kappa + \varphi}): \hat{\mathbb{E}} f_k(\mathcal{M}_h, \mathcal{M}_x | x + \alpha) 
> \max_{i: i \neq k} \hat{\mathbb{E}} f_k(\mathcal{M}_h, \mathcal{M}_x | x + \alpha) \nonumber
\end{equation}
with probability $\geq \eta$.
Corollary \ref{prop2} does hold.
\end{proof}

\section{Effective Monte Carlo Estimation of $\hat{\mathbb{E}} f(x)$}

Recall that the Monte Carlo estimation is applied to estimate the expected value $\hat{\mathbb{E}} f(x) = \frac{1}{n} \sum_n f(x)_n$, where $n$ is the number of invocations of $f(x)$ with independent draws in the noise, i.e., $\frac{1}{m} Lap(0, \frac{\Delta_{\mathcal{R}}}{\epsilon_1})$ and $\frac{2}{m} Lap(0, \frac{\Delta_{\mathcal{R}}}{\epsilon_1})$ in our case. When $\epsilon_1$ is small (indicating a strong privacy protection), it causes a \textit{notably large distribution shift between training and inference, given independent draws of the Laplace noise}.

 In fact, let us denote a single draw in the noise as $\chi_1 = \frac{1}{m} Lap(0, \frac{\Delta_{\mathcal{R}}}{\epsilon_1})$ used to train the function $f(x)$, the model converges to the point that the noise $\chi_1$ and $2 \chi_2$ need to be correspondingly added into $x$ and $h$ in order to make correct predictions. $\chi_1$ can be approximated as $Lap(\chi_1, \varrho)$, where $\varrho \rightarrow 0$. It is clear that independent draws of the noise $\frac{1}{m} Lap(0, \frac{\Delta_{\mathcal{R}}}{\epsilon_1})$ have \textit{distribution shifts} with the fixed noise $\chi_1 \approxeq Lap(\chi_1, \varrho)$. These distribution shifts can also be large, when noise is large. We have experienced that these distribution shifts in having independent draws of noise to estimate $\hat{\mathbb{E}} f(x)$ can notably degrade the inference accuracy of the scoring function, when privacy budget $\epsilon_1$ is small resulting in a large amount of noise injected to provide strong privacy guarantees. 

To address this, one solution is to increase the number of invocations of $f(x)$, i.e., $n$, to a huge number per prediction. However, this is impractical in real-world scenarios. We propose a novel way to draw independent noise following the distribution of $\chi_1 + \frac{1}{m} Lap(0, \frac{\Delta_{\mathcal{R}}}{\epsilon_1}/\psi)$ for the input $x$ and $2\chi_2 + \frac{2}{m} Lap(0, \frac{\Delta_{\mathcal{R}}}{\epsilon_1}/\psi)$ for the affine transformation $h$, where $\psi$ is a hyper-parameter to control the distribution shifts. This approach works well and does not affect the DP bounds and the certified robustness condition, since: \textbf{(1)} Our mechanism achieves both DP and certified robustness in the training process; and \textbf{(2)} It is clear that $\hat{\mathbb{E}} f(x) = \frac{1}{n} \sum_n f(x)_n = \frac{1}{n} \sum_n g\big(a(x + \chi_1 + \frac{1}{m} Lap_n(0, \frac{\Delta_{\mathcal{R}}}{\epsilon_1}/\psi), \theta_1) + 2\chi_2 + \frac{2}{m} Lap_n(0, \frac{\Delta_{\mathcal{R}}}{\epsilon_1}/\psi), \theta_2\big)$, where $Lap_n(0, \frac{\Delta_{\mathcal{R}}}{\epsilon_1}/\psi)$ is the $n$-th draw of the noise. When $n \rightarrow \infty$, $\hat{\mathbb{E}} f(x)$ will converge to $\frac{1}{n} \sum_n g\big(a(x + \chi_1, \theta_1) + 2\chi_2, \theta_2\big)$, which aligns well with the convergence point of the scoring function $f(x)$. Injecting $\chi_1$ and $2\chi_2$ to $x$ and $h$ during the estimation of $\hat{\mathbb{E}} f(x)$ yields better performance, without affecting the DP and the composition robustness bounds.

\section{Approximation Error Bounds}

To compute how much error our polynomial approximation approaches (i.e., truncated Taylor expansions), $\widetilde{\mathcal{R}}_{B_t}(\theta_1)$ (Eq. \ref{poly1}) and $\mathcal{L}_{\overline{B}_t}\big(\theta_2\big)$, incur, we directly apply Lemma 4 in \citep{Phan0WD16}, Lemma 3 in \citep{zhang2012functional}, and the well-known error bound results in \citep{Apostol}. Note that $\widetilde{\mathcal{R}}_{B_t}(\theta_1)$ is the 1st-order Taylor series and $\mathcal{L}_{\overline{B}_t}\big(\theta_2\big)$ is the 2nd-order Taylor series following the implementation of \citep{TensorFlowSoftMax}. Let us closely follow \citep{Phan0WD16,zhang2012functional,Apostol} to adapt their results into our scenario, as follows:

Given the truncated function $\widetilde{\mathcal{R}}_{B_t}(\theta_1) = \sum_{x_i \in B_t} \sum_{j = 1}^d \sum_{l=1}^{2} \sum_{r = 0}^{1} \frac{\mathbf{F}^{(r)}_{lj}(0)}{r!}\big(\theta_{1j}h_i\big)^r$, the original Taylor polynomial function $\widehat{\mathcal{R}}_{B_t}(\theta_1) = \sum_{x_i \in B_t} \sum_{j = 1}^d \sum_{l=1}^{\infty} \sum_{r = 0}^{1} \frac{\mathbf{F}^{(r)}_{lj}(0)}{r!}\big(\theta_{1j}h_i\big)^r$, the average error of the approximation is bounded as
\begin{align}
\frac{1}{|B_t|}|\widehat{\mathcal{R}}_{B_t}(\widetilde{\theta}_1) - \widehat{\mathcal{R}}_{B_t}(\widehat{\theta}_1)| & \leq \frac{4e \times d}{(1+e)^2}
\label{Ebound1}
\\
\frac{1}{|B_t|}|\widehat{\mathcal{L}}_{B_t}(\widetilde{\theta}_2) - \widehat{\mathcal{L}}_{B_t}(\widehat{\theta}_2)| & \leq \frac{e^2 + 2e - 1}{e(1+e)^2} \times K
\label{Ebound2}
\end{align}
where $\widehat{\theta}_1 = \arg \min_{\theta_1} \widehat{\mathcal{R}}_{B_t}(\theta_1)$, $\widetilde{\theta}_1 = \arg \min_{\theta_1} \widetilde{\mathcal{R}}_{B_t}(\theta_1)$, $\widehat{\mathcal{L}}_{B_t}(\theta_2)$ is the original Taylor polynomial function of $\sum_{x_i \in B_t} \mathcal{L}\big(f(\overline{x}_i, \theta_2), y_i\big)$,
$\widehat{\theta}_2 = \arg \min_{\theta_2} \widehat{\mathcal{L}}_{B_t}(\theta_2)$, $\widetilde{\theta}_2 = \arg \min_{\theta_2} {\mathcal{L}}_{B_t}(\theta_2)$.

\begin{proof}
Let $U = \max_{\theta_1} \big(\widehat{\mathcal{R}}_{B_t}(\theta_1) - \widetilde{\mathcal{R}}_{B_t}(\theta_1)\big)$ and $S = \min_{\theta_1} \big(\widehat{\mathcal{R}}_{B_t}(\theta_1) - \widetilde{\mathcal{R}}_{B_t}(\theta_1)\big)$. 

We have that $U \geq \widehat{\mathcal{R}}_{B_t}(\widetilde{\theta}_1) - \widetilde{\mathcal{R}}_{B_t}(\widetilde{\theta}_1)$ and $\forall \theta_1^*: S \leq \widehat{\mathcal{R}}_{B_t}(\theta_1^*) - \widetilde{\mathcal{R}}_{B_t}(\theta_1^*)$. Therefore, we have
\begin{align}
&\widehat{\mathcal{R}}_{B_t}(\widetilde{\theta}_1) - \widetilde{\mathcal{R}}_{B_t}(\widetilde{\theta}_1) - \widehat{\mathcal{R}}_{B_t}(\theta_1^*) + \widetilde{\mathcal{R}}_{B_t}(\theta_1^*) \leq U - S \\
\Leftrightarrow & \widehat{\mathcal{R}}_{B_t}(\widetilde{\theta}_1) - \widehat{\mathcal{R}}_{B_t}(\theta_1^*) \leq U - S + \big(\widetilde{\mathcal{R}}_{B_t}(\widetilde{\theta}_1) - \widetilde{\mathcal{R}}_{B_t}(\theta_1^*) \big)
\end{align}
In addition, $\widetilde{\mathcal{R}}_{B_t}(\widetilde{\theta}_1) - \widetilde{\mathcal{R}}_{B_t}(\theta_1^*) \leq 0$, it is straightforward to have: 
\begin{equation}
\widehat{\mathcal{R}}_{B_t}(\widetilde{\theta}_1) - \widehat{\mathcal{R}}_{B_t}(\theta_1^*) \leq U - S
\end{equation}
If $U \geq 0$ and $S \leq 0$ then we have:
\begin{equation}
|\widehat{\mathcal{R}}_{B_t}(\widetilde{\theta}_1) - \widehat{\mathcal{R}}_{B_t}(\theta_1^*)| \leq U - S 
\label{AppError1}
\end{equation} 

Eq. \ref{AppError1} holds for every $\theta_1^*$, including $\widehat{\theta}_1$. Eq. \ref{AppError1} shows that the error incurred by truncating the Taylor series approximate function depends on the maximum and minimum values of $\widehat{\mathcal{R}}_{B_t}(\theta_1) - \widetilde{\mathcal{R}}_{B_t}(\theta_1)$. This is consistent with \citep{Phan0WD16,zhang2012functional}. To quantify the magnitude of the error, we rewrite $\widehat{\mathcal{R}}_{B_t}(\theta_1) - \widetilde{\mathcal{R}}_{B_t}(\theta_1)$ as: 
\begin{align}
\widehat{\mathcal{R}}_{B_t}(\theta_1) - \widetilde{\mathcal{R}}_{B_t}(\theta_1) & = \sum_{j=1}^d \big(\widehat{\mathcal{R}}_{B_t}(\theta_{1j}) - \widetilde{\mathcal{R}}_{B_t}(\theta_{1j})\big) \\ 
& = \sum_{j = 1}^d \Big(\sum_{i = 1}^{|B_t|} \sum_{l=1}^{2} \sum_{r = 3}^{\infty} \frac{\mathbf{F}^{(r)}_{lj}(z_{lj})}{r!}\big(g_{lj}(x_i, \theta_{1j}) - z_{lj}\big)^r \Big)
\end{align}
where $g_{1j}(x_i, \theta_{1j}) = \theta_{1j}h_i$ and $g_{2j}(x_i, \theta_{1j}) = \theta_{1j}h_i$.

By looking into the remainder of Taylor expansion for each $j$ (i.e., following \citep{Phan0WD16,Apostol}), with $z_j \in [z_{lj} - 1, z_{lj} + 1]$, $\frac{1}{|B_t|}\big(\widehat{\mathcal{R}}_{B_t}(\theta_{1j}) - \widetilde{\mathcal{R}}_{B_t}(\theta_{1j})\big)$ must be in the interval $\Big[\sum_l \frac{\min_{z_j} \mathbf{F}^{(2)}_{lj}(z_j)(z_j - z_{lj})^2}{2!}, \sum_l \frac{\max_{z_j} \mathbf{F}^{(2)}_{lj}(z_j)(z_j - z_{lj})^2}{2!}\Big]$. 
If $\sum_l \frac{\max_{z_j} \mathbf{F}^{(2)}_{lj}(z_j)(z_j - z_{lj})^2}{2!} \geq 0$ and $\sum_l \frac{\min_{z_j} \mathbf{F}^{(2)}_{lj}(z_j)(z_j - z_{lj})^2}{2!} \leq 0$, then we have that $|\frac{1}{|B_t|}\big(\widehat{\mathcal{R}}_{B_t}(\theta_{1}) - \widetilde{\mathcal{R}}_{B_t}(\theta_{1})\big)| \leq \sum_{j =1}^{d} \sum_l \frac{\max_{z_j} \mathbf{F}^{(2)}_{lj}(z_j)(z_j - z_{lj})^2 - \min_{z_j} \mathbf{F}^{(2)}_{lj}(z_j)(z_j - z_{lj})^2}{2!}$. This can be applied to the case of our auto-encoder, as follows:

For the functions $\mathbf{F}_{1j}(z_j) = x_{ij}\log (1 + e^{-z_j})$ and $\mathbf{F}_{2j}(z_j) = (1-x_{ij})\log (1 + e^{z_j})$, we have $\mathbf{F}_{1j}^{(2)}(z_j) = \frac{x_{ij}e^{-z_j}}{(1 + e^{-z_j})^2}$ and $\mathbf{F}_{2j}^{(2)}(z_j) = (1-x_{ij})\frac{e^{z_j}}{(1+ e^{z_{j}})^2}$. It can be verified that $\arg \min_{z_j}\mathbf{F}_{1j}^{(2)}(z_j) = \frac{-e}{(1+e)^2} < 0$, $\arg \max_{z_j}\mathbf{F}_{1j}^{(2)}(z_j) =\frac{e}{(1+e)^2} > 0$, $\arg \min_{z_j}\mathbf{F}_{2j}^{(2)}(z_j) = 0$, and $\arg \max_{z_j}\mathbf{F}_{2j}^{(2)}(z_j) = \frac{2e}{(1+e)^2} > 0$. Thus, the average error of the approximation is at most:
\begin{equation}
\frac{1}{|B_t|}|\widehat{\mathcal{R}}_{B_t}(\widetilde{\theta}_1) - \widehat{\mathcal{R}}_{B_t}(\widehat{\theta}_1)| \leq \Big[\big( \frac{e}{(1+e)^2} - \frac{-e}{(1+e)^2}\big) 
+ \frac{2e}{(1+e)^2}\Big] \times d = \frac{4e \times d}{(1+e)^2}
\end{equation}

Consequently, Eq. \ref{Ebound1} does hold. Similarly, by looking into the remainder of Taylor expansion for each label $k$, Eq. \ref{Ebound2} can be proved straightforwardly. In fact, by using the 2nd-order Taylor series with $K$ categories, we have that: $\frac{1}{|B_t|}|\widehat{\mathcal{L}}_{B_t}(\widetilde{\theta}_2) - \widehat{\mathcal{L}}_{B_t}(\widehat{\theta}_2)| \leq \frac{e^2 + 2e - 1}{e(1+e)^2} \times K$. \vspace{-5pt}
\end{proof}

\section{Model Configurations}

The MNIST database consists of handwritten digits \citep{Lecun726791}. Each example is a 28 $\times$ 28 size gray-level image. The CIFAR-10 dataset consists of color images belonging to 10 classes, i.e., airplanes, dogs, etc. The dataset is split into 50,000 training samples and 10,000 test samples \citep{krizhevsky2009learning}. Tiny Imagenet $(64\times 64 \times 3)$ has 200 classes. Each class has 500 training images, 50 validation images, and 50 test images. We used the first thirty classes with data augmented, including horizontal flip and random brightness, in the Tiny ImageNet dataset in our experiment. In general, the dataset is split into 45,000 training samples and 1,500 test samples \citep{Imagenet,hendrycks2018benchmarking}. The experiments were conducted on a server of 4 GPUs, each of which is an NVIDIA TITAN Xp, 12 GB with 3,840 CUDA cores. 
All the models share the same structure, consisting of 2 and 3 convolutional layers, respectively for MNIST and CIFAR-10 datasets, and a ResNet18 model for the Tiny ImageNet dataset.

Both fully-connected and convolution layers can be applied in the representation learning model $a(x, \theta_1)$. Given convolution layer, the computation of each feature map needs to be DP; since each of them independently reads a local region of input neurons. Therefore, the sensitivity $\Delta_{\mathcal{R}}$ can be considered the maximal sensitivity given any single feature map in the first affine transformation layer. In addition, each hidden neuron can only be used to reconstruct a unit patch of input units. That results in $d$ (Lemma \ref{lemma3}) being the size of the unit patch connected to each hidden neuron, e.g., $d = 9$ given a $3\times 3$ unit patch, and $\beta$ is the number of hidden neurons in a feature map.

\textit{MNIST:} We used two convolutional layers (32 and 64 features). Each hidden neuron connects with a 5x5 unit patch. A fully-connected layer has 256 units. The batch size $m$ was set to 2,499, $\xi = 1$, $\psi = 2$. I-FGSM, MIM, and MadryEtAl were used to draft $l_\infty(\mu)$ adversarial examples in training, with $T_\mu = 10$. Learning rate $\varrho_t$ was set to $1e-4$. Given a predefined total privacy budget $\epsilon$, $\epsilon_2$ is set to be $0.1$, and $\epsilon_1$ is computed as: $\epsilon_1 = \frac{\epsilon - \epsilon_2}{(1 + 1/\gamma + 1/\gamma_\mathbf{x})}$. This will guarantee that $(\epsilon_1 + \epsilon_1/\gamma_\mathbf{x} + \epsilon_1/\gamma + \epsilon_2) = \epsilon$. $\Delta_\mathcal{R} = (14^2 + 2)\times25$ and $\Delta_{\mathcal{L}2} = 2\times256$. The number of Monte Carlo sampling for certified inference $n$ is set to 2,000.

\textit{CIFAR-10:} We used three convolutional layers (128, 128, and 256 features). Each hidden neuron connects with a 4x4 unit patch in the first layer, and a 5x5 unit patch in other layers. One fully-connected layer has 256 neurons. The batch size $m$ was set to 1,851, $\xi = 1.5$, $\psi = 10$, and $T_\mu = 3$.
The ensemble of attacks $A$ includes I-FGSM, MIM, and MadryEtAl. We use data augmentation, including random crop, random flip, and random contrast. Learning rate $\varrho_t$ was set to $5e-2$. In the CIFAR-10 dataset, $\epsilon_2$ is set to $(1 + r/3.0)$ and $\epsilon_1 = (1 + 2r/3.0)/(1 + 1/\gamma + 1/\gamma_\mathbf{x})$, where $r \geq 0$ is a ratio to control the total privacy budget $\epsilon$ in our experiment. For instance, given $r = 0$, we have that $\epsilon = (\epsilon_1 + \epsilon_1/\gamma_\mathbf{x} + \epsilon_1/\gamma + \epsilon_2) = 2$. $\Delta_\mathcal{R} = 3\times(14^2+2)\times16$ and $\Delta_{\mathcal{L}2} = 2\times256$. $\mathbb{N}$ and $\mathbb{M}$ are set to $1$ and $4$ in the distributed training. The number of Monte Carlo sampling for certified inference $n$ is set to 1,000.

\textit{Tiny ImageNet:} We used a ResNet-18 model. Each hidden neuron connects with a 7x7 unit patch in the first layer, and 3x3 unit patch in other layers. The batch size $m$ was set to 4,500, $\xi = 1.5$, $\psi = 10$, and $T_\mu = 10$. The ensemble of attacks $A$ includes I-FGSM, MIM, and MadryEtAl. Learning rate $\varrho_t$ was set to $1e-2$. In the Tiny ImageNet dataset, $\epsilon_2$ is set to $1$ and $\epsilon_1 = (1 + r)/(1 + 1/\gamma + 1/\gamma_\mathbf{x})$, where $r \geq 0$ is a ratio to control the total privacy budget $\epsilon$ in our experiment. $\Delta_\mathcal{R} = 3\times(32^2+2)\times49$ and $\Delta_{\mathcal{L}2} = 2\times256$. $\mathbb{N}$ and $\mathbb{M}$ are set to $1$ and $20$ in the distributed training.
The number of Monte Carlo sampling for certified inference $n$ is set to 1,000.

\section{Complete and Detailed Experimental Results}

\begin{figure*}[h]
\centering
$\begin{array}{c@{\hspace{0.1in}}c@{\hspace{0.1in}}c@{\hspace{0.1in}}c}
\includegraphics[width=2.3in]{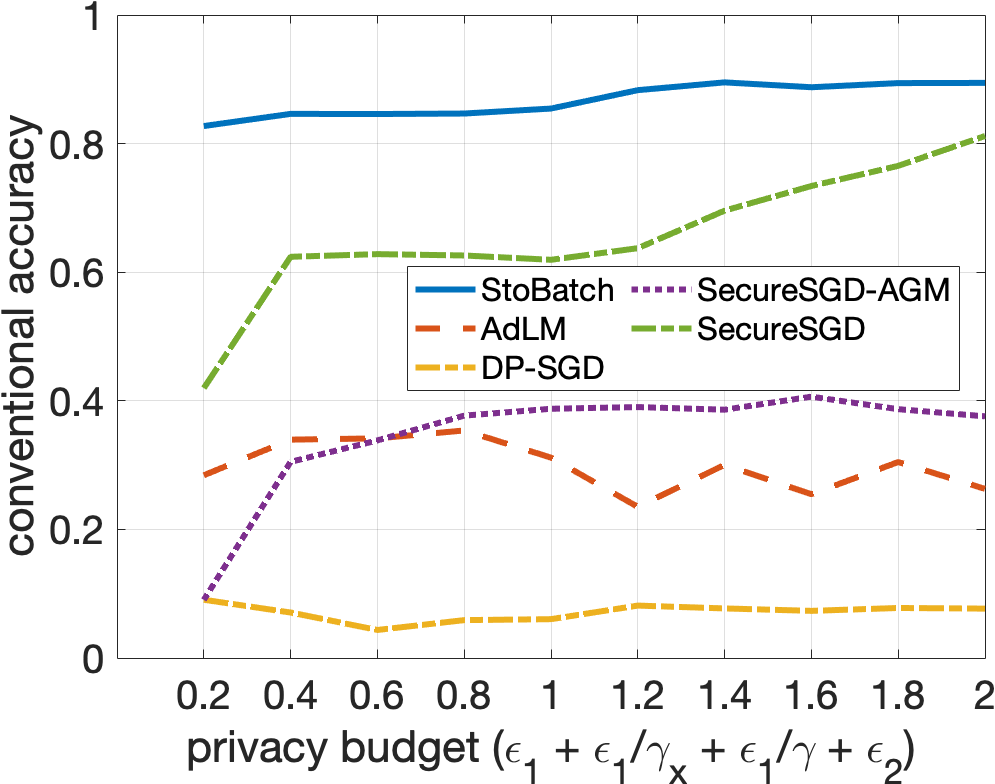} & \includegraphics[width=2.3in] {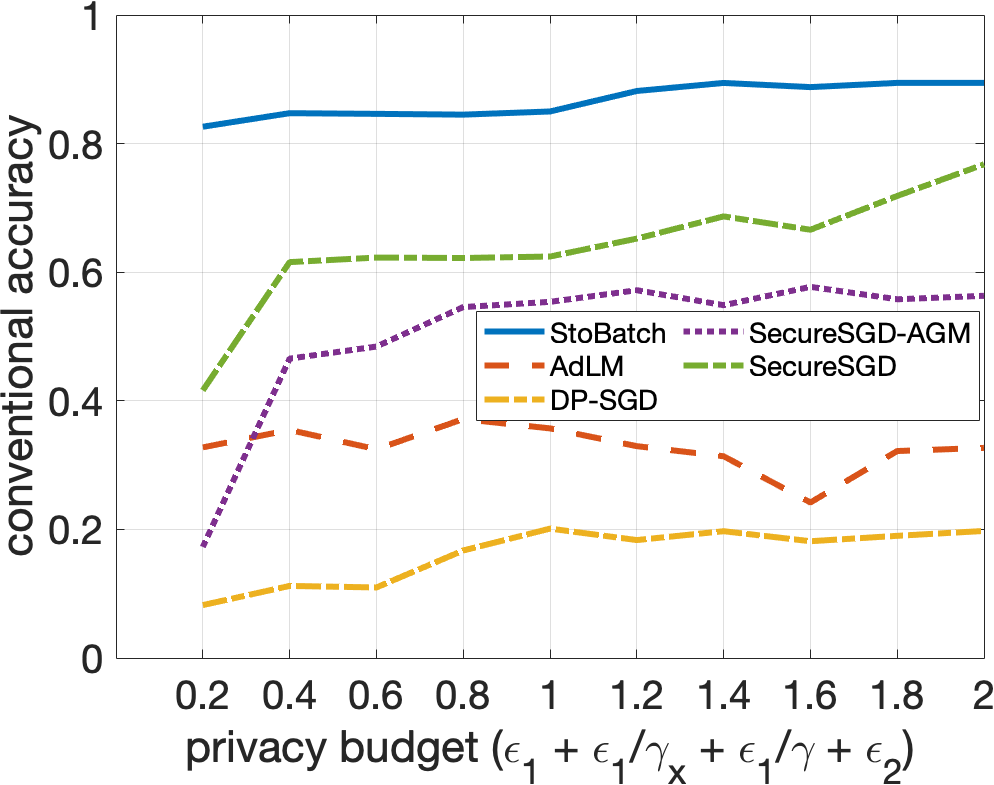} \\ [0.0cm] 
\mbox{(a) I-FGSM attacks} & \mbox{(b) FGSM attacks} 
\\ [0.0cm]  \includegraphics[width=2.3in]{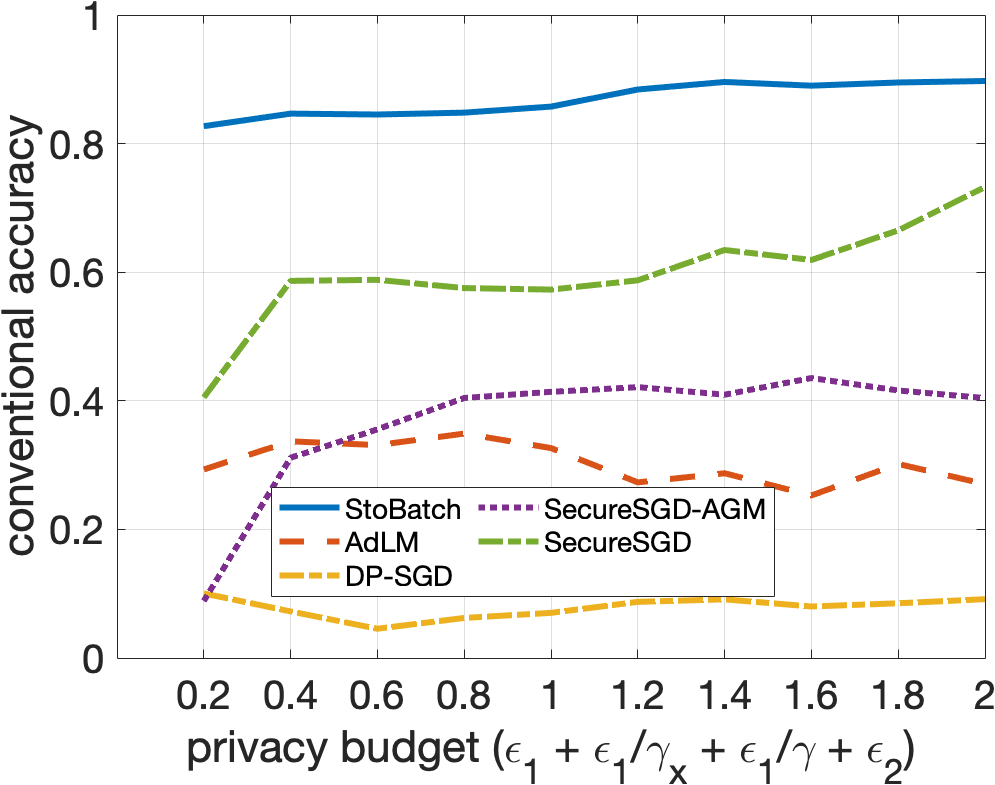} & \includegraphics[width=2.3in]{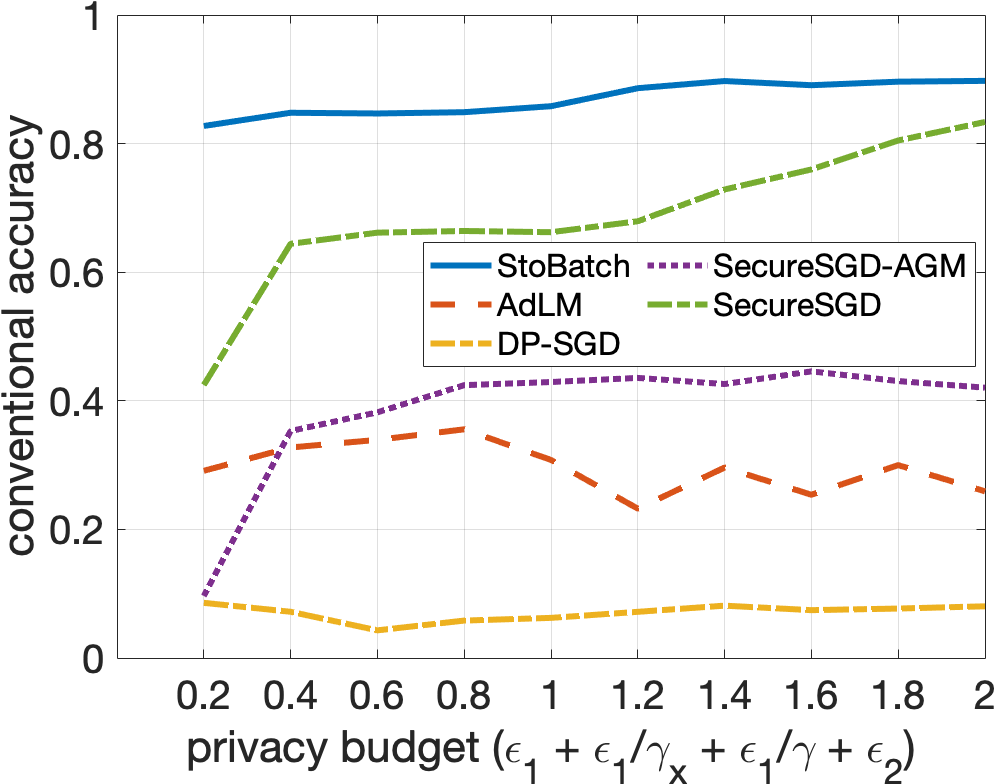} \\ [0.0cm]
\mbox{(c) MIM attacks} & \mbox{(d) MadryEtAl attacks}
\end{array}$
\caption{Conventional accuracy on the MNIST dataset given $\epsilon$, under $l_\infty(\mu_a = 0.2)$ and $T_a = 10$.}
\label{MNIST02Full}
\end{figure*}

\textbf{Results on the MNIST Dataset.} Figure \ref{MNIST02Full} illustrates the conventional accuracy of each model as a function of the privacy budget $\epsilon$ on the MNIST dataset under $l_\infty(\mu_a)$-norm attacks, with $\mu_a = 0.2$ (a pretty strong attack). It is clear that our StoBatch outperforms AdLM, DP-SGD, SecureSGD, and SecureSGD-AGM, in all cases, with $p < 1.32e-4$. On average, we register a 22.36\% improvement over SecureSGD ($p < 1.32e-4$), a 46.84\% improvement over SecureSGD-AGM ($p < 1.83e-6$), a 56.21\% improvement over AdLM ($p < 2.05e-10$), and a 77.26\% improvement over DP-SGD ($p < 5.20e-14$), given our StoBatch mechanism.
AdLM and DP-SGD achieve the worst conventional accuracies. There is no guarantee provided in AdLM and DP-SGD. Thus, the accuracy of the AdLM and DPSGD algorithms seem to show no effect against adversarial examples, when the privacy budget is varied. This is in contrast to our StoBatch model, the SecureSGD model, and the SecureSGD-AGM model, whose accuracies are proportional to the privacy budget.

When the privacy budget $\epsilon = 0.2$ (a tight DP protection), there are significant drops, in terms of conventional accuracy, given the baseline approaches. By contrast, our StoBatch mechanism only shows a small degradation in the conventional accuracy (6.89\%, from 89.59\% to 82.7\%), compared with a 37\% drop in SecureSGD (from 78.64\% to 41.64\%), and a 32.89\% drop in SecureSGD-AGM (from 44.1\% to 11.2\%) on average, when the privacy budget $\epsilon$ goes from 2.0 to 0.2. At $\epsilon = 0.2$, our StoBatch mechanism achieves 82.7\%, compared with 11.2\% and 41.64\% correspondingly for SecureSGD-AGM and SecureSGD. This is an important result, showing the ability to offer tight DP protections under adversarial example attacks in our model, compared with existing algorithms. 

\begin{figure*}[h]
\centering
$\begin{array}{c@{\hspace{0.1in}}c@{\hspace{0.1in}}c@{\hspace{0.1in}}c}
\includegraphics[width=2.3in]{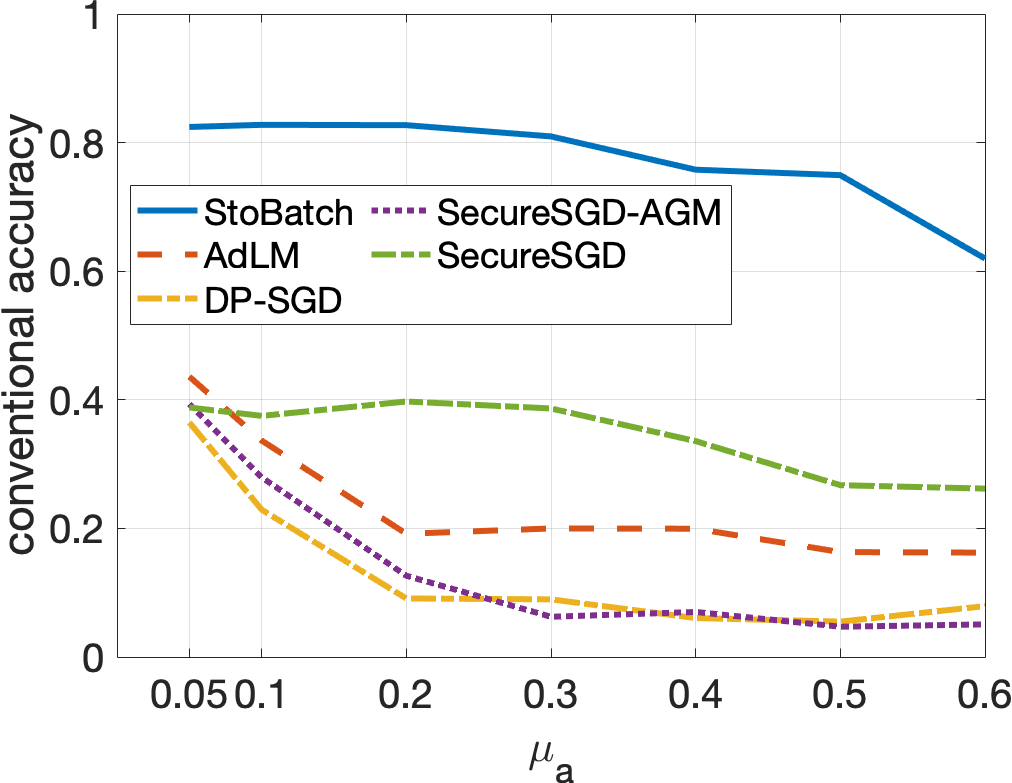} & \includegraphics[width=2.3in]{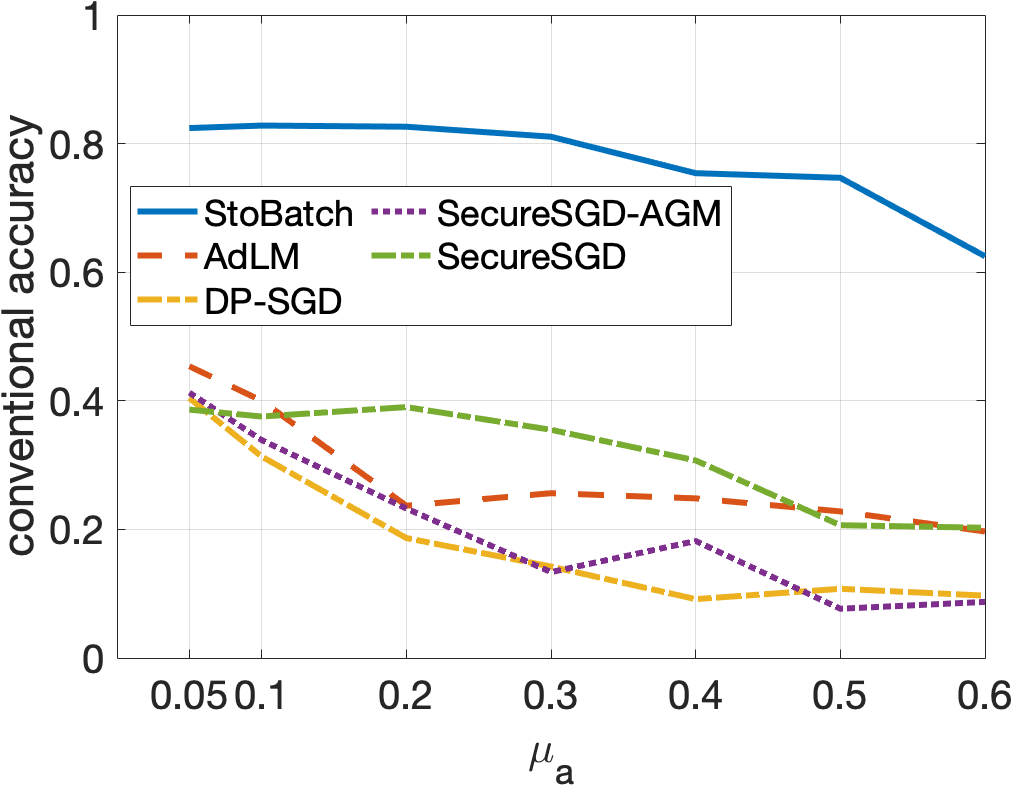} \\ [0.0cm]
\mbox{(a) I-FGSM attacks} & \mbox{(b) FGSM attacks} \\ [0.0cm] \includegraphics[width=2.3in]{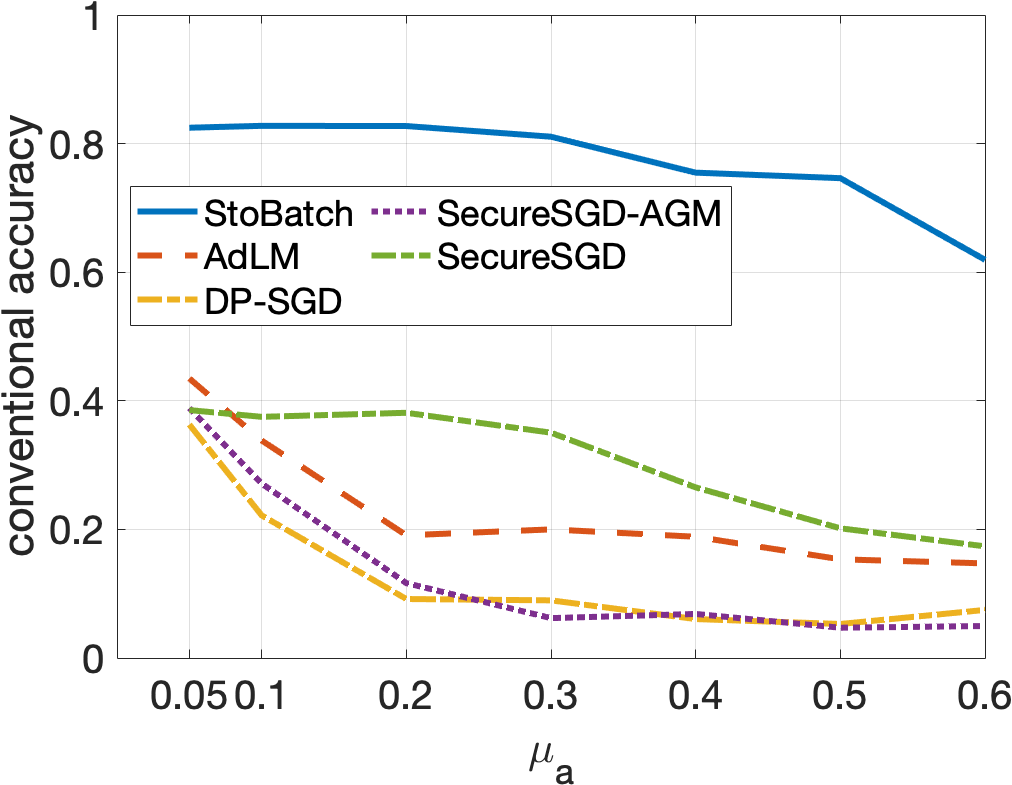} & \includegraphics[width=2.3in]{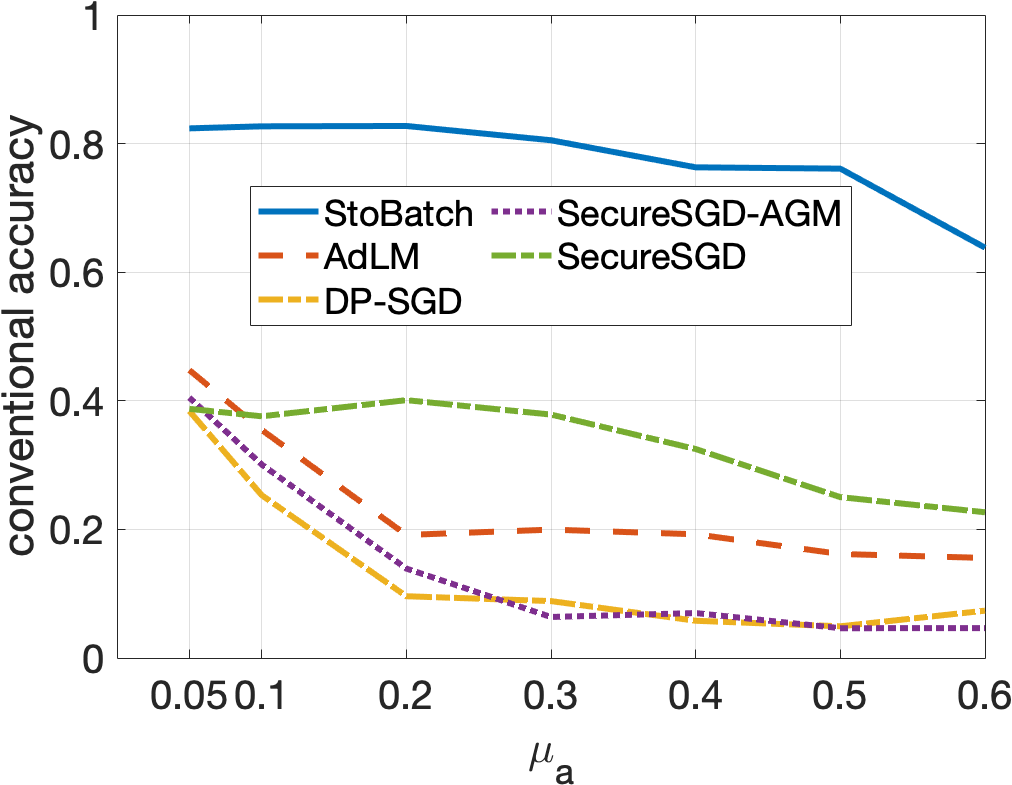} \\ [0.0cm] \mbox{(c) MIM attacks} & \mbox{(d) MadryEtAl attacks}
\end{array}$
\caption{Conventional accuracy on the MNIST dataset given $\mu_a$ ($\epsilon = 0.2$, tight DP protection) and $T_a = 10$.}
\label{MNISTAttack02Full}
\end{figure*}

$\bullet$ Figure \ref{MNISTAttack02Full} presents the conventional accuracy of each model as a function of the attack size $\mu_a$ on the MNIST dataset, under a strong DP guarantee, $\epsilon = 0.2$. Our StoBatch mechanism outperforms the baseline approaches in all cases. On average, our StoBatch model improves 44.91\% over SecureSGD ($p<7.43e-31$), a 61.13\% over SecureSGD-AGM ($p<2.56e-22$), a 52.21\% over AdLM ($p<2.81e-23$), and a 62.20\% over DP-SGD ($p<2.57e-22$). 
More importantly, our StoBatch model is resistant to different adversarial example algorithms with different attack sizes. 
When $\mu_a \geq 0.2$, AdLM, DP-SGD, SecureSGD, and SecureSGD-AGM become defenseless. We further register significantly drops in terms of accuracy, when $\mu_a$ is increased from $0.05$ (a weak attack) to $0.6$ (a strong attack), i.e., $19.87\%$ on average given our StoBatch, across all attacks, compared with 27.76\% (AdLM), 29.79\% (DP-SGD), 34.14\% (SecureSGD-AGM), and 17.07\% (SecureSGD).

\begin{figure*}[h]
\centering
$\begin{array}{c@{\hspace{0.1in}}c@{\hspace{0.1in}}c@{\hspace{0.1in}}c}
\includegraphics[width=2.3in]{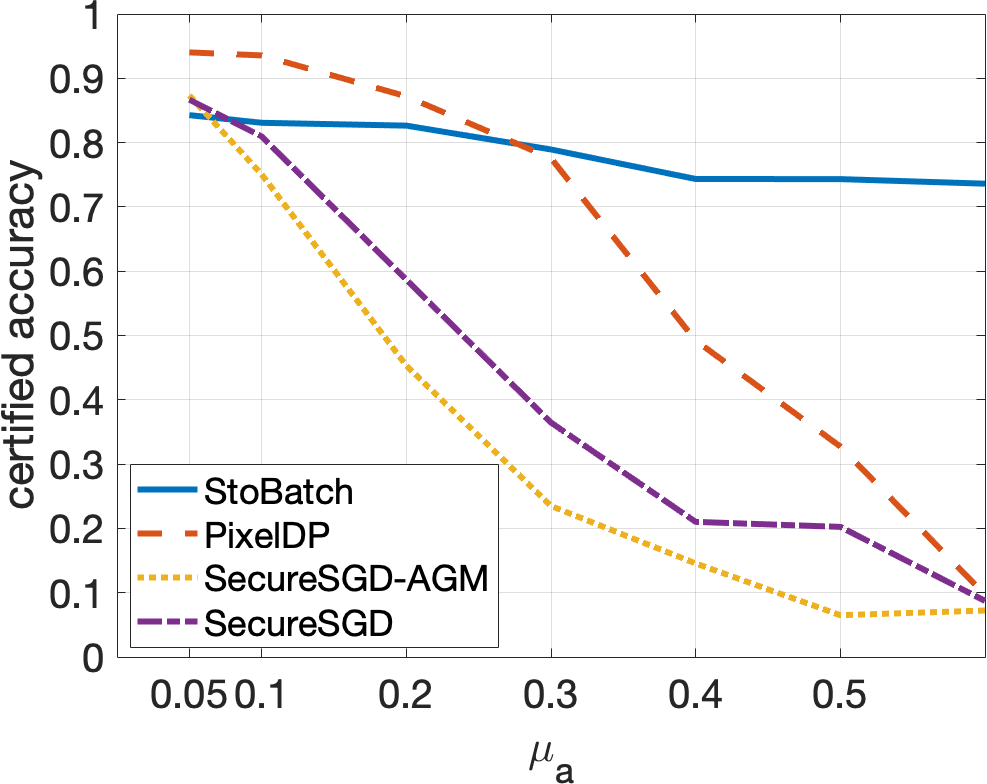} & \includegraphics[width=2.3in]{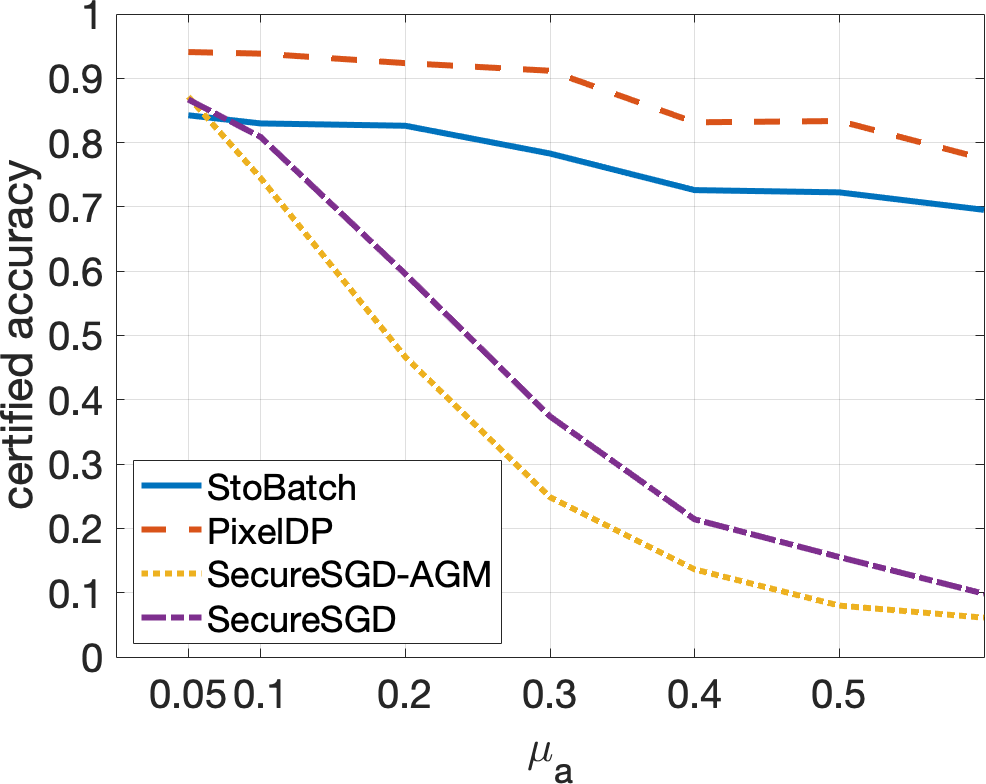} \\ [0.0cm]
\mbox{(a) I-FGSM attacks} & \mbox{(b) FGSM attacks} \\ [0.0cm] \includegraphics[width=2.3in]{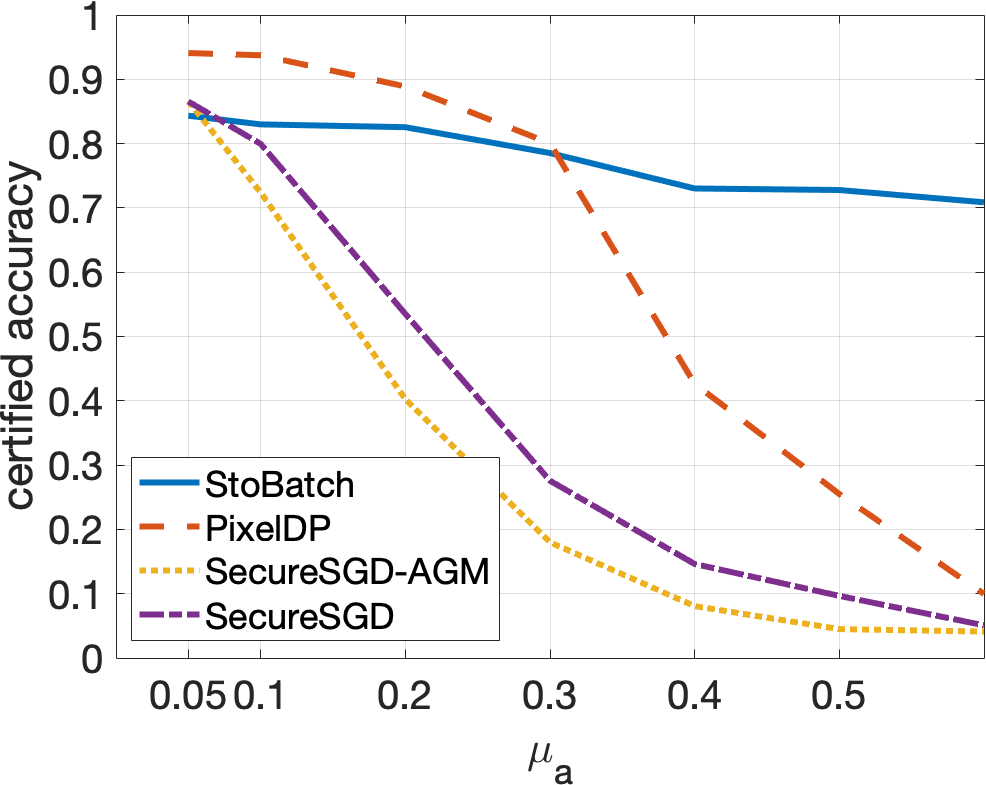} & \includegraphics[width=2.3in]{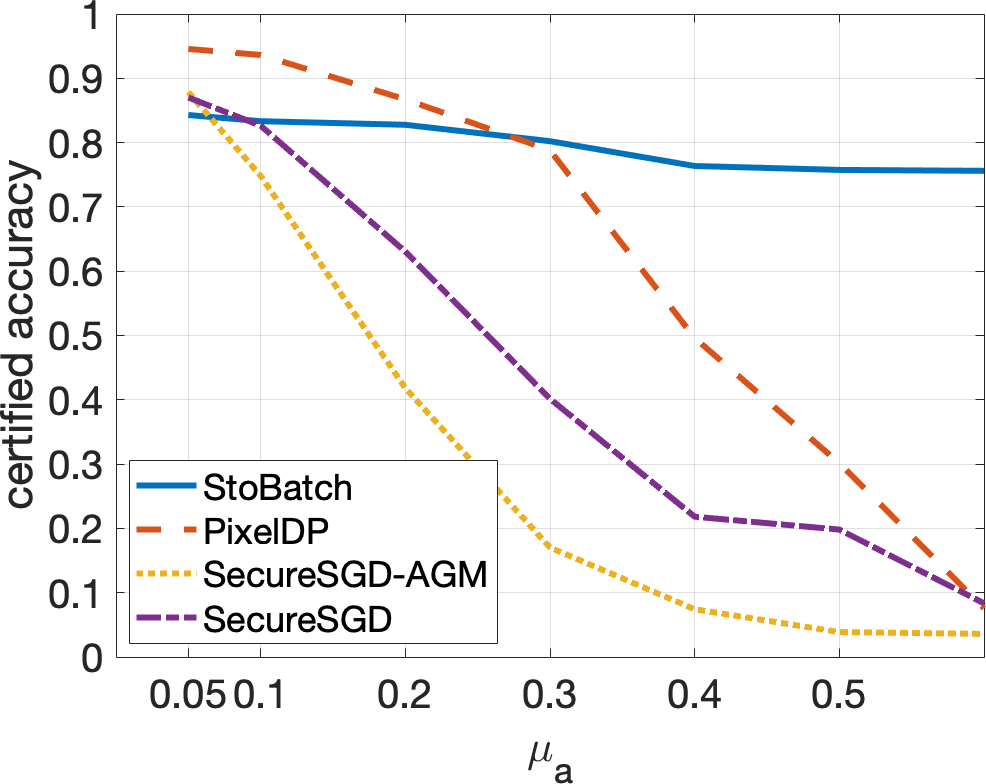} \\ [0.0cm]
\mbox{(c) MIM attacks} & \mbox{(d) MadryEtAl attacks}
\end{array}$
\caption{Certified accuracy on the MNIST dataset. $\epsilon$ is set to 1.0 (tight DP protection) and $T_a = 10$.}
\label{MNIST2Full}
\end{figure*}

$\bullet$ Figure \ref{MNIST2Full} demonstrates the certified accuracy as a function of $\mu_a$. The privacy budget is set to $1.0$, offering a reasonable privacy protection. In PixelDP, the construction attack bound $\epsilon_r$ is set to $0.1$, which is a pretty reasonable defense. With (small perturbation) $\mu_a \leq 0.2$, PixelDP achieves better certified accuracies under all attacks; since PixelDP does not preserve DP to protect the training data, compared with other models. 
Meanwhile, our StoBatch model outperforms all the other models when $\mu_a \geq 0.3$, indicating a stronger defense to more aggressive attacks. More importantly, our StoBatch has a consistent certified accuracy to different attacks given different attack sizes, compared with baseline approaches. In fact, when $\mu_a$ is increased from $0.05$ to $0.6$, our StoBatch shows a small drop ($11.88\%$ on average, from $84.29\% (\mu_a = 0.05)$ to $72.41\% (\mu_a = 0.6)$), compared with a huge drop of the PixelDP, i.e., from $94.19\% (\mu_a = 0.05)$ to $9.08\% (\mu_a = 0.6)$ on average under I-FGSM, MIM, and MadryEtAl attacks, and to $77.47\% (\mu_a = 0.6)$ under FGSM attack. Similarly, we also register significant drops in terms of certified accuracy for SecureSGD (78.74\%, from 86.74\% to 7.99\%) and SecureSGD-AGM (81.97\%, from 87.23\% to 5.26\%) on average.
This is promising.

\begin{figure*}[h]
\centering
$\begin{array}{c@{\hspace{0.1in}}c@{\hspace{0.1in}}c@{\hspace{0.1in}}c}
\includegraphics[width=2.3in]{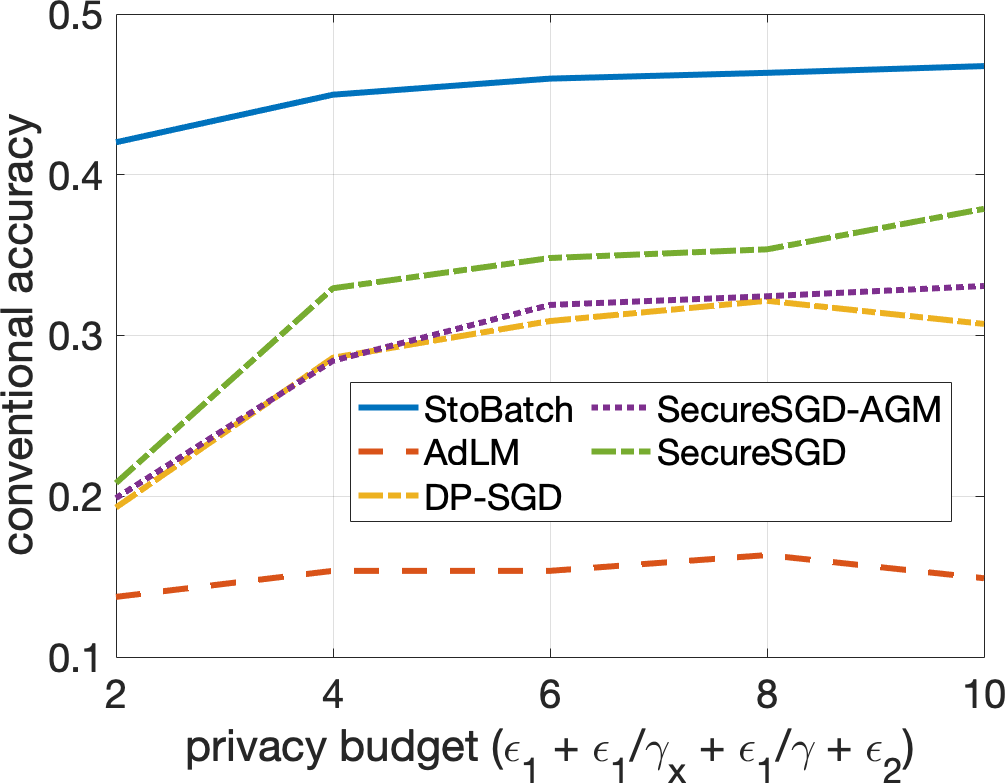} & \includegraphics[width=2.3in]{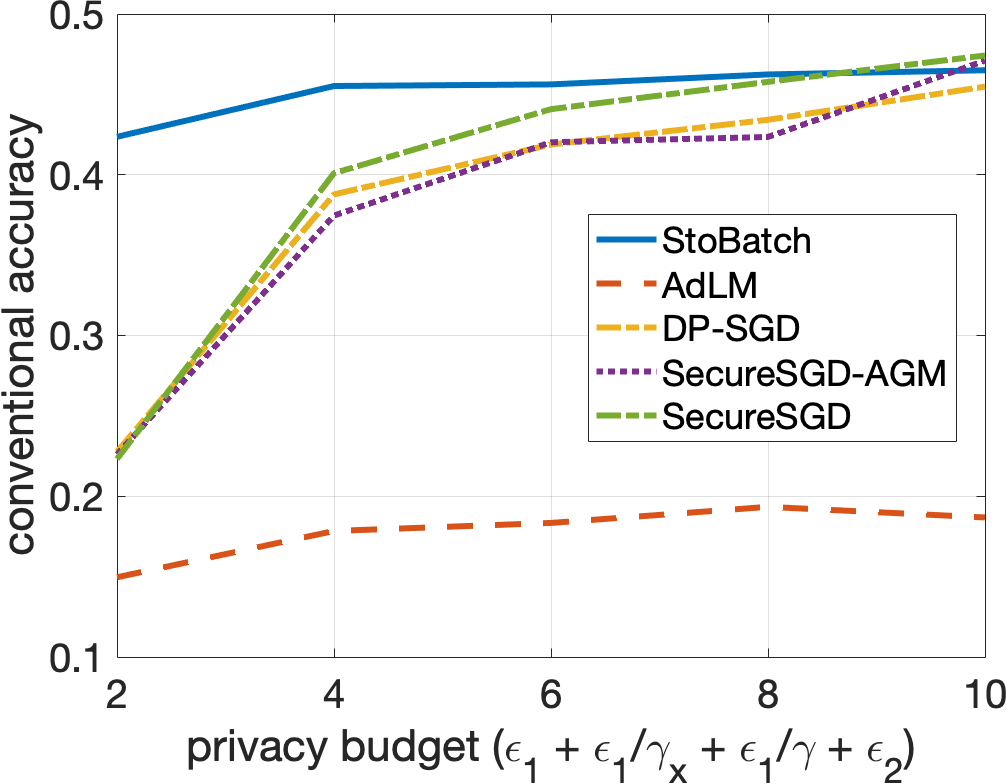} \\ [0.0cm]
\mbox{(a) I-FGSM attacks} & \mbox{(b) FGSM attacks} \\ [0.0cm] 
\includegraphics[width=2.3in]{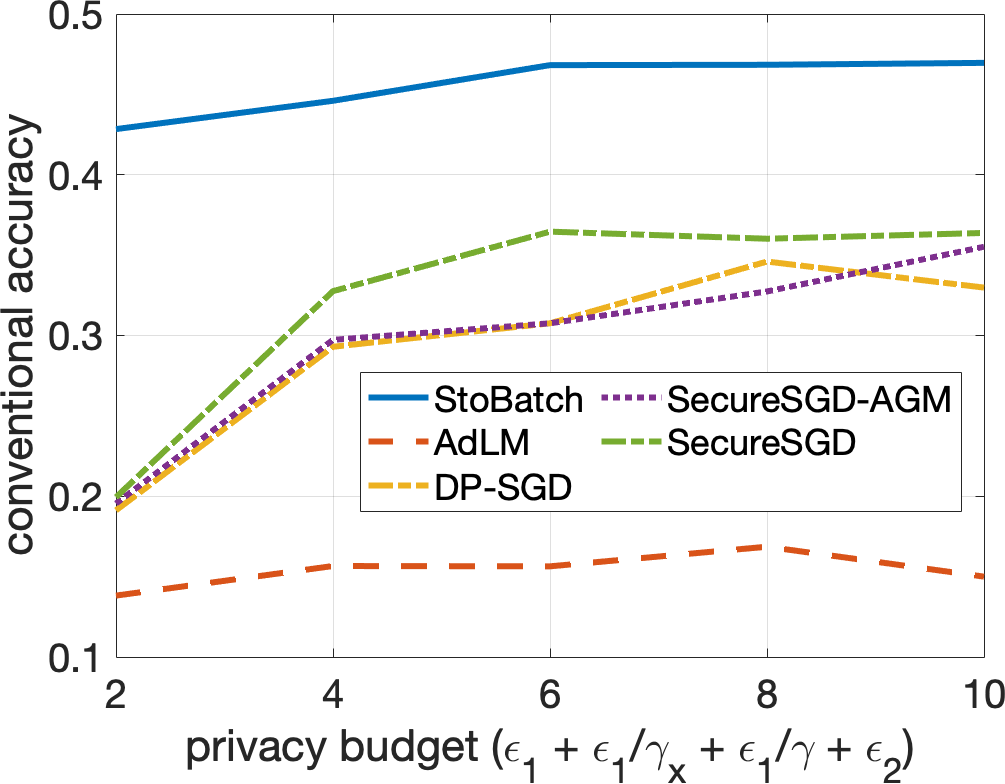} & \includegraphics[width=2.3in]{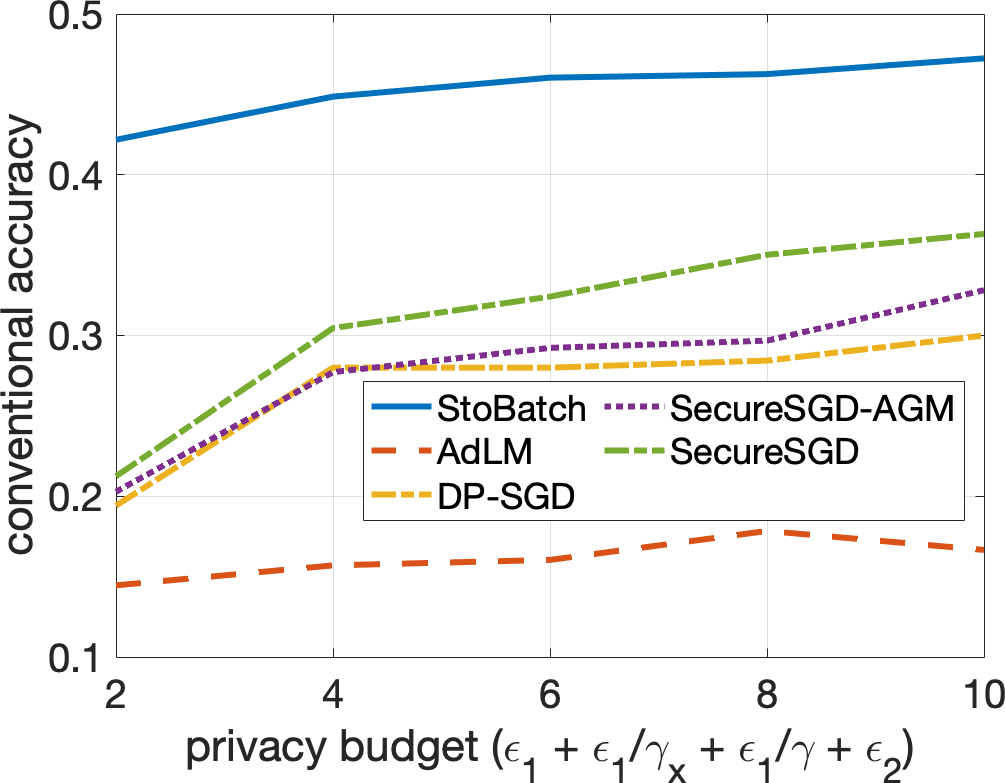} \\ [0.0cm] \mbox{(c) MIM attacks} & \mbox{(d) MadryEtAl attacks}
\end{array}$
\caption{Conventional accuracy on the CIFAR-10 dataset given $\epsilon$, under $l_\infty(\mu_a = 0.2)$ and $T_a = 3$.}
\label{CIFAR02Full}
\end{figure*}

\textbf{Results on the CIFAR-10 Dataset} further strengthen our observations. In Figure \ref{CIFAR02Full}, our StoBatch clearly outperforms baseline models in all cases ($p < 6.17e-9$), especially when the privacy budget is small ($\epsilon < 4$), yielding strong privacy protections. On average conventional accuracy, our StoBatch mechanism has an improvement of 10.42\% over SecureSGD ($p < 2.59e-7$), an improvement of 14.08\% over SecureSGD-AGM ($p < 5.03e-9$), an improvement of 29.22\% over AdLM ($p < 5.28e-26$), and a 14.62\% improvement over DP-SGD ($p < 4.31e-9$).
When the privacy budget is increased from 2 to 10, the conventional accuracy of our StoBatch model increases from 42.02\% to 46.76\%, showing a 4.74\% improvement on average. However, the conventional accuracy of our model under adversarial example attacks is still low, i.e., 44.22\% on average given the privacy budget at 2.0. This opens a long-term research avenue to achieve better robustness under strong privacy guarantees in adversarial learning.

\begin{figure*}[h]
\centering
$\begin{array}{c@{\hspace{0.1in}}c@{\hspace{0.1in}}c@{\hspace{0.1in}}c}
\includegraphics[width=2.3in]{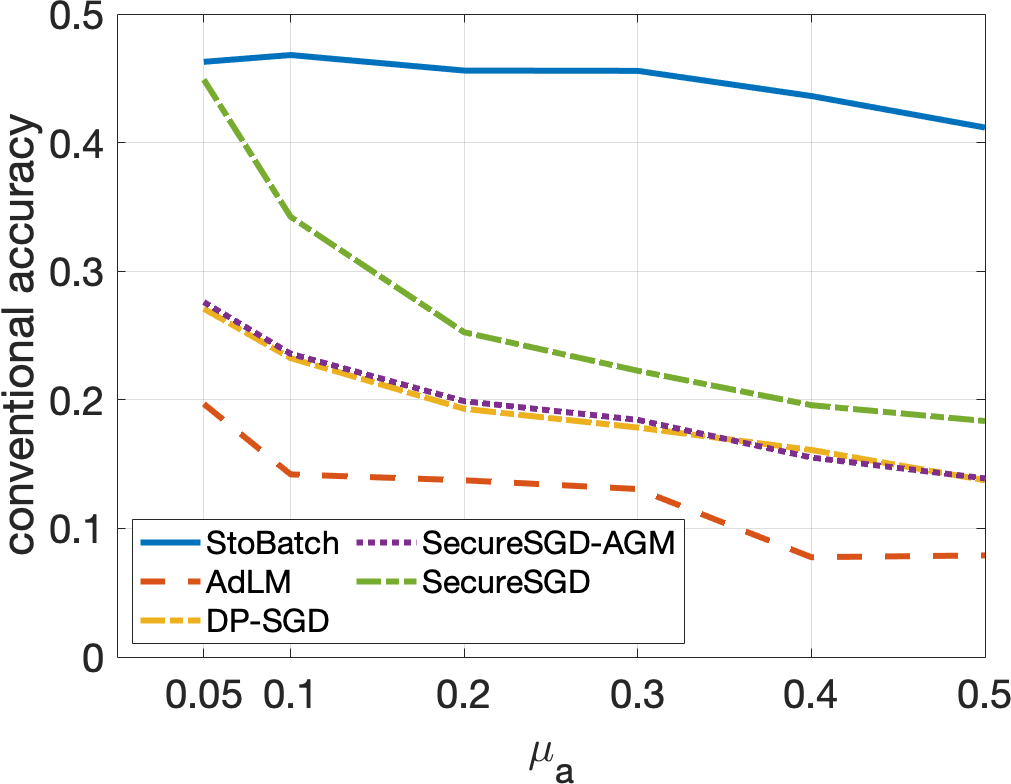} & \includegraphics[width=2.3in]{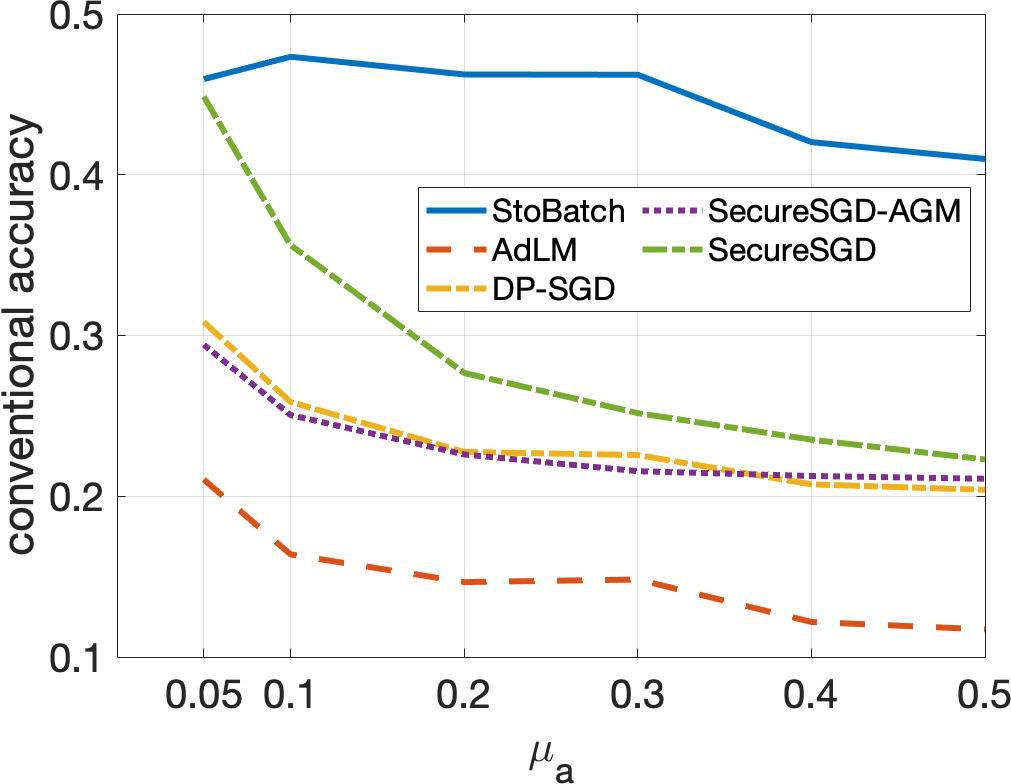} \\ [0.0cm]
\mbox{(a) I-FGSM attacks} & \mbox{(b) FGSM attacks} \\ [0.0cm] \includegraphics[width=2.3in]{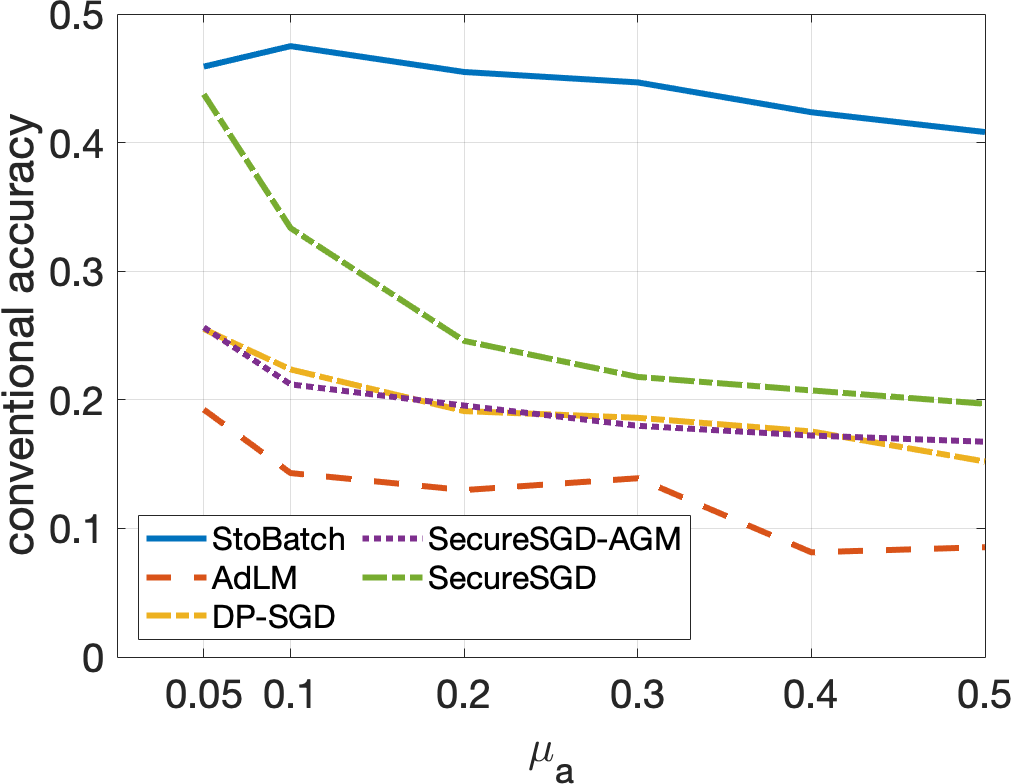} & \includegraphics[width=2.3in]{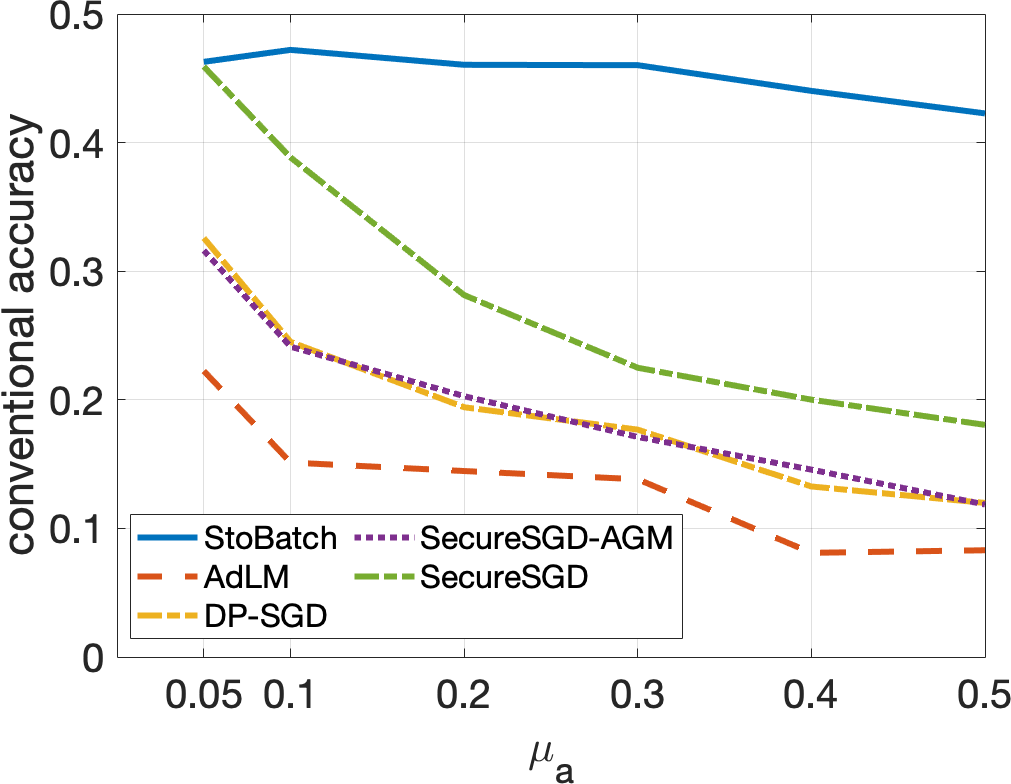} \\ [0.0cm] \mbox{(c) MIM attacks} & \mbox{(d) MadryEtAl attacks}
\end{array}$
\caption{Conventional accuracy on the CIFAR-10 dataset given $\mu_a$ ($\epsilon = 2$, tight DP protection) and $T_a = 3$.}
\label{CIFARAttack02Full}
\end{figure*}

\begin{figure*}[h]
\centering
$\begin{array}{c@{\hspace{0.1in}}c@{\hspace{0.1in}}c@{\hspace{0.1in}}c}
\includegraphics[width=2.3in]{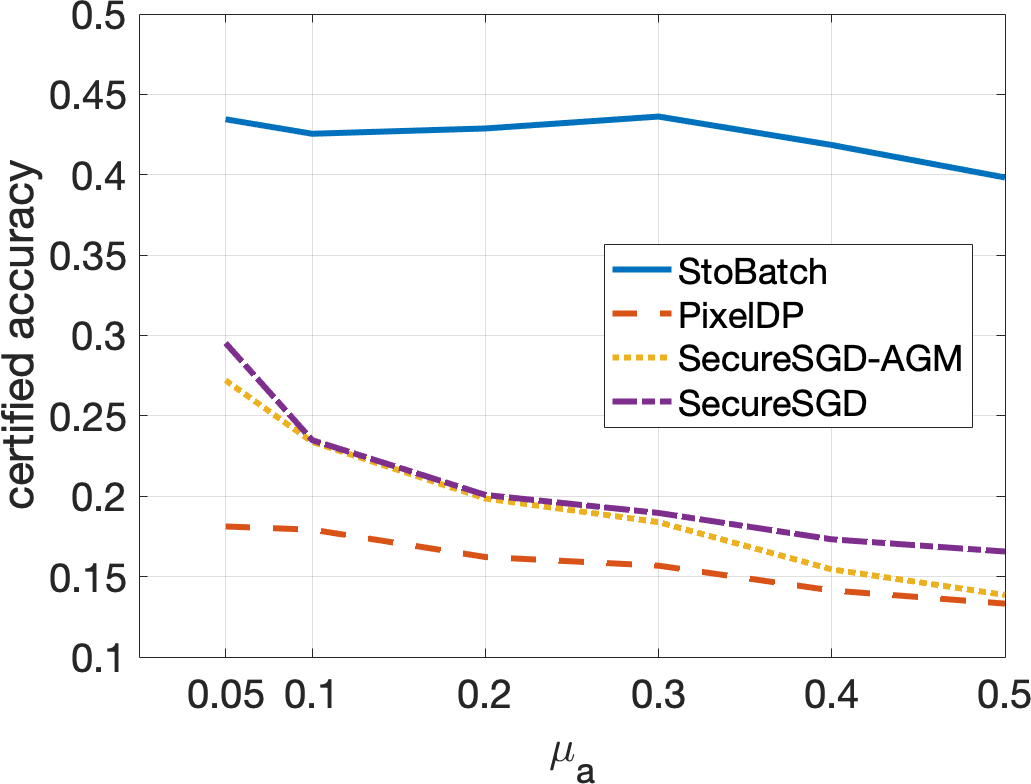} & \includegraphics[width=2.3in]{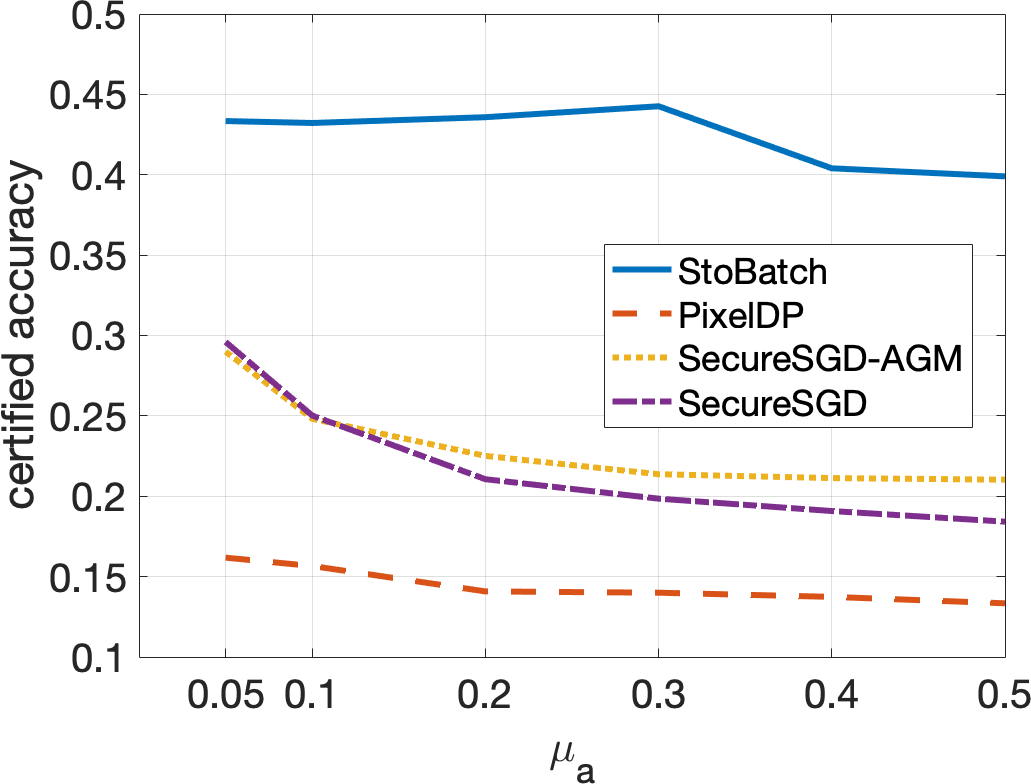} \\ [0.0cm]
\mbox{(a) I-FGSM attacks} & \mbox{(b) FGSM attacks}  \\ [0.0cm] \includegraphics[width=2.3in]{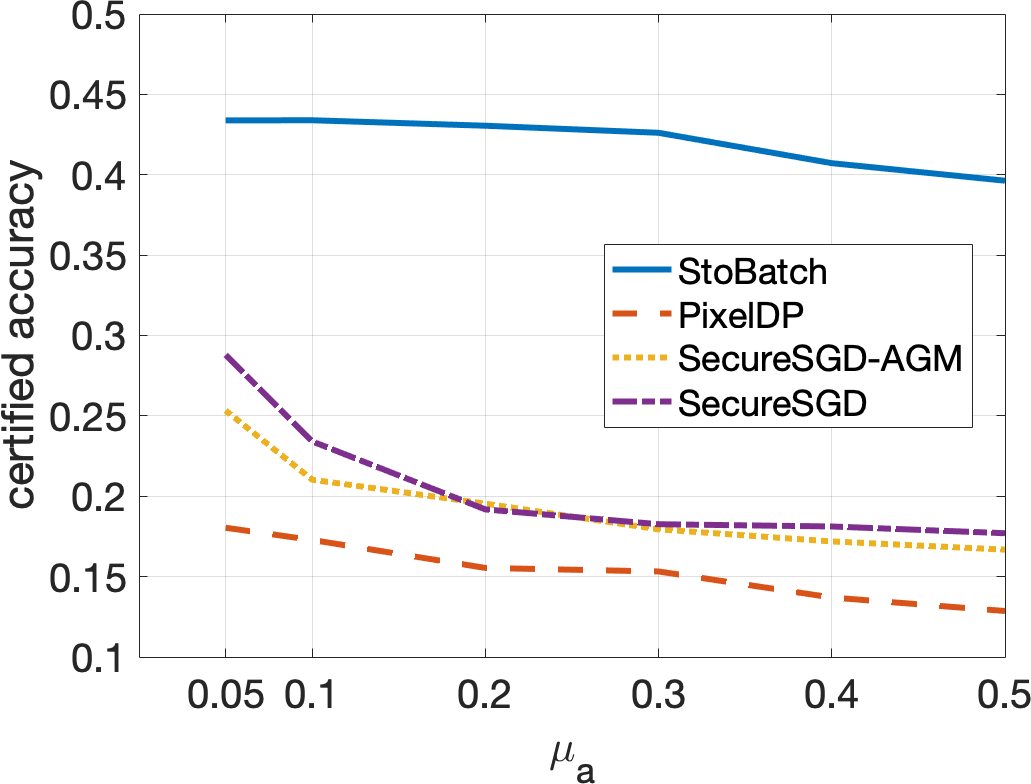} & \includegraphics[width=2.3in]{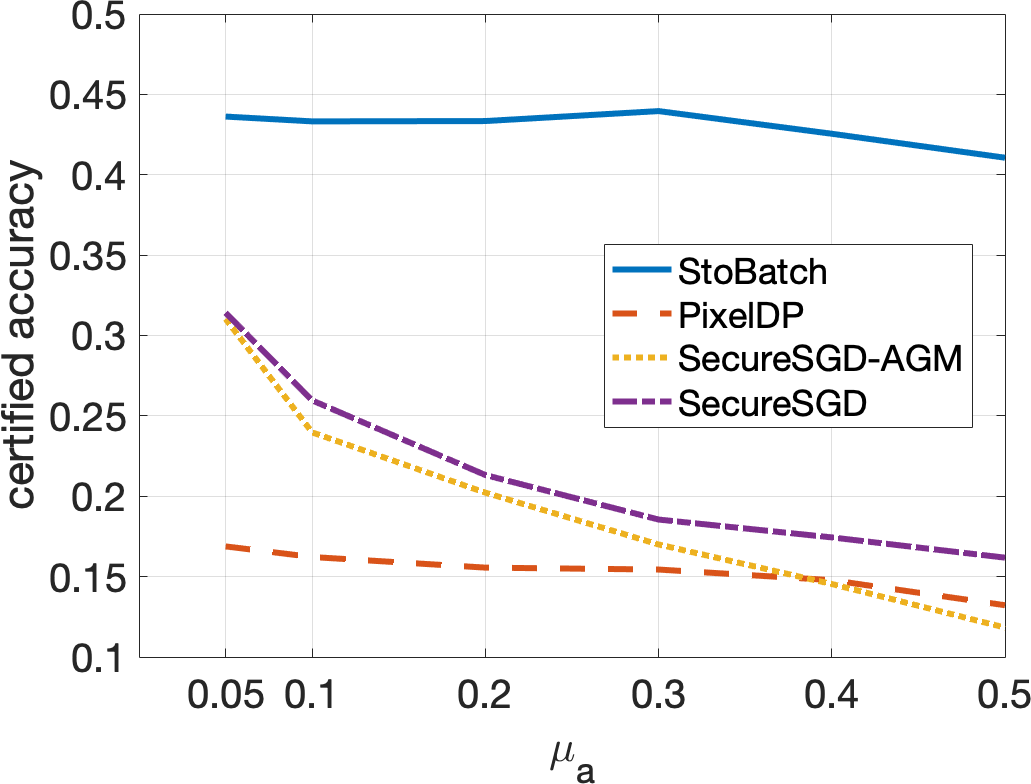} \\ [0.0cm]  \mbox{(c) MIM attacks} & \mbox{(d) MadryEtAl attacks}
\end{array}$
\caption{Certified accuracy on the CIFAR-10 dataset. $\epsilon$ is set to 2 (tight DP protection) and and $T_a = 3$.}
\label{CIFAR2Full}
\end{figure*}

$\bullet$ The accuracy of our model is consistent given different attacks with different adversarial perturbations $\mu_a$ under a rigorous DP protection ($\epsilon = 2.0$), compared with baseline approaches (Figure \ref{CIFARAttack02Full}). In fact, when the attack size $\mu_a$ increases from 0.05 to 0.5, the conventional accuracies of the baseline approaches are remarkably reduced, i.e., a drop of 25.26\% on average given the most effective baseline approach, SecureSGD. Meanwhile, there is a much smaller degradation (4.79\% on average) in terms of the conventional accuracy observed in our StoBatch model. Our model also achieves better accuracies compared with baseline approaches in all cases ($p < 8.2e-10$).
Figure \ref{CIFAR2Full} further shows that our StoBatch model is more accurate than baseline approaches (i.e., $\epsilon_r$ is set to 0.1 in PixelDP) in terms of certified accuracy in all cases, with a tight privacy budget set to 2.0 ($p < 2.04e-18$). We register an improvement of 21.01\% in our StoBatch model given the certified accuracy over SecureSGD model, which is the most effective baseline approach ($p < 2.04e-18$).

\begin{figure*}[h]
\centering
$\begin{array}{c@{\hspace{0.1in}}c@{\hspace{0.1in}}c@{\hspace{0.1in}}c}
\includegraphics[width=2.3in]{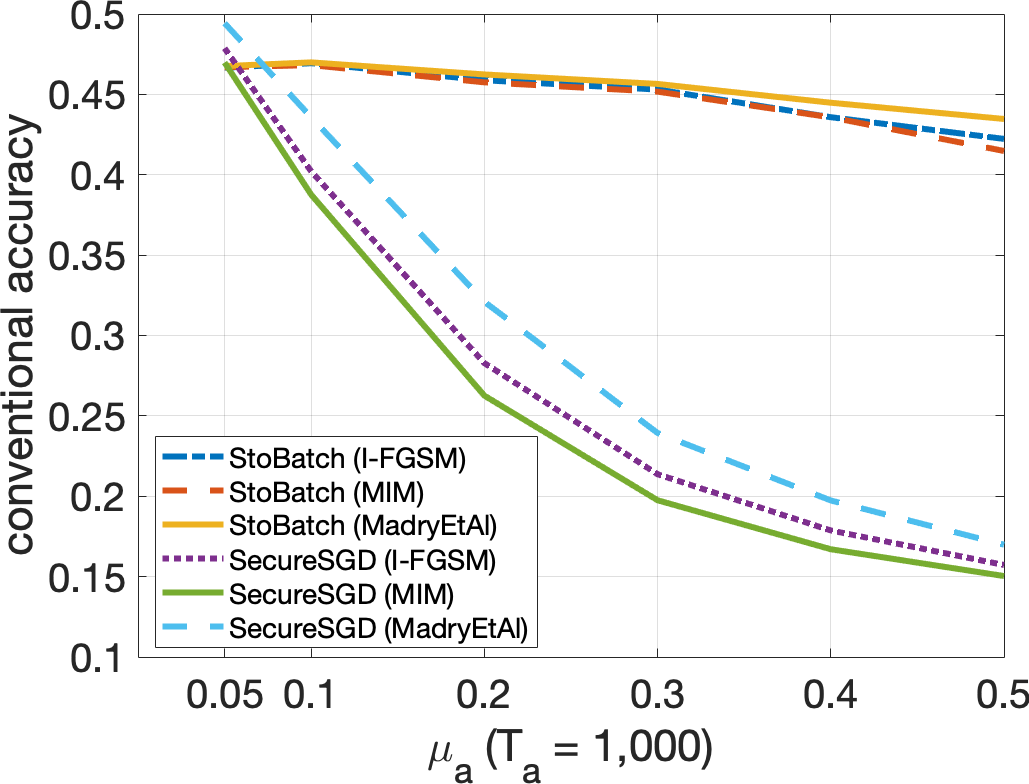} & \includegraphics[width=2.3in]{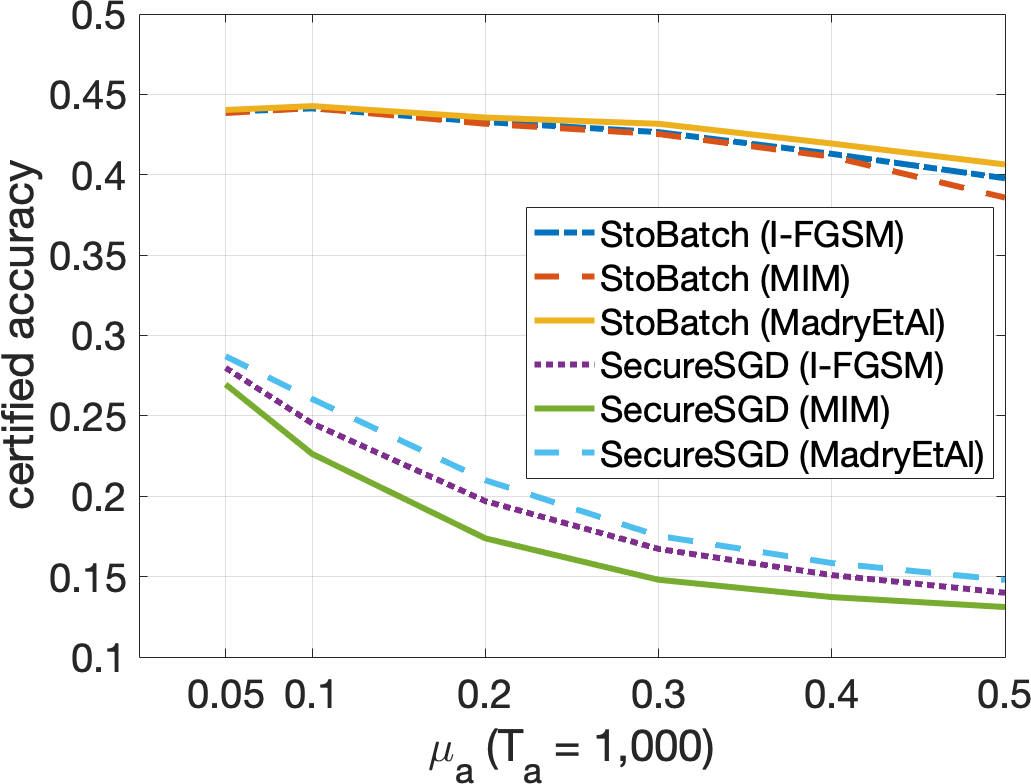} \\ [0.0cm] 
\mbox{(a) Conventional Accuracy ($T_a = 1,000$)} & \mbox{(b) Certified Accuracy ($T_a = 1,000$)} \\ [0.0cm]
\includegraphics[width=2.3in]{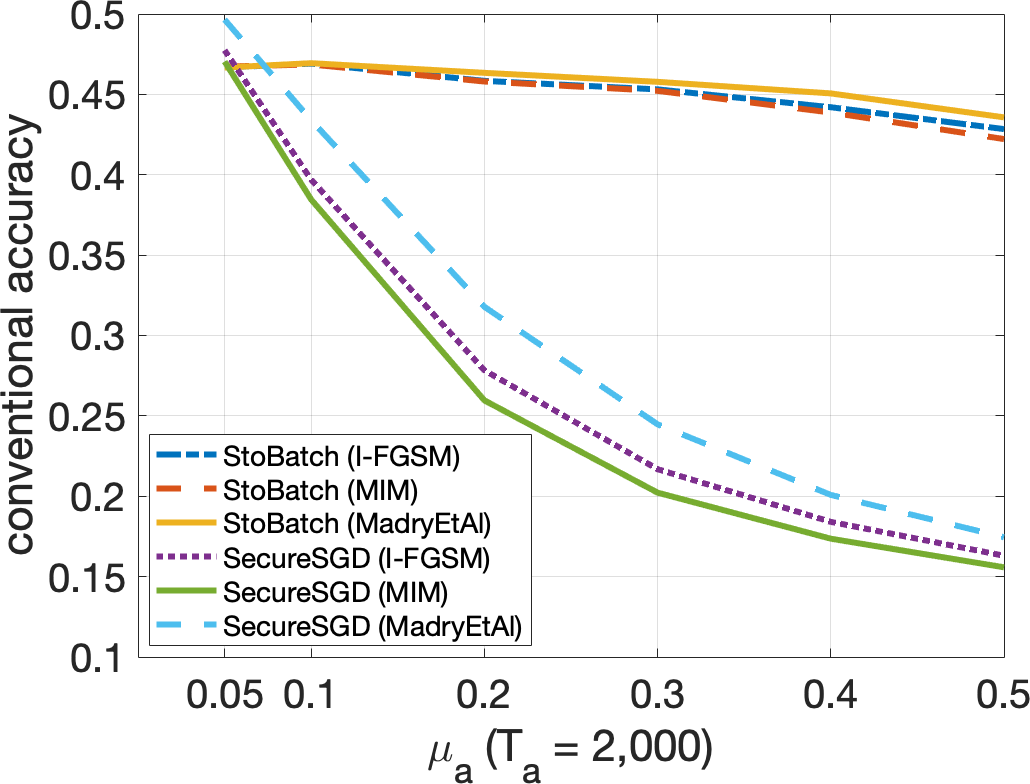} & \includegraphics[width=2.3in]{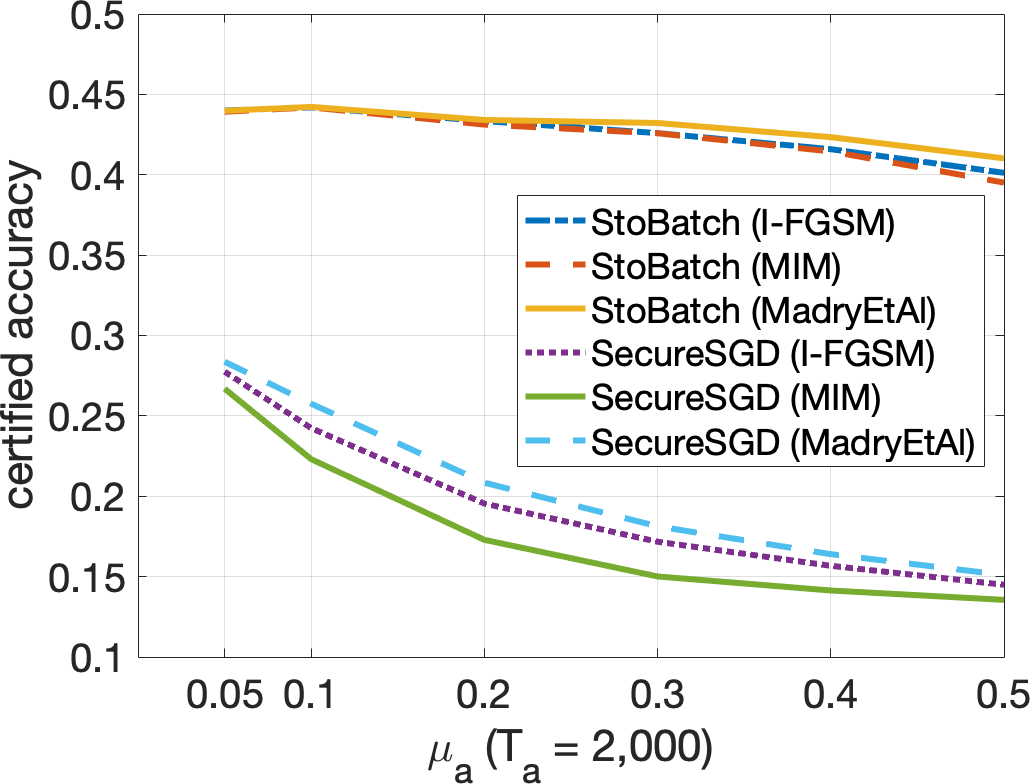} \\ [0.0cm] 
\mbox{(c) Conventional Accuracy ($T_a = 2,000$)} & \mbox{(d) Certified Accuracy ($T_a = 2,000$)}
\end{array}$
\caption{Accuracy on the CIFAR-10 dataset, under Strong Iterative Attacks ($T_a = 1,000$; $2,000$). $\epsilon$ is set to 2 (tight DP protection).}
\label{StoBatchCIFAR}
\end{figure*}

\textbf{Scalability under Strong Iterative Attacks.} First, we scale our model in terms of \textit{adversarial training} in the CIFAR-10 dataset, in which the number of iterative attack steps is increased from $T_\mu = 3$ to \textit{$T_\mu$ = 200 in training}, and up to \textit{$T_a$ = 2,000 in testing}. Note that the traditional iterative batch-by-batch DP adversarial training (Alg. \ref{DPAT}) is nearly infeasible in this setting, taking over 30 days for one training with 600 epochs. Thanks to the parallel and distributed training, our StoBatch only takes $\approxeq$ 3 days to finish the training. More importantly, our StoBatch achieves consistent conventional and certified accuracies under strong iterative attacks with $T_a = 1,000$, compared with the best baseline, i.e., SecureSGD (Figure \ref{StoBatchCIFAR}). Across attack sizes $\mu_a \in \{0.05, 0.1, 0.2, 0.3, 0.4, 0.5\}$ and steps $T_a \in \{100, 500, 1000, 2000\}$, on average, our StoBatch achieves 44.87$\pm$1.8\% and 42.18$\pm$1.8\% in conventional and certified accuracies, compared with 29.47$\pm$12.5\% and 20$\pm$6.1\% of SecureSGD ($p < 1.05e-9$).

\begin{figure*}[h]
\centering
$\begin{array}{c@{\hspace{0.1in}}c@{\hspace{0.1in}}c@{\hspace{0.1in}}c}
\includegraphics[width=2.3in]{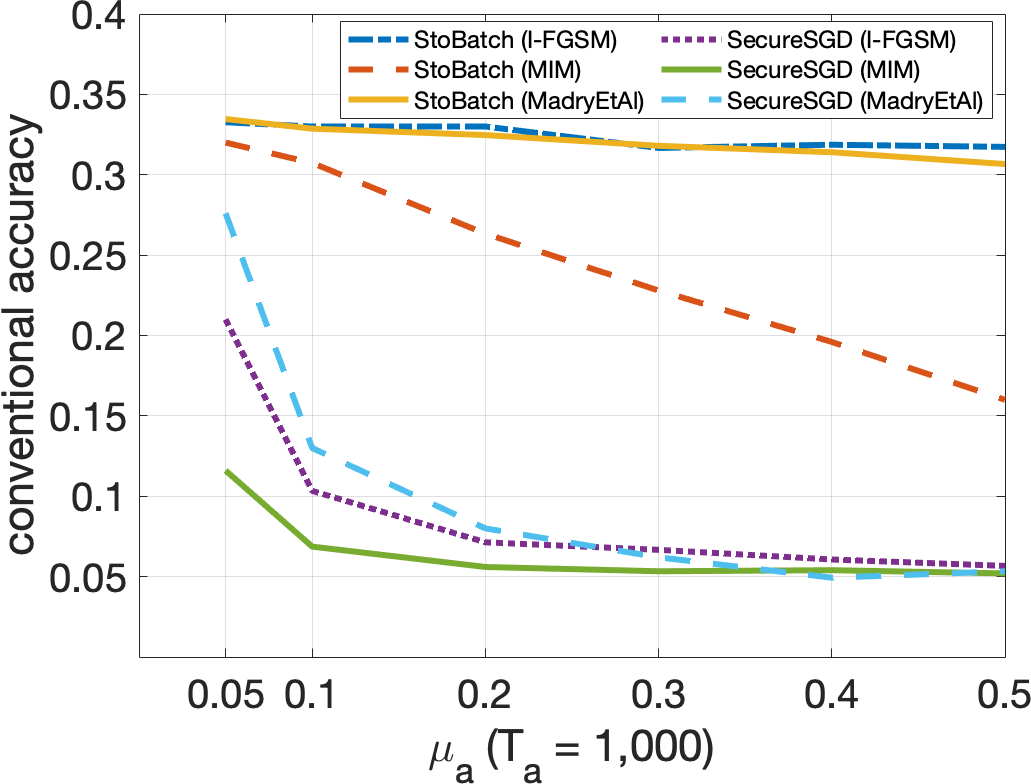} & \includegraphics[width=2.3in]{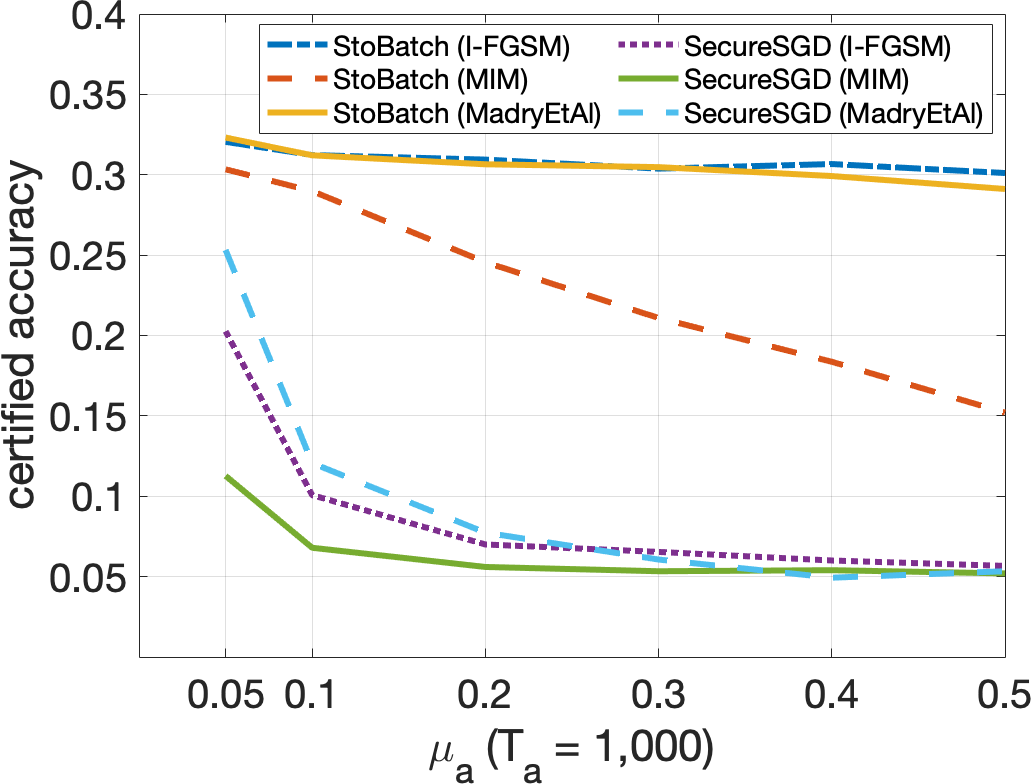} \\ [0.0cm] 
\mbox{(a) Conventional Accuracy ($T_a = 1,000$)} & \mbox{(b) Certified Accuracy ($T_a = 1,000$)} \\ [0.0cm]
\includegraphics[width=2.3in]{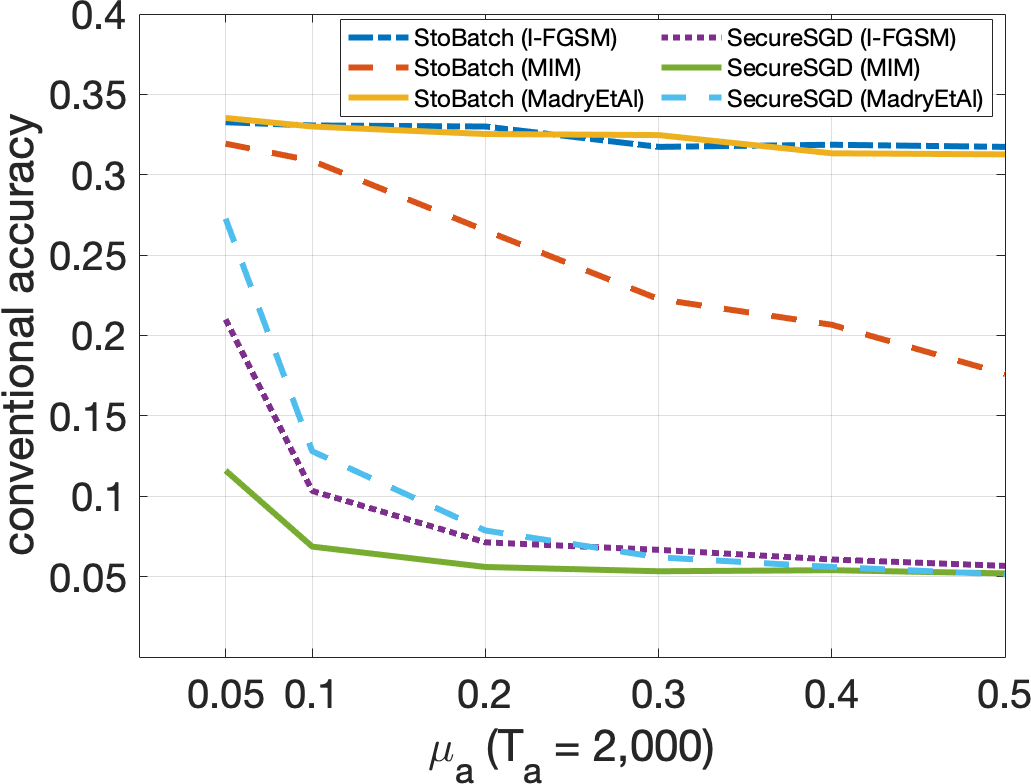} & \includegraphics[width=2.3in]{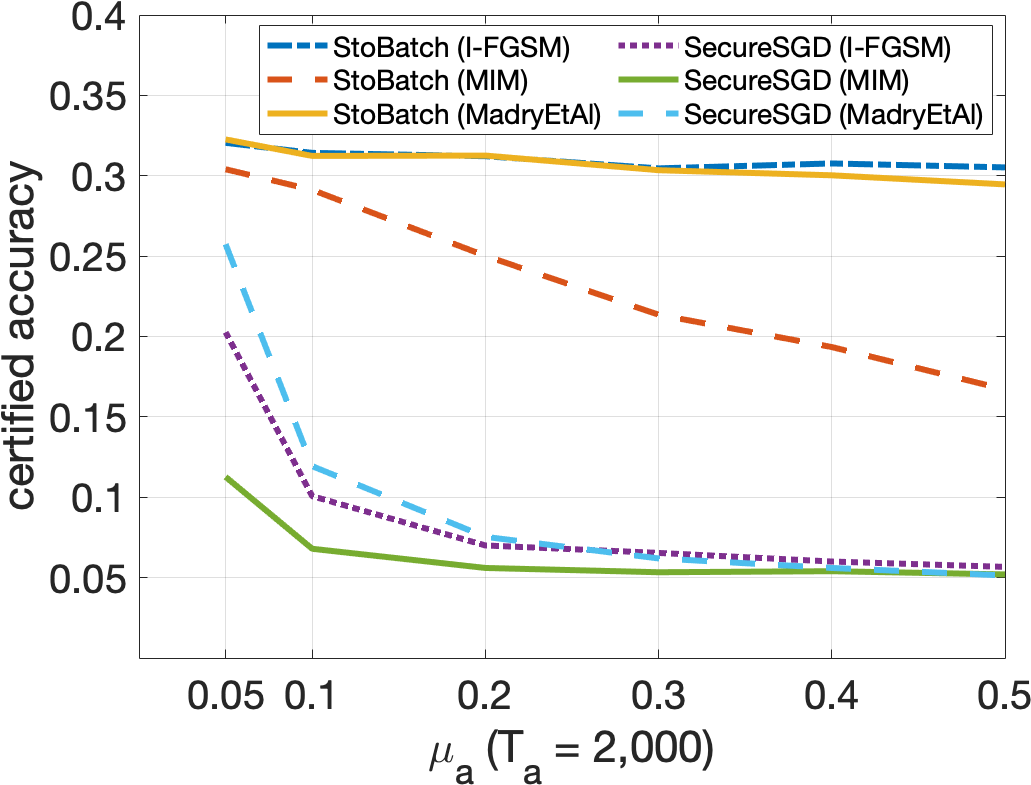} \\ [0.0cm] 
\mbox{(c) Conventional Accuracy ($T_a = 2,000$)} & \mbox{(d) Certified Accuracy ($T_a = 2,000$)}
\end{array}$
\caption{Accuracy on the Tiny ImageNet dataset, under Strong Iterative Attacks ($T_a = 1,000$; $2,000$). $\epsilon$ is set to 5.}
\label{StoBatchImageNet}
\end{figure*}

$\bullet$ We achieve a similar improvement over the \textbf{Tiny ImageNet}, i.e., following \citep{hendrycks2018benchmarking}, with a ResNet18 model, i.e., \textit{a larger dataset on a larger network} (Figure \ref{StoBatchImageNet}). On average, across attack sizes $\mu_a \in \{0.05, 0.1, 0.2, 0.3, 0.4, 0.5\}$ and steps $T_a \in \{100, 500, 1000, 2000\}$, our StoBatch achieves 29.78$\pm$4.8\% and 28.31$\pm$1.58\% in conventional and certified accuracies, compared with 8.99$\pm$5.95\% and 8.72$\pm$5.5\% of SecureSGD ($p < 1.55e-42$). 

\textbf{Key observations:} \textbf{(1)} Incorporating ensemble adversarial learning into DP preservation, tightened sensitivity bounds, a random perturbation size $\mu_t$ at each training step, and composition robustness bounds in both input and latent spaces does enhance the consistency, robustness, and accuracy of DP model against different attacks with different levels of perturbations. These are key advantages of our mechanism; \textbf{(2)} As a result, our StoBatch model outperforms baseline algorithms, 
in terms of conventional and certified accuracies in most of the cases. It is clear that existing DP-preserving approaches have not been designed to withstand against adversarial examples; and
\textbf{(3)} Our StoBatch training can help us to scale our mechanism to larger DP DNNs and datasets with distributed adversarial learning, without affecting the model accuracies and DP protections.

\begin{thebibliography}{63}
\providecommand{\natexlab}[1]{#1}
\providecommand{\url}[1]{\texttt{#1}}
\expandafter\ifx\csname urlstyle\endcsname\relax
  \providecommand{\doi}[1]{doi: #1}\else
  \providecommand{\doi}{doi: \begingroup \urlstyle{rm}\Url}\fi

\bibitem[Abadi et~al.(2016)Abadi, Chu, Goodfellow, McMahan, Mironov, Talwar,
  and Zhang]{Abadi}
Abadi, M., Chu, A., Goodfellow, I., McMahan, H.~B., Mironov, I., Talwar, K.,
  and Zhang, L.
\newblock Deep learning with differential privacy.
\newblock \emph{arXiv:1607.00133}, 2016.

\bibitem[Abadi et~al.(2017)Abadi, Erlingsson, Goodfellow, McMahan, Mironov,
  Papernot, Talwar, and Zhang]{abadi2017protection}
Abadi, M., Erlingsson, U., Goodfellow, I., McMahan, H.~B., Mironov, I.,
  Papernot, N., Talwar, K., and Zhang, L.
\newblock On the protection of private information in machine learning systems:
  Two recent approches.
\newblock In \emph{2017 IEEE 30th Computer Security Foundations Symposium
  (CSF)}, pp.\  1--6. IEEE, 2017.

\bibitem[Abbasi \& Gagn{\'{e}}(2017)Abbasi and
  Gagn{\'{e}}]{DBLP:journals/corr/AbbasiG17}
Abbasi, M. and Gagn{\'{e}}, C.
\newblock Robustness to adversarial examples through an ensemble of
  specialists.
\newblock \emph{CoRR}, abs/1702.06856, 2017.
\newblock URL \url{http://arxiv.org/abs/1702.06856}.

\bibitem[Apostol(1967)]{Apostol}
Apostol, T.
\newblock \emph{Calculus}.
\newblock John Wiley \& Sons, 1967.

\bibitem[Arfken(1985)]{tagkey1985}
Arfken, G.
\newblock In \emph{Mathematical Methods for Physicists (Third Edition)}.
  Academic Press, 1985.

\bibitem[Baidu(2020)]{FedCubeBaidu}
Baidu.
\newblock Fedcube, 2020.
\newblock URL \url{http://fedcube.baidu.com/}.

\bibitem[Balle \& Wang(2018)Balle and Wang]{pmlr-v80-balle18a}
Balle, B. and Wang, Y.-X.
\newblock Improving the {G}aussian mechanism for differential privacy:
  Analytical calibration and optimal denoising.
\newblock In Dy, J. and Krause, A. (eds.), \emph{Proceedings of the 35th
  International Conference on Machine Learning}, volume~80 of \emph{Proceedings
  of Machine Learning Research}, pp.\  394--403, Stockholmsmässan, Stockholm
  Sweden, 10--15 Jul 2018. PMLR.
\newblock URL \url{http://proceedings.mlr.press/v80/balle18a.html}.

\bibitem[Carlini \& Wagner(2017)Carlini and Wagner]{7958570}
Carlini, N. and Wagner, D.
\newblock Towards evaluating the robustness of neural networks.
\newblock In \emph{2017 IEEE Symposium on Security and Privacy (SP)}, pp.\
  39--57, May 2017.
\newblock \doi{10.1109/SP.2017.49}.

\bibitem[Chatzikokolakis et~al.(2013)Chatzikokolakis, Andr{\'e}s, Bordenabe,
  and Palamidessi]{Chatzikokolakis}
Chatzikokolakis, K., Andr{\'e}s, M.~E., Bordenabe, N.~E., and Palamidessi, C.
\newblock Broadening the scope of differential privacy using metrics.
\newblock In De~Cristofaro, E. and Wright, M. (eds.), \emph{Privacy Enhancing
  Technologies}, pp.\  82--102, 2013.

\bibitem[Cisse et~al.(2017)Cisse, Bojanowski, Grave, Dauphin, and
  Usunier]{pmlr-v70-cisse17a}
Cisse, M., Bojanowski, P., Grave, E., Dauphin, Y., and Usunier, N.
\newblock Parseval networks: Improving robustness to adversarial examples.
\newblock In Precup, D. and Teh, Y.~W. (eds.), \emph{Proceedings of the 34th
  International Conference on Machine Learning}, volume~70 of \emph{Proceedings
  of Machine Learning Research}, pp.\  854--863, International Convention
  Centre, Sydney, Australia, 06--11 Aug 2017.

\bibitem[Cohen et~al.(2019)Cohen, Rosenfeld, and Kolter]{pmlr-v97-cohen19c}
Cohen, J., Rosenfeld, E., and Kolter, Z.
\newblock Certified adversarial robustness via randomized smoothing.
\newblock In Chaudhuri, K. and Salakhutdinov, R. (eds.), \emph{Proceedings of
  the 36th International Conference on Machine Learning}, volume~97 of
  \emph{Proceedings of Machine Learning Research}, pp.\  1310--1320, Long
  Beach, California, USA, 09--15 Jun 2019.

\bibitem[Dong et~al.(2017)Dong, Liao, Pang, Hu, and
  Zhu]{DBLP:journals/corr/abs-1710-06081}
Dong, Y., Liao, F., Pang, T., Hu, X., and Zhu, J.
\newblock Discovering adversarial examples with momentum.
\newblock \emph{CoRR}, abs/1710.06081, 2017.

\bibitem[Dwork \& Roth(2014)Dwork and Roth]{Dwork:2014:AFD:2693052.2693053}
Dwork, C. and Roth, A.
\newblock The algorithmic foundations of differential privacy.
\newblock \emph{Found. Trends Theor. Comput. Sci.}, 9\penalty0
  (3\&\#8211;4):\penalty0 211--407, August 2014.
\newblock ISSN 1551-305X.
\newblock \doi{10.1561/0400000042}.
\newblock URL \url{http://dx.doi.org/10.1561/0400000042}.

\bibitem[Dwork et~al.(2006)Dwork, McSherry, Nissim, and
  Smith]{dwork2006calibrating}
Dwork, C., McSherry, F., Nissim, K., and Smith, A.
\newblock {Calibrating noise to sensitivity in private data analysis}.
\newblock \emph{Theory of Cryptography}, pp.\  265--284, 2006.

\bibitem[Fredrikson et~al.(2015)Fredrikson, Jha, and
  Ristenpart]{Fredrikson:2015:MIA}
Fredrikson, M., Jha, S., and Ristenpart, T.
\newblock Model inversion attacks that exploit confidence information and basic
  countermeasures.
\newblock In \emph{Proceedings of the 22Nd ACM SIGSAC Conference on Computer
  and Communications Security}, CCS '15, pp.\  1322--1333, 2015.
\newblock \doi{10.1145/2810103.2813677}.

\bibitem[Gao et~al.(2017)Gao, Wang, and Qi]{DBLP:journals/corr/GaoWQ17}
Gao, J., Wang, B., and Qi, Y.
\newblock Deepmask: Masking {DNN} models for robustness against adversarial
  samples.
\newblock \emph{CoRR}, abs/1702.06763, 2017.
\newblock URL \url{http://arxiv.org/abs/1702.06763}.

\bibitem[Goodfellow et~al.(2014)Goodfellow, Shlens, and
  Szegedy]{DBLP:journals/corr/GoodfellowSS14}
Goodfellow, I.~J., Shlens, J., and Szegedy, C.
\newblock Explaining and harnessing adversarial examples.
\newblock \emph{CoRR}, abs/1412.6572, 2014.

\bibitem[Goyal et~al.(2017)Goyal, Doll{\'{a}}r, Girshick, Noordhuis,
  Wesolowski, Kyrola, Tulloch, Jia, and He]{DBLP:journals/corr/GoyalDGNWKTJH17}
Goyal, P., Doll{\'{a}}r, P., Girshick, R.~B., Noordhuis, P., Wesolowski, L.,
  Kyrola, A., Tulloch, A., Jia, Y., and He, K.
\newblock Accurate, large minibatch {SGD:} training imagenet in 1 hour.
\newblock \emph{CoRR}, abs/1706.02677, 2017.

\bibitem[Grosse et~al.(2017)Grosse, Manoharan, Papernot, Backes, and
  McDaniel]{DBLP:journals/corr/GrosseMP0M17}
Grosse, K., Manoharan, P., Papernot, N., Backes, M., and McDaniel, P.~D.
\newblock On the (statistical) detection of adversarial examples.
\newblock \emph{CoRR}, abs/1702.06280, 2017.
\newblock URL \url{http://arxiv.org/abs/1702.06280}.

\bibitem[Gu \& Rigazio(2014)Gu and Rigazio]{DBLP:journals/corr/GuR14}
Gu, S. and Rigazio, L.
\newblock Towards deep neural network architectures robust to adversarial
  examples.
\newblock \emph{CoRR}, abs/1412.5068, 2014.
\newblock URL \url{http://arxiv.org/abs/1412.5068}.

\bibitem[Gu et~al.(2017)Gu, Dolan{-}Gavitt, and
  Garg]{DBLP:journals/corr/abs-1708-06733}
Gu, T., Dolan{-}Gavitt, B., and Garg, S.
\newblock Badnets: Identifying vulnerabilities in the machine learning model
  supply chain.
\newblock \emph{CoRR}, abs/1708.06733, 2017.
\newblock URL \url{http://arxiv.org/abs/1708.06733}.

\bibitem[Hendrycks \& Dietterich(2019)Hendrycks and
  Dietterich]{hendrycks2018benchmarking}
Hendrycks, D. and Dietterich, T.
\newblock Benchmarking neural network robustness to common corruptions and
  perturbations.
\newblock In \emph{International Conference on Learning Representations}, 2019.
\newblock URL \url{https://openreview.net/forum?id=HJz6tiCqYm}.

\bibitem[Hosseini et~al.(2017)Hosseini, Chen, Kannan, Zhang, and
  Poovendran]{hosseini2017blocking}
Hosseini, H., Chen, Y., Kannan, S., Zhang, B., and Poovendran, R.
\newblock Blocking transferability of adversarial examples in black-box
  learning systems.
\newblock \emph{arXiv preprint arXiv:1703.04318}, 2017.

\bibitem[Kardan \& Stanley(2017)Kardan and Stanley]{7965897}
Kardan, N. and Stanley, K.~O.
\newblock Mitigating fooling with competitive overcomplete output layer neural
  networks.
\newblock In \emph{2017 International Joint Conference on Neural Networks
  (IJCNN)}, pp.\  518--525, 2017.

\bibitem[Kolter \& Wong(2017)Kolter and
  Wong]{DBLP:journals/corr/abs-1711-00851}
Kolter, J.~Z. and Wong, E.
\newblock Provable defenses against adversarial examples via the convex outer
  adversarial polytope.
\newblock \emph{CoRR}, abs/1711.00851, 2017.
\newblock URL \url{http://arxiv.org/abs/1711.00851}.

\bibitem[Krizhevsky \& Hinton(2009)Krizhevsky and
  Hinton]{krizhevsky2009learning}
Krizhevsky, A. and Hinton, G.
\newblock Learning multiple layers of features from tiny images.
\newblock 2009.

\bibitem[Kurakin et~al.(2016{\natexlab{a}})Kurakin, Goodfellow, and
  Bengio]{DBLP:journals/corr/KurakinGB16}
Kurakin, A., Goodfellow, I.~J., and Bengio, S.
\newblock Adversarial examples in the physical world.
\newblock \emph{CoRR}, abs/1607.02533, 2016{\natexlab{a}}.

\bibitem[Kurakin et~al.(2016{\natexlab{b}})Kurakin, Goodfellow, and
  Bengio]{DBLP:journals/corr/KurakinGB16a}
Kurakin, A., Goodfellow, I.~J., and Bengio, S.
\newblock Adversarial machine learning at scale.
\newblock \emph{CoRR}, abs/1611.01236, 2016{\natexlab{b}}.

\bibitem[Lecun et~al.(1998)Lecun, Bottou, Bengio, and Haffner]{Lecun726791}
Lecun, Y., Bottou, L., Bengio, Y., and Haffner, P.
\newblock Gradient-based learning applied to document recognition.
\newblock \emph{Proceedings of the IEEE}, 86\penalty0 (11):\penalty0
  2278--2324, 1998.
\newblock \doi{10.1109/5.726791}.

\bibitem[Lecuyer et~al.(2018)Lecuyer, Atlidakis, Geambasu, Hsu, and
  Jana]{Lecuyer2018}
Lecuyer, M., Atlidakis, V., Geambasu, R., Hsu, D., and Jana, S.
\newblock Certified robustness to adversarial examples with differential
  privacy.
\newblock In \emph{arXiv:1802.03471}, 2018.
\newblock URL \url{https://arxiv.org/abs/1802.03471}.

\bibitem[Lee \& Kifer(2018)Lee and Kifer]{Lee:2018:CDP:3219819.3220076}
Lee, J. and Kifer, D.
\newblock Concentrated differentially private gradient descent with adaptive
  per-iteration privacy budget.
\newblock In \emph{Proceedings of the 24th ACM SIGKDD International Conference
  on Knowledge Discovery \& Data Mining}, pp.\  1656--1665, 2018.

\bibitem[Li et~al.(2018)Li, Chen, Wang, and
  Carin]{DBLP:journals/corr/abs-1809-03113}
Li, B., Chen, C., Wang, W., and Carin, L.
\newblock Second-order adversarial attack and certifiable robustness.
\newblock \emph{CoRR}, abs/1809.03113, 2018.
\newblock URL \url{http://arxiv.org/abs/1809.03113}.

\bibitem[Liu et~al.(2018)Liu, Ma, Aafer, Lee, Zhai, Wang, and Zhang]{Trojannn}
Liu, Y., Ma, S., Aafer, Y., Lee, W.-C., Zhai, J., Wang, W., and Zhang, X.
\newblock Trojaning attack on neural networks.
\newblock In \emph{25nd Annual Network and Distributed System Security
  Symposium, {NDSS} 2018, San Diego, California, USA, February 18-221, 2018}.
  The Internet Society, 2018.

\bibitem[Madry et~al.(2018)Madry, Makelov, Schmidt, Tsipras, and
  Vladu]{madry2018towards}
Madry, A., Makelov, A., Schmidt, L., Tsipras, D., and Vladu, A.
\newblock Towards deep learning models resistant to adversarial attacks.
\newblock In \emph{International Conference on Learning Representations}, 2018.
\newblock URL \url{https://openreview.net/forum?id=rJzIBfZAb}.

\bibitem[Matyasko \& Chau(2017)Matyasko and Chau]{7965869}
Matyasko, A. and Chau, L.~P.
\newblock Margin maximization for robust classification using deep learning.
\newblock In \emph{2017 International Joint Conference on Neural Networks
  (IJCNN)}, pp.\  300--307, 2017.

\bibitem[McMahan et~al.(2016)McMahan, Moore, Ramage, and
  y~Arcas]{DBLP:journals/corr/McMahanMRA16}
McMahan, H.~B., Moore, E., Ramage, D., and y~Arcas, B.~A.
\newblock Federated learning of deep networks using model averaging.
\newblock \emph{CoRR}, abs/1602.05629, 2016.

\bibitem[Metzen et~al.(2017)Metzen, Genewein, Fischer, and
  Bischoff]{metzen2017detecting}
Metzen, J.~H., Genewein, T., Fischer, V., and Bischoff, B.
\newblock On detecting adversarial perturbations.
\newblock In \emph{Proceedings of 5th International Conference on Learning
  Representations (ICLR)}, 2017.
\newblock URL \url{https://arxiv.org/abs/1702.04267}.

\bibitem[Pang et~al.(2020)Pang, Shen, Zhang, Ji, Vorobeychik, Luo, Liu, and
  Wang]{pang2020tale}
Pang, R., Shen, H., Zhang, X., Ji, S., Vorobeychik, Y., Luo, X., Liu, A., and
  Wang, T.
\newblock A tale of evil twins: Adversarial inputs versus poisoned models.
\newblock In \emph{Proceedings of ACM SAC Conference on Computer and
  Communications (CCS)}, 2020.

\bibitem[Papernot \& McDaniel(2017)Papernot and
  McDaniel]{papernot2017extending}
Papernot, N. and McDaniel, P.
\newblock Extending defensive distillation.
\newblock \emph{arXiv preprint arXiv:1705.05264}, 2017.

\bibitem[Papernot et~al.(2016{\natexlab{a}})Papernot, McDaniel, Jha,
  Fredrikson, Celik, and Swami]{7467366}
Papernot, N., McDaniel, P., Jha, S., Fredrikson, M., Celik, Z.~B., and Swami,
  A.
\newblock The limitations of deep learning in adversarial settings.
\newblock In \emph{2016 IEEE European Symposium on Security and Privacy}, pp.\
  372--387, March 2016{\natexlab{a}}.
\newblock \doi{10.1109/EuroSP.2016.36}.

\bibitem[Papernot et~al.(2016{\natexlab{b}})Papernot, McDaniel, Wu, Jha, and
  Swami]{7546524}
Papernot, N., McDaniel, P., Wu, X., Jha, S., and Swami, A.
\newblock Distillation as a defense to adversarial perturbations against deep
  neural networks.
\newblock In \emph{2016 IEEE Symposium on Security and Privacy (SP)}, pp.\
  582--597, May 2016{\natexlab{b}}.
\newblock \doi{10.1109/SP.2016.41}.

\bibitem[Papernot et~al.(2018)Papernot, Song, Mironov, Raghunathan, Talwar, and
  Erlingsson]{papernot2018scalable}
Papernot, N., Song, S., Mironov, I., Raghunathan, A., Talwar, K., and
  Erlingsson, {\'U}.
\newblock Scalable private learning with pate.
\newblock \emph{arXiv preprint arXiv:1802.08908}, 2018.

\bibitem[Phan et~al.(2016)Phan, Wang, Wu, and Dou]{Phan0WD16}
Phan, N., Wang, Y., Wu, X., and Dou, D.
\newblock Differential privacy preservation for deep auto-encoders: an
  application of human behavior prediction.
\newblock In \emph{AAAI'16}, pp.\  1309--1316, 2016.

\bibitem[Phan et~al.(2017{\natexlab{a}})Phan, Wu, and Dou]{PhanMLJ2017}
Phan, N., Wu, X., and Dou, D.
\newblock Preserving differential privacy in convolutional deep belief
  networks.
\newblock \emph{Machine Learning}, 2017{\natexlab{a}}.
\newblock \doi{10.1007/s10994-017-5656-2}.

\bibitem[Phan et~al.(2017{\natexlab{b}})Phan, Wu, Hu, and Dou]{NHPhanICDM17}
Phan, N., Wu, X., Hu, H., and Dou, D.
\newblock Adaptive laplace mechanism: Differential privacy preservation in deep
  learning.
\newblock In \emph{IEEE ICDM'17}, 2017{\natexlab{b}}.

\bibitem[Phan et~al.(2019)Phan, Vu, Liu, Jin, Dou, Wu, and Thai]{PhanIJCAI}
Phan, N., Vu, M.~N., Liu, Y., Jin, R., Dou, D., Wu, X., and Thai, M.~T.
\newblock Heterogeneous gaussian mechanism: Preserving differential privacy in
  deep learning with provable robustness.
\newblock In \emph{Proceedings of the 28th International Joint Conference on
  Artificial Intelligence (IJCAI'19)}, pp.\  4753--4759, 10--16 August 2019.

\bibitem[Raghunathan et~al.(2018)Raghunathan, Steinhardt, and
  Liang]{DBLP:journals/corr/abs-1801-09344}
Raghunathan, A., Steinhardt, J., and Liang, P.
\newblock Certified defenses against adversarial examples.
\newblock \emph{CoRR}, abs/1801.09344, 2018.
\newblock URL \url{http://arxiv.org/abs/1801.09344}.

\bibitem[Rudin(1976)]{WalterRudin}
Rudin, W.
\newblock \emph{Principles of Mathematical Analysis}.
\newblock McGraw-Hill, 1976.

\bibitem[Salman et~al.(2019)Salman, Yang, Li, Zhang, Zhang, Razenshteyn, and
  Bubeck]{DBLP:journals/corr/abs-1906-04584}
Salman, H., Yang, G., Li, J., Zhang, P., Zhang, H., Razenshteyn, I.~P., and
  Bubeck, S.
\newblock Provably robust deep learning via adversarially trained smoothed
  classifiers.
\newblock \emph{CoRR}, abs/1906.04584, 2019.

\bibitem[Shafahi et~al.(2018)Shafahi, Huang, Najibi, Suciu, Studer, Dumitras,
  and Goldstein]{NIPS2018_7849}
Shafahi, A., Huang, W.~R., Najibi, M., Suciu, O., Studer, C., Dumitras, T., and
  Goldstein, T.
\newblock Poison frogs! targeted clean-label poisoning attacks on neural
  networks.
\newblock In \emph{Advances in Neural Information Processing Systems 31}, pp.\
  6103--6113. 2018.

\bibitem[Shokri \& Shmatikov(2015)Shokri and Shmatikov]{ShokriVitaly2015}
Shokri, R. and Shmatikov, V.
\newblock Privacy-preserving deep learning.
\newblock In \emph{CCS'15}, pp.\  1310--1321, 2015.

\bibitem[{Song} et~al.(2019){Song}, {Shokri}, and
  {Mittal}]{2019arXiv190510291S}
{Song}, L., {Shokri}, R., and {Mittal}, P.
\newblock {Privacy Risks of Securing Machine Learning Models against
  Adversarial Examples}.
\newblock \emph{arXiv e-prints}, art. arXiv:1905.10291, May 2019.

\bibitem[TensorFlow()]{TensorFlowSoftMax}
TensorFlow.
\newblock URL
  \url{https://github.com/tensorflow/tensorflow/blob/r1.4/tensorflow/python/ops/nn_impl.py}.

\bibitem[\text{Operator norm}(2018)]{WikipediaOperatornorm}
\text{Operator norm}.
\newblock Operator norm, 2018.
\newblock URL \url{https://en.wikipedia.org/wiki/Operator\_norm}.

\bibitem[TinyImageNet()]{Imagenet}
TinyImageNet.
\newblock URL \url{https://tiny-imagenet.herokuapp.com}.

\bibitem[Tram{\`e}r et~al.(2017)Tram{\`e}r, Kurakin, Papernot, Boneh, and
  McDaniel]{tramer2017ensemble}
Tram{\`e}r, F., Kurakin, A., Papernot, N., Boneh, D., and McDaniel, P.
\newblock Ensemble adversarial training: Attacks and defenses.
\newblock \emph{arXiv preprint arXiv:1705.07204}, 2017.

\bibitem[Wang et~al.(2016)Wang, Guo, Zhang, II, Xing, Giles, and
  Liu]{DBLP:journals/corr/WangGZOXGL16}
Wang, Q., Guo, W., Zhang, K., II, A. G.~O., Xing, X., Giles, C.~L., and Liu, X.
\newblock Learning adversary-resistant deep neural networks.
\newblock \emph{CoRR}, abs/1612.01401, 2016.

\bibitem[Wu et~al.(2019)Wu, Zhao, Sun, Zhang, Su, Zeng, and
  Liu]{DBLP:journals/corr/abs-1905-12883}
Wu, B., Zhao, S., Sun, G., Zhang, X., Su, Z., Zeng, C., and Liu, Z.
\newblock {P3SGD:} patient privacy preserving {SGD} for regularizing deep cnns
  in pathological image classification.
\newblock In \emph{CVPR}, 2019.

\bibitem[Xie et~al.(2019)Xie, Wu, van~der Maaten, Yuille, and
  He]{Xie_2019_CVPR}
Xie, C., Wu, Y., van~der Maaten, L., Yuille, A.~L., and He, K.
\newblock Feature denoising for improving adversarial robustness.
\newblock In \emph{The IEEE Conference on Computer Vision and Pattern
  Recognition (CVPR)}, June 2019.

\bibitem[Xu et~al.(2017)Xu, Evans, and Qi]{DBLP:journals/corr/XuEQ17}
Xu, W., Evans, D., and Qi, Y.
\newblock Feature squeezing: Detecting adversarial examples in deep neural
  networks.
\newblock \emph{CoRR}, abs/1704.01155, 2017.
\newblock URL \url{http://arxiv.org/abs/1704.01155}.

\bibitem[Xu et~al.(2020)Xu, Shi, Liu, Zhao, and Chen]{ADADP}
Xu, Z., Shi, S., Liu, X.~A., Zhao, J., and Chen, L.
\newblock An adaptive and fast convergent approach to differentially private
  deep learning.
\newblock In \emph{INFOCOM}, 2020.

\bibitem[Yu et~al.(2019)Yu, Liu, Pu, Gursoy, and Truex]{Yu2019}
Yu, L., Liu, L., Pu, C., Gursoy, M., and Truex, S.
\newblock Differentially private model publishing for deep learning.
\newblock In \emph{2019 IEEE Symposium on Security and Privacy (SP)}, pp.\
  326--343, 2019.

\bibitem[Zhang et~al.(2012)Zhang, Zhang, Xiao, Yang, and
  Winslett]{zhang2012functional}
Zhang, J., Zhang, Z., Xiao, X., Yang, Y., and Winslett, M.
\newblock Functional mechanism: regression analysis under differential privacy.
\newblock \emph{PVLDB}, 5\penalty0 (11):\penalty0 1364--1375, 2012.

\end{thebibliography}
\end{document}